\documentstyle[12pt,a4]{article}

\newcommand{\qslash}{\mbox{$\not{\hspace{-0.8mm}q}$}}
\newcommand{\pslash}{\mbox{$\not{\hspace{-0.8mm}p}$}}

\newcommand{\Dslash}{\mbox{$\not{\hspace{-1.1mm}D}$}}
\newcommand{\dslash}{\mbox{$\not{\hspace{-1.1mm}\partial}$}}
\newcommand{\vslash}{\mbox{$\not{\hspace{-0.8mm}v}$}}
\newcommand{\etaslash}{\mbox{$\not{\hspace{-0.8mm}\eta}$}}
\newcommand{\epsslash}{\mbox{$\not{\hspace{-0.8mm}\epsilon}$}}

\newcommand{\kslash}{\mbox{$\not{\hspace{-0.8mm}k}$}}
\newcommand{\lslash}{\mbox{$\not{\hspace{-0.8mm}l}$}}

\newcommand{\nc}{\newcommand}

\nc{\be}{\begin{equation}}
\nc{\ee}{\end{equation}}
\nc{\bea}{\begin{eqnarray}}
\nc{\eea}{\end{eqnarray}}

\begin{document}

\begin{titlepage}
\begin{flushright}
IC/95/15 \\ hep-ph/9502241
\end{flushright}
 \begin{center}
  {\Huge An Introduction to the}

 \vspace*{0.5cm}

  {\Huge  Heavy Quark Effective Theory }

 \vspace*{1cm}

 {\large
  F. Hussain and G. Thompson}

\vspace*{0.5cm}
{\large International Centre for Theoretical Physics, Trieste, Italy\\}
\vspace*{0.5cm}
\vspace*{1cm}

\begin{abstract}
These lecture notes begin with a brief survey of the physics of heavy quark
systems. This discusion motivates the introduction of the Heavy Quark
Effective theory (HQET) which captures a great deal of the intuition
developed. A derivation of the HQET from QCD is presented as well as an
analysis of its special properties. The effective theory can be seen to
amount to a one-dimensional field theory in the quark sector. The heavy
quark flavour- and spin- symmetry of the effective lagrangian is an
offspring of this. Other topics covered include the question of covariance of
the theory, the construction of interpolating fields for the heavy hadron
states, the application of LSZ reduction theorems to determine the (reduced)
number of form factors in flavour changing transitions, a complete
verification of Luke's theorem, plus the matching conditions between QCD and
the HQET beyond tree level.

These are an expanded version of lectures presented during the Trieste
1994 Summer school on High Energy physics.

 \end{abstract}

 \end{center}
\end{titlepage}
\tableofcontents
\section{Introduction}
QCD as the theory of strong interactions has been with us for over
twenty years. It has been remarkably successful in describing high energy
physics. The discovery of asymptotic freedom has allowed
for many perturbative calculations of physical quantities within QCD
when combined with the parton model, such as Drell-Yan
processes, deep inelastic scattering and various scaling relations just to
name a few. These successes notwithstanding QCD has proved to be
intractable in the infrared regime by direct analytic means. For example
we do not have a handle on confinement. One, at
this point, normally needs to employ lattice gauge theory or to pass to
QCD inspired effective theories such as non-linear sigma models or potential
models of various kinds. Put rather blankly-QCD in this regime is ${\bf hard}$.

However, during the last few years there has been a resurging interest
in heavy quark physics within the context of QCD. The reason for this is
that one has been able to extract general principles from particular models of
heavy flavour transitions. In the work of \cite{hmw}
one can see the importance played by the velocity of the heavy quarks (as
opposed to their momemtum); indeed hidden in the analysis is a peaking about a
velocity super selection rule \cite{hkt}. It is to the credit of
Isgur and Wise \cite{iw232} that they were able to extract the physical,
model independent implication, namely, that for infinitely heavy quarks,
velocity is the only important parameter.

Further,

(i) there is a heavy quark flavour symmetry,

(ii)
there is a two-fold spin degeneracy (because
the spin coupling is $\propto 1/m_Q$, which tends to zero), and

(iii)
 at the zero recoil point, or equivalently at
maximum momentum transfer, the elastic transition is absolutely
normalized (following Voloshin and Shifman \cite{vs}).

\noindent Our understanding of the dynamics has not improved. However,
the new symmetries in heavy quark theory give rise to numerous
predictions for free!

The heavy quark effective theory HQET, as it has come to be known,
captures the physics of the heavy quark systems which brings to light
these new symmetries. HQET, as we will see, is an expansion of the QCD
action in inverse powers of the heavy quark mass. Heavy quarks $Q$, for
present purposes will be $(c,b,t)$ with masses $m_{Q}\approx(1.5,5,175)$
Gev. The scale is set by $\Lambda_{QCD} \simeq 330$ Mev and the expansion
parameter is typically $\Lambda_{QCD}/m_{Q}$. See Table 1. While the top
quark seems to be a theorists delight,
the expansion parameter being about $10^{-2}$, it unfortunately
decays far too fast to be useful.

\begin{table}[htbp]
\centering
\begin{tabular}{lr} \hline
spin $0^{-}$ & spin $1^{-}$ \\ \hline\hline
$m_{D}= 1.86$ Gev & $m_{D^{*}} = 2.01$ Gev \\
$D^{0} = c \bar{u}$ or $D^{+} = c\bar{d}$ & $D^{* 0} = c\bar{u}$ or
$D^{* +}= c\bar{d}$ \\ \hline
$m_{B} = 5.28$ Gev & $m_{B^{*}} = 5.32$ Gev \\
$B^{-} = b \bar{u}$ or $B^{0} = b\bar{d}$ & $B^{*-} = b\bar{u}$ or
$B^{*0} = b\bar{d}$ \\ \hline
\end{tabular}
\caption{Charm and bottom mesons.}
\end{table}

Working within the context of lattice theory Eichten and Hill \cite{eh}
had suggested a non-covariant effective action for the infinitely heavy
quark limit of QCD, based on previous work of the related, but somewhat
different, non-relativistic limit of QCD by Feinberg \cite{f}
and by Caswell and Lepage \cite{cl}. This action was covariantised by
Georgi \cite{g} and
subsequently a great deal of effort has been expended on determining the
implications of this theory, both on the lattice and in the continuum,
\cite{iw66}-\cite{grinrev}. One of the important developments
has been the observation by Luke \cite{l} that in the case of heavy meson
transitions the Voloshin-Shifman normalization at zero recoil remains
valid even when $1/m_Q$ corrections to the lowest order effective theory
are taken into account. This no-correction theorem has been generalised
and proven in various ways in the literature \cite{bb1}, \cite{bb2}.

A derivation of the effective theory which also gives a systematic
expansion to any order in $1/m_Q$ was given in
\cite{kt1}, \cite{kt2} and \cite{mannel0}. Previously,
corrections to the infinitely heavy quark effective theory had been
obtained by matching the results to QCD. As in \cite{kt1} the
heavy quark effective theory  is obtained from QCD directly by a series
of Foldy-Wouthuysen type of field redefinitions, the matching, at tree
level is immediate. The same is true for the derivation given in
\cite{g}. That one needs to `improve' the relationship between QCD and
the HQET at one and higher loop level comes about because regularization
does not respect the derivations.

These notes are based on an unpublished manuscript
written with J. G. Koerner and S. Balk and on a talk delivered at the
1992 DESY workshop given by the second author. As the reader will see, we
have also benefited from the review articles of M. Neubert \cite{Neub} and
B. Grinstein \cite{grinrev}. The notes have turned out rather long on
account of the fact
that we attempt to give a rather complete account of the properties of the
heavy quark effective theory  at ${\cal O} (1)$ and ${\cal O} (1/m_Q)$. It has
been our intention to give a critical account of the assumptions that go
into the formulation of the HQET. We have tried to be clear on what is
derived and what is intuited, pinpointing the assumptions that are being
fed into the analysis. Our first aim then is to make precise what we
can, with confidence, say about heavy quark physics. The following
picture emerges.

\subsection{Physics of a Heavy Hadron}
The classical picture of a heavy hadron that one has is in many ways
similar to that of the Hydrogen atom. The typical momentum carried by
the light degrees of freedom (light quarks plus glue) inside
a hadron $\Lambda_{QCD}$ is of the order of the proton mass divided by
three $
  \sim 330$Mev. This is a measurement of how far the light quarks are ``off''
shell or, rather more correctly, this is about what their `constituent'
masses are. Typically, then, the light quarks and their gluonic cloud are
carrying momentum $\Lambda_{QCD}$. In heavy hadrons the `mass' of the heavy
quark $m_{Q}$ is much greater than the typical scale, $m_{Q} \gg
\Lambda_{QCD}$, and the heavy quark is carrying most of the momentum of the
heavy hadron. The interactions of the heavy quark with the light degrees
of freedom will also only change the momentum of the heavy quark by the order
of $\Lambda_{QCD}$, so that the heavy quark is then almost on mass shell.
Indeed one expects $M_{Q} \approx m_{Q} + 0 (\Lambda_{QCD})$.

While the change in momentum of the heavy quark is of order $\Lambda_{QCD}$,
its change in velocity, $\Lambda_{QCD}/m_{Q} \ll 1$, is negligible as the
mass
of the heavy quark goes to infinity. The picture one has then is of the
heavy quark moving with constant velocity and this velocity is that of
the heavy hadron. It is important to emphasise that it is {\em not}
momentum that is being equated but, rather, velocity. In the rest frame
of the heavy hadron the heavy quark is almost at rest; it is only
slightly recoiling from the emission and absorption of soft gluons.
This picture does not depend on the actual value of $m_{Q}$ but just that it
satisfies $m_{Q}\gg \Lambda_{QCD}$. As the mass of the heavy quark is
taken to
be bigger and bigger the recoil is less and less  until ultimately, in the
limit $m_{Q} \rightarrow \infty$, the heavy quark does not recoil at all
from the emission and absorption of soft gluons. In this limit the
heavy quark acts, therefore, like a static colour source. It is clear
that in this limit the binding is independent of the flavour and hence
the difference between the mass of the heavy hadron and the heavy quark,
$\bar{\Lambda}=M_{Q}-m_{Q}$, is a universal, flavour independent, constant.

In many ways we have also just decribed the Hydrogen atom. Take the
proton to be a fundamental particle. The typical momentum imparted to
the proton by the electron and the photon cloud is very much smaller
than the mass of the proton ($m_{p}$). The proton can be taken to be a
static photon source which binds the electron to form the Hydrogen atom.
The fact that the Schr$\ddot{o}$dinger equation for the electron in a
$1/r$ potential describes quantitatively the Hydrogen atom so well is
indicative of the success of this picture. One of the things that is
missed by this `non-relativistic' analysis is the hyperfine splitting of
energy levels. But such corrections are rather small compared to the
enery level, $\Delta E /E \ll 1$. One can derive the Schr$\ddot{o}$dinger
equation from the fully relativistic and interacting Dirac theory for
the Hydrogen atom and systematically incorporate corrections such as the
Thomas term for the spin-orbit coupling.

Likewise for the heavy hadron, it is immaterial, in a first approximation,
what the spin state of the heavy quark is and, in analogy to
the above discussion, one can give a systematic derivation of corrections
to this picture. The corrections will be of the form of a series in
$1/m_{Q}$ with the spin coupling coming in at next to leading order. One
of our aims in these notes is to give a derivation from QCD of this
`effective' theory whose lowest term is analogous to the static proton
action in QED.

We do not need, however, to get into any highly abstract
formal considerations in order to exhibit the features mentioned above
within the context of QCD \cite{grin}. Let us simply take the $m_{Q}
\rightarrow
\infty$ limit of the lowest order propagator and connected three point
function for the heavy quark near its `mass-shell', that is with
$p_{Q}=m_{Q}v + k$, where $v$ is the velocity of the heavy hadron and the
components of $k$ are bounded by $\Lambda_{QCD}$. The first of these behaves
as, \be
\frac{i}{\pslash_{Q}-m_{Q}}= i\frac{(1+\vslash)}{2v.k} + O(k/m_{Q})
 \, . \label{exp1}
\ee
This asymptotic form already exhibits some of the features of the
physical arguments. Recall that QCD with massless quarks has a flavour
symmetry. This symmetry is (softly) broken by the introduction of different
masses for the different flavours. Clearly there is no dependence on the
mass at lowest order in (\ref{exp1}) and so, just as for massless quarks,
there is no dependence, at this order, on the flavour of the quark.

The second important feature is that the lowest order term is diagonal in
spin. The $(1+ \vslash)$ projector picks out the particle state in the
field while $(1-\vslash)$ projects out the anti-quark component of the
field. The rest of the propagator is proportional to the unit matrix and
so does not depend on the spin of either the quark or anti-quark
components.

The three point Greens function (we amputate the gluon leg)
\be
\frac{i}{\pslash_{Q}- m_{Q} }(ig \gamma^{\nu}) \frac{i}{ \pslash_{Q} +
\qslash - m_{Q}}
\ee
with soft gluon momentum $q$ (also bounded by $\Lambda_{QCD}$) goes like
\be
igv^{\nu}\frac{(1+ \vslash )}{2}\frac{i}{v.k}\frac{i}{v.(k+q)} +
O(\Lambda_{QCD}/m_{Q}) \, . \ee
This expression also picks out the quark component and does not depend
on spin to leading order.

To summarise: the heavy quark symmetry is the statement that

\vspace{1cm}
{\bf HEAVY HADRON$_{1}$(spin$_{1}$)} $\equiv$ {\bf HEAVY
HADRON$_{2}$(spin$_{2}$)}
\subsection{Analogies and Differences}
We can get some more information as regards the physics of heavy quark
theory by pushing the analogy with the Hydrogen atom further. At some
point the analogy must breakdown and we will have to come to grips with
that as well.

Recall that the Schr$\ddot{o}$dinger equation makes no reference at all to
the nucleon and so it {\em is} invariant under the symmetry described
above. For the moment let us think of the different isotopes of Hydrogen
(p+e), Deuterium (pn+e) and Tritium (pnn+e), as arising from
different `flavours' of the proton. As far as the electron is concerned
the photonic field is the same regardless of the flavour. Treating
the nuclei as infinitely massive the Schr$\ddot{o}$dinger equation
describes these systems well. The spectra of the three (ignoring reduced
mass corrections) are identical. This coincides with the heavy quark
symmetry which allows one to exchange heavy quarks without effecting the
spectra. Of course the Schr$\ddot{o}$dinger equation is blind to the spin of
the
electron relative to the proton. At this order of approximation this is
the other part of the heavy quark symmetry, namely that the spin of the
heavy quark is immaterial.

What the Schr$\ddot{o}$dinger equation misses is
the hyperfine splitting of the energy levels. A more correct account follows
from an analysis of the Dirac equation. Still treating the nuclei as
very heavy the predictions of the Dirac equation differ from those of
the Schr$\ddot{o}$dinger equation at order $1/m_{p}$. The correction of
interest is a Pauli-term with the typical $\bar{\psi}\sigma^{\mu \nu}F_{\mu
\nu}\psi$ coupling. The corrections can be systematically derived by a
series of Foldy-Wouthuysen transformations. We will apply these to QCD.

Now it is time to mention the major difference between QCD and QED. Field
theoretically we know that $\alpha_{e}$ ($e^{2}/4\pi$) and $\alpha_{g}$
($g^{2}/4\pi$) run. For
large distances $\alpha_{e}$ tends to zero while $\alpha_{g}$ grows. The
reason that one can perform with some reliability the calculation of the
spectra of the Hydrogen atoms is that the coupling constant is small. Now
the typical distance is the Bohr radius $a=h^{2}/(2\pi e)^{2}m_{e} \approx
.5 \times 10^{-10}cm = 5 \times 10^{4}$ fermi so the binding occurs at
``large'' distances compared to the typical length scales of the nucleons
which are of the order of fermi's. However, in QCD although, again,
binding is a long distance phenomena, the coupling constant is large and
perturbation theory is therefore not reliable.

Recall that, in the static proton approximation, the potential for an
electron in the photonic field of the proton depends on distance as
$1/r$. One obtains this behaviour in the Born approximation by
consideration of figure 1.

The corresponding amplitude is
\be
\bar{u}_{P}(p+k)\gamma_{\mu}u_{P}(p) D^{\mu \nu}(k) \bar{u}_{e}(q-k)
\gamma_{\nu}u_{e}(q) \label{coulomb}
\ee
where $p'-p = k$, and $p=m_{P}v$ and $v=(1, \vec{0})$. Apply now the Gordon
decomposition,
\be
\bar{u}_{P}(p+k) \gamma^{\mu}u_{P}(p) = \bar{u}_{P}(p+k)\left(
\frac{(2p+k)^{\mu}}{2m_{P}} + i \frac{\sigma^{\mu \nu}k_{\nu}}{2m_{P}}
\right) u_{P}(p) \label{gordon}
\ee
on the proton side to arrive at
\be
\bar{u}_{P}(p+k)\left(
\frac{(2p+k)^{\mu}}{2m_{P}} + i \frac{\sigma^{\mu \nu}k_{\nu}}{2m_{P}}
\right) u_{P}(p) D_{\mu \sigma}(k) \bar{u}_{e}(q-k)
\gamma^{\sigma}u_{e}(q) \, . \label{scattering}
\ee
In a covariant gauge the photon propagator is
\be
D^{\mu \nu}(k) = \frac{(\eta^{\mu \nu} -k^{\mu}k^{\nu}/k^{2})}{k^{2}}
\, .
\ee
Substituting this into (\ref{coulomb}), and remembering that
$\bar{u}(p+k) \kslash u(p) =0$, one obtains
\be
\bar{u}_{P}(p+k)\gamma_{\mu}u_{P}(p) \frac{\eta^{\mu \nu}}{k^{2}}
\bar{u}_{e}(q-k)\gamma_{\nu}u_{e}(q) \, .
\ee
Now, as $p^{2}=p'^{2}=m_{P}^{2}$, we find that $2m_{P}k_{0} + k_{0}^{2} -
\vec{k}^{2} =0$. In the infinite mass
limit, so as not to have an infinite momentum transfer, one finds that
$k_{0}  \approx \vec{k}^{2}/2m_{P}$.
Notice that in the heavy proton limit, $k/m_{P} \rightarrow 0$, the
Gordon decomposition goes to $\bar{u}_{P}(p+k) v^{\mu}u_{P}(p)$, which is
consistent with our tree level analysis of the vertex in that limit. As
$k_{0} \rightarrow 0$, (\ref{scattering}) reduces to
\be
\frac{2m_{P}}{-\vec{k}^{2}}\bar{u}_{e}(q-k) \gamma^{0} u_{e}(q) + \, \,
O(1/m_{P}) \, . \label{scat}
\ee
This reproduces the usual result of the coulomb scattering of an
electron in a given external electromagnetic field (generated by a heavy
nucleus) $A_{\mu}(x) = (ie/4\pi r, \vec{0})$.

It is only the $0$-component of the gauge field which enters finally so
that one is tempted, as we will be tempted in due course, to ask what
happens in the gauge $A_{0}=0$? Physics is gauge invariant, so that going
to this gauge changes nothing. Indeed the photon propagator in the gauge
$v.A=0$ is
\be
D_{\mu \nu}(k,v) = (\eta_{\mu \nu} - \frac{v_{\mu}k_{\nu} +
v_{\nu}k_{\mu}}{v.k}  + \frac{k_{\mu}k_{\nu}}{(v.k)^{2}}  ). \frac{1}{k^{2}}
\ee
and when sandwiched between the currents only the $\eta_{\mu \nu}$
part survives. The calculation is the same as in the covariant gauge and
one reproduces (\ref{scat}).

A problem arises when one passes to the heavy proton limit first. The
naive limit allows us to replace (\ref{gordon}) with
\be
\bar{u}_{P}(p+k) v^{\mu}u_{P}(p) \label{gord}
\ee
but, in the gauge $v.A=0$, the photon propagator satisfies $v^{\mu}D_{\mu
\nu}(k,v)=0 $, so that at first sight, it seems that there will be no
scattering
at all! The resolution of this puzzle is that in non-covariant gauges spurious
poles in the propagator mix terms of different orders in $1/m_{P}$. Let
us first dispense with the term $\bar{u}_{P}(p+k) \sigma_{\mu \nu}
k^{\nu} u_{P}(p) D^{\mu \sigma}(k,v) \bar{u}_{e}(q-k)
\gamma_{\sigma}u_{e}(q)$ in (\ref{scattering}). By current conservation
and anti-symmetry of $\sigma_{\mu \nu}$ the only part of the propagator
that survives is that proportional to $\eta_{\mu \nu}$. This is the same
as for the covariant gauges and can safely be ignored in the limit. In
passing to the limit (\ref{gord}) we have dropped a term
\be
\bar{u}_{P}(p+k) \frac{k_{\mu}}{2m_{P}}u_{P}(p) \label{missing}
\ee
which is clearly $1/m_{P}$ down. However,
\be
k_{\mu}D^{\mu \nu}(k,v) = \frac{k^{\nu}-v^{\nu}v.k}{(v.k)^{2}} \, .
\ee
We can disgard the $k^{\mu}$ piece by the usual argument, to find
\be
k_{\mu}D^{\mu \nu}(k,v) \sim -\frac{v^{\nu}}{v.k} \approx
-\frac{2m_{P}v^{\nu}}{ \vec{k}^{2}} \, .
\ee
While (\ref{missing}) is indeed $1/m_{P}$ down it may not be neglected
as the photon propagator is not just of order $1$ but there are
components of it that are of order $m_{P}$.

The upshot of this discussion is that one cannot simply apply the heavy
quark limit blindly. We have seen that while one is, at the level that
we are working, allowed to use the covariant gauges one cannot employ
the axial gauge without mixing different orders in the heavy quark
(proton) mass.

\subsection{Physics of Flavour Changing Transitions}
We will be interested, in these notes, in calculating the flavour
changing electro-weak transitions of one heavy hadron into another. The
transitions of prime interest will be where the heavy quark in the first
interacts with the electro-weak particle to flavour change into the
heavy quark of the second. The light quarks will essentially be
spectators. The reason for this is that in the heavy mass limit the heavy
quark, as we have seen, is essentially on-shell and acts as a colour
source for the light degrees of freedom. In particular the spin of the
heavy quark decouples from the dynamics. Thus the dynamics of the weak
transition of a heavy hadron are essentially determined by the point like
interaction of the weak current with the heavy quark. The picture which
emerges of the transition is as follows.

In the infinite mass limit in the rest frame of the heavy hadron, the heavy
quark $Q$ is at rest and is surrounded by the light cloud with no spin
interaction between them. In the transition the heavy quark $Q$ emits a
$W$ meson and becomes another heavy quark $Q'$ moving with some velocity
(the velocity of the final heavy hadron). The light cloud has thus to
adjust its velocity to keep up with the heavy quark $Q'$ in forming the
new hadron. It is this adjustment, or overlap, of the light degrees of
freedom which gives rise to the form factors. We immediately see from
this simple, atomic physics, picture the physical reason why the unique
form factor in heavy meson decays, for example, is normalised to one at
zero recoil (or $q^{2}_{max}$). At this kinematical point, the daughter
heavy quark is produced at rest in the initial rest frame. It is clear
that nothing has changed as far as the light degrees of freedom are
concerned because there is no flavour dependence in the static colour
force due to the heavy quark and the light degrees of freedom do not feel
the effect of the change in the heavy quark mass (both are infinite)
unless it moves. Thus there is a complete overlap of the light
wavefunctions before and after the transition at this kinematical point
and hence the form factor is one. Put bluntly there is no `dynamics' in the
transition at this point.

\section{Effective Theory to Lowest Order}
It is our aim in this section to derive an effective field theory from QCD
which captures the essence of the heavy mass limit.
 In this limit the mass of the heavy hadron, the bound
state\footnote{Subsequently, unless otherwise stated, whenever we say
the heavy particle (or heavy hadron) we mean the bound state containing
the heavy quark.
$M_{Q}$ is always the mass of the Q-hadron while $m_{Q}$ is the Q quark
mass.}, is approximately the mass of its heavy constituents, but it is
 important to remember that these do not make up the total momentum of
the heavy  hadron; although taken to be nearly massless, the light
degrees of freedom are allowed to carry off momentum. One is tempted
to say that, if there is just one heavy quark inside the hadron, the 4-velocity
of
the bound state is the 4-velocity of
its heavy quark. This is not quite true, but something can be said about
the heavy quarks 4-momentum projected in the direction of flight of the
heavy hadron, as will be made concrete below.

\subsection{Basics of the Effective Theory}
Georgi \cite{g} describes the situation rather vividly. If the bound state,
made up of a light quark system and one heavy quark, is moving with velocity
$v^{\mu}$ $(v^{2}=1, v^{0}>0)$, then the $4-$momentum of the bound state
is
\be
P_{Q}^{\mu} = M_{Q}v^{\mu} = m_{Q}v^{\mu} + k^{\mu} \, , \label{eq:mom}
\ee
where $M_{Q}$ is the mass of the bound state, essentially the same as the
mass of
the heavy quark $m_{Q}$ (as indicated by the second equality). Clearly
$k^{\mu} = (M_{Q}-m_{Q})v^{\mu}$, which in the weak binding limit is
$k^{\mu} = \Sigma_{q}m_{q}v^{\mu} $ where $m_{q}$ is the light quark
mass.  If the
bound state undergoes scattering into a new state of the same heavy
quark, with momentum
\be
P_{Q}'^{\mu} = m_{Q}v'^{\mu} + k'^{\mu} \, ,
\ee
($v'^{2}=1$), the momentum transfer is $m_{Q}(v-v')^{\mu}+
(k-k')^{\mu}$. In the limit as $m_{Q} \rightarrow \infty$, for fixed
$k^{\mu} -
k'^{\mu}$, then necessarily $v'^{\mu}=v^{\mu}$ to keep the momentum
transfer finite. Conversely, to alter the velocity of the heavy hadron
takes an infinite amount of $4-$momentum. So one is able to follow the
velocity of the heavy hadron. Soft gluons will not alter the heavy
bound state velocity. Only very hard gluons or electroweak transitions can
do this.

To sharpen this picture, consider the heavy quark's behaviour during this
scattering event. The heavy quark is off shell, but as the
physical mass is (in the limit $m_{Q} \rightarrow \infty$) identifiable
with the heavy quark mass,
the heavy quark cannot be `far' off shell. It carries most of the four
momentum of the system. Decompose the hadron $4-$momentum, before the
scattering, as a sum of the heavy quark $4-$momentum $p_{Q}$ and that
carried by the light degrees of freedom $p_{q}$
\be
P_{Q}=m_{Q}v + k = p_{Q} + p_{q} \, .
\ee
The heavy quark four momentum is then $p_{Q}=m_{Q}v + (k-p_{q})$.
Repeating the argument above, now for finite $p_{q}$ as well,
establishes that the heavy quark $4$-momentum after the scattering event
is $p'_{Q}=m_{Q}v + (k'-p'_{q})$ ($p'_{q}$ finite).

{}From this we learn that the velocity of the heavy quark is {\em not} fixed,
but rather the projection of the heavy quark's four momentum in the
direction of the four momentum of the hadron is the heavy quark
mass, as for finite $(k-p_{q})$, $v.(k-p_{q})$ is very small
compared to the mass. Thus
\be
v.p_{Q} \simeq m_{Q} \, . \label{eq:mompr}
\ee
The idea then, is to construct an effective field theory that is able to
keep track of heavy hadrons with {\em given} velocity and which also
encodes the fact that the heavy quark momentum satisfies
(\ref{eq:mompr}) ``on shell" .

Georgi \cite{g}, following
Eichten and Hill \cite{eh}, did precisely
this in the limit that the heavy quark masses are infinite. The action
that he finds in this case is
\be
S_Q^0(v) = \int \bar{Q} (i \vslash v.D-m_Q)Q  \, ,
\label{eq:bn}
\ee
for a given heavy quark with a given hadron velocity $v^{\mu}$, and
with the covariant form of the projection (\ref{eq:mompr}) as the
kernel\footnote{ It is
not quite this as we have not rotated out the mass, nor have we summed
over all possible velocities, but the actions are essentially equivalent.},
where
$D$ is the covariant derivative $D_{\mu}=\partial _{\mu}-ig A_{\mu}$. Here
$A=A^a T^a$ and $T^a $ are the usual Hermitian SU(3) Lie algebra
generators. The heavy quark effective theory at this order will be called
HQET$_{0}$.

We have seen, in the introduction, that the heavy quark propagator can be
expanded as
\be
 \frac{i}{\pslash_{Q}-m_{Q}}=i\frac{(1+\vslash)/2}{v.k}+O(k/m_{Q}).
\ee
This is precisely what one gets for the zeroth order term on expanding,
in powers of $O(k/m_{Q})$, the tree-level heavy quark
propagator
\be
\frac{i}{\vslash v.p_{Q}-m_{Q}}
\label{hqprop}
\ee
arising from the effective action (\ref{eq:bn}). In fact one can also
write
\bea
\frac{i}{\vslash
v.p_{Q}-m_{Q}}&=&i\frac{\frac{1}{2}(1+\vslash)}{v.p_{Q}-m_{Q}}
-i\frac{\frac{1}{2}(1-\vslash)}{v.p_{Q}+m_{Q}}\nonumber\\
&=&i\frac{\frac{1}{2}(1+\vslash)}{v.k}
-i\frac{\frac{1}{2}(1-\vslash)}{v.k+2m_{Q}}\,.
\label{hqprop1}
\eea
and then one notes that the heavy quark mass pole is in the first term and as
$m_{Q}\rightarrow\infty$ the second term goes further and further away
from the pole. All the higher order, in $(1/m_{Q})$, terms arise from the
second factor in the propagator (\ref{hqprop1}). Subsequently in Green's
functions when considering terms of zeroth order we shall often take the
heavy quark propagator to consist only of the first term in the above
equation (\ref{hqprop1}).

Similarly, the 3-point function (a heavy quark and a gluon) is given by
\be
G_{\mu}^{(2,1)}(p_{Q},q)=
\frac{i}{\pslash_{Q}-m_{Q}}(ig\gamma^{\nu})
\frac{i}{\pslash_{Q}+\qslash-m_{Q}}\Delta_{\nu\mu}(q)\,,
\ee
where $\Delta_{\nu\mu}(q)$ is the gluon propagator. Now consider soft
gluons, i.e. $q$ is also of order $\Lambda_{QCD}$. Then we can expand
as before to obtain
\be
G_{\mu}^{(2,1)}(p_{Q},q)=
(\frac{1+\vslash}{2})\frac{i}{v.k}(igv^{\nu})(\frac{1+\vslash}{2})
\frac{i}{v.(k+q)}\Delta_{\nu\mu}(q)+O(\Lambda_{QCD}/m_{Q})\,.
\ee

\vspace{1cm}
\noindent \underline{Exercise:} Check the form of the two and three
point functions in the heavy quark limit. Hint:
\be
(\frac{1+\vslash}{2})\gamma^{\nu}(\frac{1+\vslash}{2})=
(\frac{1+\vslash}{2})v^{\nu}(\frac{1+\vslash}{2})\,.
\ee
\vspace{1cm}

This corresponds precisely to the vertex, $i\vslash v^{\nu}$, obtained from
the effective Lagrangian (\ref{eq:bn}), when sandwiched between the
leading order term in the heavy quark propagator (\ref{hqprop1}). Thus in the
limit $m_{Q}\rightarrow\infty$ the
only part of the gluon field which contributes is that along the
$v$-direction. These results can be easily extended to arbitrary tree
level diagrams, provided that we have only one heavy quark and all other
particles carry small momenta.

The theory defined by (\ref{eq:bn}) has some rather remarkable
properties. All of the simple characteristics and enlarged symmetries that the
theory enjoys may be traced back to the fact that the operator $(i\vslash
v.D-m_{Q})$ that appears in (\ref{eq:bn}) depends only on a one-dimensional
derivative and only on one linear combination of the Dirac
matrices, namely $\vslash$. For all intents and purposes this action defines
a one dimensional field theory in the heavy quark sector (whence solvable in
that sector). This should be kept in mind.

\subsection{Flavour Symmetry}
The flavour symmetry at this order, valid in the static limit, may
be seen as follows. The combined action for a $b$ and $c$ quark with hadron
velocities $v_{b}$ and $ v_{c}$ respectively is
\be
S_b^0(v_b) +  S_c^0(v_c) = \int \bar{b} (i\vslash_{b}
v_{b}.D-m_b) b +
\int \bar{c} (i \vslash_{c} v_{c}.D-m_c) c \, \,
  . \label{eq:bn2}
\ee
The flavour mixing symmetry is between $b$ and $c$ quarks in hadrons moving
with
the {\em same} velocity, so that the relevant form of (\ref{eq:bn2}) is
\be
S_b^0(v) +  S_c^0(v) = \int \bar{b}  (i \vslash v.D-m_b)b
 +
\int \bar{c} (i \vslash v.D-m_c)c \, \, , \label{eq:bn3}
\ee
where $v=v_{b}=v_{c}$. The symmetry transformation is
\bea
\delta c & =&  \epsilon \exp{(i \vslash v.x (m_{b}-m_{c}) )}
 b \, ,
\nonumber \\
\delta b & = & - \epsilon \exp{(i \vslash v.x (m_{c}-m_{b}) )} c
\, , \label{eq:tr}
\eea
where $\epsilon$ is an infinitesimal parameter.
If one includes the top quark (or other heavy flavour quarks) there is a
corresponding set of transformations leaving the action invariant.

\vspace{1cm}
\noindent \underline{Exercise:} Derive the corresponding invariance for
$F$ heavy flavours.
\vspace{1cm}

In order
to be rid of the cumbersome exponentials in (\ref{eq:tr}) one may, if one
wishes, define new fields
\be
\hat{Q} = \exp{(i \vslash v.x m_{Q})} Q \, , \label{eq:hf}
\ee
in terms of which the action (\ref{eq:bn}) has the same form but the
mass is set to zero \cite{g}
\be
S_{Q}^0(v) = \int \bar{\hat{Q}} (i \vslash v.D)\hat{Q}  \, . \label{eq:hfa}
\ee
This property that one may define fields in terms of which the mass is
transformed away is characteristic of the one-dimensional first order
operator. Eq.(\ref{eq:tr}) is the infinitesimal form of a more general
global symmetry, here a U(2) symmetry (in general U(F) for F heavy
flavours). Thinking of the $(b,c)$ as an U(2) doublet the action
\be
 \sum_{Q=b,c} \int \bar{\hat{Q}} (i \vslash v.D)\hat{Q}  \,  \label{eq:sumact}
\ee
is invariant under
\bea
 (\hat b,\hat c)\to (\hat b,\hat c) U \, &,&  \,
\left( \begin{array}{c} \bar {\hat b} \\ \bar
{\hat c} \end{array} \right) \to U^{\dag } \left( \begin{array}{c}
\bar {\hat b} \\ \bar
{\hat c} \end{array} \right) \mbox{ , }
\label{eq:utrafo}
\eea
where $U\in $U(2).

\subsection{Particle Properties}
A second characteristic of the one dimensional nature of the field
theory is the fact that one may create `covariant' velocity projection
operators to project out quark and anti-quark states.
Let
\be
P_{\pm}(v) = \frac{1}{2}(1 \pm \vslash ) \, ,
\ee
with the properties
\be
P_{+}^{2}= P_{+} \, , \; \; \; P_{-}^{2} = P_{-} \, , \; \; \;
P_{\pm}P_{\mp}=0  \, , \; \; \; P_{+} + P_{-} = 1 \, .
\ee
Now decompose the heavy quark field as
\be
Q = Q_{+} + Q_{-} \, , \; \; \; Q_{\pm} =
P_{\pm} Q \, ,
\ee
so that $Q_{+}$ satisfies $P_{-}Q_{+}= 0$, which is the Dirac
equation $(\pslash - m_{Q})Q_{+}= 0$ for a quark with momentum
$m_{Q}v^{\mu}$
and $Q_{-}$ satisfies the Dirac equation for an anti-quark
with the same momentum. As operators these annihilate precisely those
states. Note that these projection operators act in much the same way as
the parity projection operators do in the rest frame of a `normal' particle,
which when boosted become Dirac equations \cite{hkt}. In any case the
action (\ref{eq:bn}) decomposes into quark and anti-quark pieces
\be
S_{Q}^0(v) = \int \bar{Q} _{+} (i v.D-m_{Q}) Q_{+} -
\bar{Q} _{-} (i
v.D + m_{Q}) Q_{-} \, . \label{eq:bn+-}
\ee

These projections are also reminiscent of chirality, and do not tell us
about the dynamics of the wave equation itself (except of course that it
has two orthogonal sets of solutions).  The free wave equation for
$Q_{+}$, for example, is
\be
(iv.\partial - m_{Q}) Q_{+} = 0 \, , \label{eq:bn+}
\ee
with the solutions
\be
Q_{+ \alpha}(x) = P_{+ \alpha}\,^{\beta}F_{+ \beta}(x^{\perp})
 e^{-im_{Q}v.x}  \, , \label{eq:sol+}
\ee
where $F_{+}$ is an arbitrary Dirac spinor depending only
on $x^{\perp \mu}= x^{\mu} - v^{\mu}v.x$ which satisfies $x^{\perp}.v=0$,
so that $F_{+ \beta}(x^{\perp})$ is unaffected by $v.\partial$.

Physically, the situation is then somewhat
underdetermined. Certainly the wave function satisfies the Dirac equation
(its `chirality' is fulfilled) but what happens orthogonally to the line
of flight of the hadron is not known\footnote{In previous works we have
pointed
out, rather euphemestically,  that it is the heavy quark `label' that
satisfies the Dirac equation with velocity $v$ and not that the heavy
quark itself has only that velocity. We hope our remarks here clarify
the situation.}. The reason for this is not too surprising. A
one-dimensional field theory can only determine the dynamics in that
direction. There is a similar solution for $\tilde{Q}_{-}$ with similar
conclusions.

An alternative way of saying this is the following. The fact that the
wave equation has such arbitrariness in its solution is an indication
that the action has a rather large invariance. We may as well exhibit
this for (\ref{eq:bn}). Let
\be
Q \rightarrow M(x^{\perp}) Q \, , \; \; \; \;
\bar{Q} \rightarrow  \bar{Q}M^{-1}(x^{\perp}) \, ,
\label{eq:tra}
\ee
with $M$ diagonal in colour space, but a general matrix in spinor space. To
guarantee invariance of the action one only needs that $[ \vslash, M] = 0
$, which is partially solved by $M= P_{+}M_{-} + P_{-}M_{+}$ where $M_{+}$
and $M_{-}$ are functions. These invariances of the action account for
the arbitrariness in the wave functions (\ref{eq:sol+}). A complete
decomposition of the invariances is given in the appendix.

\subsection{Covariance and Spin (I)}
How covariant is the construction thus far? The actions (\ref{eq:bn}) and
(\ref{eq:bn+-}) depend explicitly on a fixed four-vector, so that arbitrary
Lorentz boosts cannot be a symmetry of the action. Having chosen a preferred
$4-$velocity we wish to consider the subgroup of the Lorentz group which
keeps this time-like direction fixed. In the heavy hadron's rest frame
$v^{\mu}= (1,\underline{0})$ the actions are clearly invariant under the
$O(3)$ group of three space rotations. Let a general infinitesimal
Lorentz transformation be parameterised by $\eta_{\mu \nu} + \lambda_{\mu
\nu}$,
where $\eta_{\mu \nu}$ is the metric tensor.

Then the little group is defined to be generated by all those $\lambda_{\mu
\nu}$
which satisfy
\be
\lambda_{\mu}\,^{\nu}v_{\nu} = 0 \, .
\ee
It is easy to see that, for $\lambda_{\mu \nu}$ satisfying this
condition, the heavy quark action is invariant under the transformations
\bea
\delta Q_{\pm}& =& \frac{1}{4}\lambda^{\mu \nu} \gamma_{\mu} \gamma_{\nu}
Q_{\pm} \, , \nonumber \\
\delta \bar{Q}_{\pm} & =& -\frac{1}{4}\bar{Q}_{\pm} \lambda^{\mu \nu}
\gamma_{\mu} \gamma_{\nu} \, , \label{eq:spin}
\eea
with the consistency property that the particle projection operators
$P_{\pm}$ commute with the spin transformation, $[P_{\pm}, \lambda^{\mu
 \nu} \gamma_{\mu} \gamma_{\nu}]= 0 $, so that the transformed fields
remain unambiguously eigenstates of $P_{\pm}$, i.e.
\be
P_{\mp} \delta Q_{\pm} = 0 \, .
\ee

The spin transformations (\ref{eq:spin}) require some explanation. There
is no transformation (Lorentz rotation) of the co-ordinates $x^{\mu}$.
This comes about because the operator $(i\vslash v.D
-m_{Q})$, as we noted before, acts only in the direction of
$v$, so any Lorentz transformation $x^{\mu} \rightarrow x^{\mu} +
\lambda^{\mu}\,_{\nu} x^{\nu}$ with $\lambda_{\mu}\,^{\nu}v_{\nu} = 0$ is by
itself a symmetry of the action. For the fields this translates into
\bea
\delta Q_{\pm}(x)& =& -\lambda^{\mu}\,_{ \nu} x^{\nu} \partial_{\mu}
Q_{\pm}(x) \,,\nonumber\\
\delta \bar{Q}_{\pm}(x)& = &
-\lambda^{\mu}\,_{ \nu} x^{\nu} \partial_{\mu} \bar{Q}_{\pm}(x) \,
,\label{eq:lb}\\
\delta A_{\rho}& = & -\lambda^{\mu}\,_{ \nu} x^{\nu} \partial_{\mu}A_{\rho}
\eea
which is a symmetry of the action.
The two sets of transformations
(\ref{eq:spin}) and (\ref{eq:lb}) together form the conventional Lorentz
transformations for the spinors.

\vspace{1cm}
\noindent \underline{Exercise:} Check the invariance of the heavy quark
action (\ref{eq:bn}) under the above transformation rules.
\vspace{1cm}

Note also that the transformations (\ref{eq:spin}), the solutions given
after (\ref{eq:tra}), and
\be
\delta Q= \gamma_{5} \epsslash ^{\perp}(v)M_{0} Q \, ,
\; \; \; \; \;
\delta \bar{Q} =    \bar{Q} \epsslash
^{\perp}(v) \gamma_{5}
M_{0} \; , \label{eq:symm}
\ee
with $M_0$ a function of $x^{\perp}$ and
$\epsslash^{\perp}(v)=\epsslash - \vslash v.\epsilon $
\footnote{For a general tensor
of rank $n$, $A_{\mu_{1},...,\mu_{n}}$, we set
$A^{\perp}_{\mu_{1},...,\mu_{n}}(v)=
\eta_{\mu_{1}}^{\perp\nu_{1}}(v)...\eta_{\mu_{n}}^{\perp\nu_{n}}(v)%
A_{\nu_{1},...,\nu_{n}}$
with $\eta_{\mu \nu}^{\perp}(v)=\eta_{\mu \nu}^{\perp}-v_{\mu}v_{\nu}$.}
exhaust
the symmetry of the action (\ref{eq:bn}) in four dimensions. Indeed, the
transformation (\ref{eq:spin}) may also be generalized as
\bea
\delta Q & =& \etaslash ^{\perp }(v) \gamma_5 \vslash
Q M_1 \, , \nonumber \\
\delta \bar{Q} & =& -\bar{Q}
\etaslash ^{\perp }(v) \gamma_5 \vslash M_1 \, ,
\eea
where $\eta $ is defined by
$\lambda^{\mu \nu} \gamma_{\mu} \gamma_{\nu} =
\etaslash ^{\perp }(v) \gamma_5 \vslash $ and $M_1$ is a function.

\subsection{Field Choices and Interactions}\label{fci}
As only one component of the gauge field appears in the action
$S_{Q}^{0}(v)$, namely $v.A$, if we pick the gauge $v.A=0$ then the heavy
quarks associated with that hadron velocity decouple from the glue. Such
non-covariant
gauges may be quite difficult to work with, and in any case are of
limited use for they provide no simplification for heavy particles with
other velocities. In fact, as we saw in the introduction,
this gauge can, in certain situations be dangerous\footnote{Though for
properties of transitions at equal
velocity, that is at maximum momentum transfer, such gauges have been very
useful \cite{kt1}. }. We would like to see this decoupling, therefore, in a
more general fashion. Just as it was possible to define
fields for which the mass does not appear in the action (\ref{eq:hfa}),
we can similarly define fields $\tilde{Q}$ and
$\bar{\tilde{Q}}$ for which the gauge interaction is
absent. These are given by
\bea
Q(x) &=&
  W\left[
 {\textstyle{x \atop v}}
   \right]
 \tilde{Q}(x)
  \nonumber \\
\bar{Q}(x) &=& \bar{\tilde{Q}}(x)
  W\left[ {\textstyle{x \atop v}}
   \right] ^{-1}
 \, , \label{eq:un}
\eea
where the Wilson line
\bea
  W\left[
 {\textstyle{x \atop v}}
 \right]
 &=& P exp\left[
 ig \int_{-\infty}^{v\cdot x}ds A\cdot v\right]
 \nonumber
\eea
is a path-ordered exponential, wherein the path is a straight line from
$\infty$ to $x$ along the $v$-direction.

In terms of these fields (\ref{eq:bn}) becomes
\be
S_{Q}^{0}(v) = \int  \bar{\tilde{Q}}   (i\vslash v.\partial -m_{Q}) \tilde{Q}
 \mbox{ . }
\ee
So $S_{Q}^{0}$ represents free heavy quarks in any gauge. Let us explain
how this comes about. To say we can pick a gauge $v.A=0$ is to say that
given an arbitrary gauge field $A$ there exists a gauge transformation,
which takes us to that gauge. Specifically, if
\be
A'=\xi ^{-1}A\xi  + \xi ^{-1}\frac{i}{g}\partial \xi  \, ,  \label{eq:Ap}
\ee
and  $v.A'=0$ then
\be
0 = \xi ^{-1}( v.A  + \frac{i}{g}v.\partial) \xi  \, , \label{eq:path}
\ee
is an equation for $\xi $, solved by
\be
\xi (v.x) =
  W\left[ {\textstyle{x \atop v}}
 \right]
\ee
with the boundary condition that, at $v.x=-\infty$, $\xi $ is the identity
(actually we could just as well take (\ref{eq:path}) to be the defining
equation of the path-ordered exponential). Substituting (\ref{eq:un})
into (\ref{eq:bn}) we find
\bea
S_{Q}^0(v) &=& \int \bar{Q} (i \vslash v.D-m_{Q})Q
\nonumber \\
&=& \int \bar{\tilde{Q}} \xi ^{-1}(i \vslash v.D-m_{Q}) \xi \tilde{Q}
\nonumber \\
&=& \int \bar{\tilde{Q}} (i \vslash v.\partial -m_{Q}) \tilde{Q}
+ i \int \bar{\tilde{Q}} \vslash \xi ^{-1}((v.\partial \xi )-ig v.A\xi
)\tilde{Q}
 \nonumber \\
&=& \int \bar{\tilde{Q}} (i \vslash v.\partial-m_{Q}) \tilde{Q} \,
. \label{eq:noint}
\eea
We see that,in changing variables according to
(\ref{eq:un}), the gauge field
$A$ is altered to the gauge field $A'$ which satisfies the gauge
condition $v.A'=0$, giving the triviality of the action.
So there is no gauge condition on $A$. It is just that the
combination which makes up $A'$ satisfies (\ref{eq:path}). The fields
$\tilde{Q}$ are gauge inert, meaning that they do not vary under gauge
transformations, as can be seen from their definition. This ensures that
the action (\ref{eq:noint}) is gauge invariant as it should be.

Let us note that this is rather an extreme situation. The effective
theory is totally disinterested in the coupling of gluons to the heavy
quarks. This shows us that the action (\ref{eq:bn}) really models the
situation described above, namely, that while keeping track of the heavy
particle velocity, soft gluons are not important for questions concerning
transitions.

In terms of the effective fields (\ref{eq:un}) the action (\ref{eq:bn})
of HQET$_{0}$ takes the free form described above (\ref{eq:noint}). The
symmetry (\ref{eq:lb}) is thus directly generalised to the free case by
\be
\delta \tilde{Q}(x) = -\lambda^{\mu}\,_{ \nu} x^{\nu} \partial_{\mu}
\tilde{Q}(x) \, \, , \; \; \; \; \; \delta \bar{\tilde{Q}}(x) =
-\lambda^{\mu}\,_{ \nu} x^{\nu} \partial_{\mu} \bar{\tilde{Q}}(x) \, .
\ee

For completeness, let us note that the Feynman propagator for the
non-interacting fields $\tilde{Q}$ is
\bea
\tilde{S}_{F}(x-y;v)&=& -i \frac{(1+\vslash)}{2} \theta(v.x-v.y)
\delta_{\perp}^{3}(x-y) e^{-im_{Q}(v.x-v.y)} \nonumber \\
 && \; \; \; \; -i \frac{(1-\vslash)}{2} \theta(v.y-v.x)
\delta_{\perp}^{3}(y-x) e^{-im_{Q}(v.y-v.x)}
\eea
where the argument of the delta function $\delta_{\perp}^{3}(x)$ is the
component of $x$ not in the $v$-direction. To pass to the Feynman
propagator for the interacting field $Q$ is particularly simple
in view of (\ref{eq:un}),
\be
S_{F}(x,y;v;A) =  S_{F}(x-y;v)W\left[ {\textstyle{x  \atop v}}\right]
W\left[ {\textstyle{y \atop v}} \right]^{-1} \, .
\ee

Notice that these only propagate quarks forward in ``time'' (we mean $v.x$
which is the usual time component in the rest frame) while only
anti-quarks propagate backwards in ``time''. One immediate consequence of
this is that there are no heavy quark loops in the HQET$_{0}$, as could be
inferred directly from the fact that they are certainly not there in the
$v.A=0$ gauge.

\vspace{1cm}
\noindent \underline{Exercise:} Show that in QED one can calculate any
Green's function involving $\tilde{Q}$ and $A_{\mu}$ non-perturbatively
by going over to the free spinor variable $Q$. Hint: Calculate a Wilson
loop in pure Maxwell theory. Why does this procedure not work in QCD ?

\subsection{Derivation of Lowest Order Action}
Questions that naturally arise are, how does one get to the effective
action (\ref{eq:bn}) directly from QCD and what are the corrections that
are needed for finite $m_{Q}$? As recognised in \cite{hkstw}, this action is
familiar in a slightly different but related context. Namely it is a
generalisation of the so-called `Bloch-Nordsieck' model \cite{bn,bs} that
one arrives at in studying the infrared behaviour of QED as the electron
goes on shell. The similarities of the two contexts in which this action
arises will not have escaped the readers attention. Indeed we know how
to derive the non-relativistic action, and we may use this knowledge to
get at the heavy quark effective theory.

Recall that to obtain the relevant non-relativistic information about the
Dirac equation one makes use of Foldy-Wouthuysen transformations. These
are canonical unitary transformations which eliminate the interaction
terms between the positive and negative frequency components of the
fermion spinors in the Hamiltonian. A peek at (\ref{eq:bn},\ref{eq:bn+-}),
and the
subsequent discussion, shows us that this is precisely the situation
that we would like to get to.

In
field theory the analogous set of transformations that give a systematic
derivation of the non-relativistic limit of the relativistic Hamiltonian
in non-Abelian theories was presented by Feinberg \cite{f} (see also
\cite{ef,cl}). The non-relativistic limit picks out a preferred frame
(usually the rest frame, but this need not be so). Likewise, in the heavy
quark limit, one would like to fix or pick out the ``velocity" of the $b$- (or
other heavy) quark that matches that of the
$b$- (or heavy) hadron. Furthermore we would like to arrive at a systematic
expansion (in the inverse mass) giving corrections to this picture. This
is achieved by making use of the Foldy-Wouthuysen transformations in the
context of Lagrangian field theory.

Here we derive (\ref{eq:bn}) by ignoring $1/m_{Q}$ corrections, which
will be taken up again in later sections. Consider the action
\be
S_{Q} = \int  \bar \psi_{Q} (i \Dslash -m_{Q})\psi_{Q} \, , \label{eq:ac}
\ee
where the subscript $Q$, as before, indicates a heavy quark. We make no
distinction here between renormalized and unrenormalized quantities, as it
is not relevant for our present purposes and would clutter up the formulae.
Following \cite{kt1} we perform the change of variables
\bea
\psi_{Q} &= & e^{\textstyle i (\Dslash -\vslash v.D)/(2m_{Q})}
Q    \nonumber \\
&=& Q  + i (\Dslash -\vslash v.D)/(2m_{Q}) Q
+ O(1/m_{Q}^{2}) +\dots \, , \nonumber \\
\bar{\psi}_{Q} &= & \bar{Q} e^{\textstyle
 -i (\stackrel{\leftarrow}{\Dslash }-\vslash v.
     \stackrel{\leftarrow}{D}) /(2m_{Q})} \nonumber \\
&=& \bar{Q} -  \bar{Q} i
(\stackrel{\leftarrow}{\Dslash }-\vslash v.
     \stackrel{\leftarrow}{D}) /(2m_{Q}) + O(1/m_{Q}^{2}) +\dots
 \, , \label{eq:psih}
\eea
where $\stackrel{\leftarrow}{D}$ is defined by
\be
 \int f \stackrel{\leftarrow}{D} g  =  -\int f \vec D g  \, .
\ee
The Jacobian of this
transformation within dimensional regularization has been shown to be
unity in \cite{kt1}. We stay within the realm of dimensional
regularization in the following.

The action may now be expressed as:
\bea
 S_{Q}& =& \int \bar{Q}[1+
 i (\Dslash -\vslash v.D) /(2m_{Q})] (i \Dslash -m_{Q})
 [ 1 +
 i (\Dslash -\vslash v.D) /(2m_{Q})] Q  \, \nonumber \\
&=& \int \bar{Q}(i \vslash v.D - m_{Q}) Q
+O(1/m_{Q})+ \dots \, \nonumber \\
&=& S^{0}_{Q}(v) + O(1/m_{Q}) +\dots \, , \label{eq:bn0}
\eea
which is the relationship we have been looking for. A feature that has
just been used and that will persist throughout is that the mass in
(\ref{eq:ac}) lowers the order of the terms in the transformation
(\ref{eq:psih}), allowing for the replacement of the covariant
derivative in (\ref{eq:ac}) with the one-dimensional covariant
derivative.

The reason for the particular exponential form of the Foldy-Wouthuysen
transformation will be explained in detail when we tackle the derivation of
the effective theory at $O(1/m_{Q})$. For now, we note that to really be
able to denote the extra terms in (\ref{eq:bn0}) as $O(1/m_{Q})$ we must
be sure that there are no components of the derivatives that appear that
lie in the $v$ direction. This is exactly what the form of the
Foldy-Wouthuysen transformations (\ref{eq:psih}) guarantees.

In this derivation we have assumed that both, the components of the momentum
of the heavy quark, and the components of the gauge field that are orthogonal
to the line of flight are small relative to the (infinite) heavy quark
mass. For this to be the case, the interactions must be such that there
are no hard gluons involved.

\subsection{Covariance and Spin (II)}
While, as we have discussed above, the effective action (\ref{eq:bn}) is
invariant under little group transformations it is clearly {\em not}
invariant under general Lorentz transformations,
\bea
\delta_{\lambda}Q(x)&=& L_{\lambda}Q(x) \nonumber\\
&=& (-\lambda^{\mu}\,_{\nu}x^{\nu}\partial_{\mu}+\frac{1}{4}
\lambda^{\mu\nu}\gamma_{\mu}\gamma_{\nu})Q(x)\,,\nonumber\\
\delta_{\lambda}\bar{Q}&=&\bar{L}_{\lambda}\bar{Q}\nonumber\\
&=&
-\lambda^{\mu}\,_{\nu}x^{\nu}\partial_{\mu}\bar{Q}-\frac{1}{4}
\bar{Q}\lambda^{\mu\nu}\gamma_{\mu}\gamma_{\nu}\nonumber\\
\delta_{\lambda}A_{\rho}&=&-\lambda^{\mu}\,_{\nu}x^{\nu}\partial_{\mu}A_{\rho}+
\lambda_{\rho}\,^{\nu}A_{\nu}\,.
\label{cov}
\eea
This circumstance prompted Georgi \cite{g} to integrate over all possible
velocities, with (\ref{eq:bn}) as the integrand, so as to restore
manifest covariance. Somewhat later this integral over velocities was
demoted to a sum although it was not clear what the domain of the sum
should be. The derivation, presented in the last section, of the
effective theory directly from the $QCD$ action, fixes once and for all
the four-velocity to that of the physical hadron.

The non-covariance of the action (\ref{eq:bn}) under the Lorentz
transformation (\ref{cov}) is actually a desired feature. This may be
seen from two points of view. Firstly, think of the Lorentz
transformation in the active sense, that is, as a boost of the hadron
from its old four velocity $v$ to a new four velocity $v'=v+\lambda.v$.
The effective action designed to describe the new situation is
$S_{Q}^{0}(v')$, which is precisely what one obtains under the
transformation (\ref{cov}), namely
$S_{Q}^{0}(v')=S_{Q}^{0}(v)-\delta_{\lambda}S_{Q}^{0}(v)$. There is no
transformation on $v$ in the variation. This was, after all, to be
expected.

What we really want are the usual covariance identities that arise from
Lorentz invariance; for example (with $v'^{\mu}=\Lambda^{\mu}\,_{\nu}v^{\nu}=
v^{\mu}+\lambda^{\mu}\,_{\nu}v^{\nu}$)
\be
\langle H_{c}(v_{c}')\vert [\bar{\tilde{c}}J_{\mu}\tilde{b}](0)\vert
H_{b}(v_{b}')\rangle = \Lambda_{\mu}\,^{\nu}\langle H_{c}(v_{c})\vert
[\bar{\tilde{c}}J_{\nu}\tilde{b}](0)\vert H_{b}(v_{b})\rangle\,.
\label{covid}
\ee
The velocity dependence on the right hand side rests in the action
$S_{Q}^{0}(v_{Q})$. Perform a Lorentz transformation on all the fields
with parameter $\lambda^{\mu}\,_{\nu}$ on the right hand side. This is
just a change of dummy variables in the path integral approach. The
states $\vert H_{Q}(v_{Q})\rangle$ map to $\vert H_{Q}(v_{Q}')\rangle$,
while the action, as we saw, becomes $S_{Q}^{0}(v_{Q}')$. Thus
(\ref{covid}) is established.

Now to the second point of view. The effective action taken to all
orders is equivalent to the QCD action, which is covariant. Hence
covariance must be  achieved in the full effective theory. But
incorporating higher order (in $1/m_{Q}$) terms in the action would hardly
help the situation. The missing piece is that (\ref{cov}) is {\em not}
the correct transformation rule for the effective fields! If
$\Dslash^{\perp}(v)=\Dslash-\vslash v.D$ transformed correctly as a
bi-spinor then the effective fields would have the transformation rule
(\ref{cov}). However, as $v$ is a fixed vector, we find, using the
defining equation (\ref{eq:psih}), to first order in $1/m_{Q}$, that
\bea
\delta_{\lambda}Q(x)&=&L_{\lambda}Q(x)
+\frac{i}{2m_{Q}}[v.\lambda.\gamma v.D+\vslash v.\lambda .D]Q(x)
\,,\nonumber\\
\delta_{\lambda}\bar{Q}&=&\bar{L}_{\lambda}\bar{Q}
-\frac{i}{2m_{Q}}\bar{Q}[v.\lambda.\gamma
v.\stackrel{\leftarrow}{D}+\vslash v.\lambda.\stackrel{\leftarrow}{D}]\,.
\label{cov1}
\eea
Here $a.\lambda.b=a_{\mu}\lambda^{\mu\nu}b_{\nu}$.
It is straightforward to establish that (\ref{cov1}), alongwith the
usual transformation of the gauge field, leaves the lowest
order action (\ref{eq:bn}) invariant upto order $1/m_{Q}$. Taking into
account higher order terms in both the action and the transformation
allows one to establish covariance at any given order in the inverse mass.

\vspace{1cm}
\noindent \underline{Exercise:} Derive (\ref{cov1}). For fun determine
the $1/m_{Q}^{2}$ corrections to the transformations.

\section{Current Induced Transitions}

For definiteness consider the transition of a bound state containing one
$b$-quark $\vert \Phi_b \rangle$ to another bound state containing one
$c$-quark $\vert \Phi_c \rangle$, induced by a
flavour changing current $J_{\mu}\equiv V_{\mu}-A_{\mu}=
\bar \psi _c \gamma_{\mu}(1 -\gamma_{5}) \psi _b \equiv
\bar \psi _c \Gamma_{\mu} \psi _b $. Then we can write the transition-matrix
element as
\be
\langle \Phi_c \vert  J_{\mu} \vert \Phi_b \rangle
\, . \label{eq:trans}
\ee
The situation may be described as follows. As the $b$ hadron comes in
from the far past, we are able to track its four velocity $v_{b}$, and after
the transition we may follow the $c$-hadron's four-velocity $v_{c}$
into the future. In this way the relevant action to consider is
(\ref{eq:bn2})
\be
S_b^0(v_b) +  S_c^0(v_c) = \int \bar{b} (i\vslash_{b}
v_{b}.D-m_b) b +
\int \bar{c} (i \vslash_{c} v_{c}.D-m_c)c \, \,  ,
 \label{eq:bn2a}
\ee
which is repeated here for convenience.

Let us embroider on the picture of the heavy quark theory that has been
developed thus far. A bound state made up of one heavy quark and some
light quarks moving with velocity $v$, such that the $4$-momentum of the
system is
\be
P^{\mu}=M_{Q}v^{\mu} \, ,
\ee
undergoes a current induced transition to a heavy bound state of a
different flavour, with a new four velocity $v'$ and four momentum
\be
P'^{\mu}= M_{Q'}v'^{\mu} \, .
\ee
Then the momentum transfer, $q^{\mu} = M_{Q}v^{\mu}-M_{Q'}v'^{\mu}$; in the
limit as the heavy quark masses go to infinity it diverges. The square of
the momentum transfer is $q^{2} = q^{2}_{max}- 2M_{Q}M_{Q'}(v.v'-1)$ and is
bounded by $q^{2}_{max}= (M_{Q}-M_{Q'})^{2}$, which is achieved at equal
velocity $v=v'$. Of course, even though $q$ has components that are large,
$q^{2}$ may well be small e.g. $q^{2}=0$ at $v=(1,\underline{0})$ and
$v'=(\frac{M_{Q}^{2}+M_{Q'}^{2}}{2M_{Q}M_{Q'}},\underline{v'}) $ with
$\underline{v'} = (\frac{M_{Q}^{2}-M_{Q'}^{2}}{2M_{Q}M_{Q'}},0,0)$ say.
The velocities are well defined when the ratio $M_{Q}/M_{Q'}$ is finite,
which we take to be the case.

\subsection{LSZ Reduction and the Effective Theory}

To get a handle on the transition (\ref{eq:trans}) we first make use of
the ideas of the reduction theorems. These tell us that as long as we
can find an `interpolating field' $\phi(x)$ corresponding to a state
$\vert \Phi (P) \rangle$ with the property (for the $S$-wave
pseudoscalar state)
\be
 \langle 0 \vert \phi(x)  \vert \Phi (P) \rangle = e^{-iP.x} \, ,
 \label{eq:lsz0}
\ee
then the transition may be expressed as\footnote{All the matrix elements
are described in terms of path integrals.}
\bea
& & L(P_b,P_c) \int d^{4}x d^{4}y
 e^{iP_{c}.x}e^{-iP_{b}.y}
 \langle 0 \vert
 \phi_{c}(x) J_{\mu}(0)  \phi^{\dagger}_{b}(y) \vert 0
 \rangle \, , \label{eq:lsz}
\eea
where
\bea
  L(P_b,P_c) &=&
 \lim_{P^{2}_{b} \rightarrow M_{b}^{2}} \lim_{P^{2}_{c} \rightarrow
 M_{c}^{2}} (P_{b}^{2}-M_{b}^{2})(P_{c}^{2}-M_{c}^{2})
 \; . \nonumber
\eea
We shall  concern ourselves with the exact form of the interpolating
fields $\phi_{Q}$
presently. For now we note that whatever their precise form, they will have
the structure
\bea
 \phi_{Q}(x) &=& N_{\Phi _{Q}} \bar{\chi}_{\beta}\,^{\alpha}(\partial)
T\bar{\psi}_{q} ^{i\beta}(x) \psi_{Q}\, _{i\alpha}(x)
  \, , \label{eq:int} \\
 \phi_{Q}^{\dagger}(x) &=& N_{\Phi _{Q}}\chi _{\alpha}\,^{\beta}(\partial)
T\bar \psi_{Q}^{i\alpha}(x)\psi_{q}\, _{i\beta}(x)  \, ,\label{eq:int2}
\eea
with $ \bar{\chi}=\gamma_0 \chi ^{\dagger}\gamma_0 $
for heavy mesons,
while for heavy baryons one has
\bea
 \phi_{Q}(x) &=& N_{\Phi _{Q}}\epsilon^{ijk}
 \bar{\chi}^{\alpha \beta \gamma}(\partial)T\psi_{Q}\, _{i\alpha}(x)
 \psi_{q} \,_{j\beta}(x) \psi_{q'}\, _{k\gamma}(x) \, , \\
 \phi_{Q}^{\dagger}(x) &=& N_{\Phi _{Q}}\epsilon_{ijk}\chi_{\alpha \beta
\gamma}(\partial)
 T\bar{\psi}_{q'} ^{i\gamma}(x) \bar{\psi}_{q} ^{j\beta}(x)
 \bar{\psi}_{Q} ^{k\alpha}(x) \, ,\label{eq:intb}
\eea
with $ \bar{\chi}= \chi ^{\dagger}\gamma_0 \gamma_0 \gamma_0 $,
where $\chi^{\alpha \beta \dots}$ are projection operators (Dirac
matrices possibly including differentiations, total and partial, like
$\stackrel{\leftrightarrow}{\partial}$) which pick out the
particle states of interest\footnote{Here, for once, we explicitly show
the trace over the colour indices $i,j,k$. They will be implicit usually.}.
The normalization constant $N_{\Phi _{Q}}$, for brevity denoted by
$N_{Q}$, is fixed by (\ref{eq:lsz0}). $\psi_{Q}$ and $\psi_{q}$ are the heavy
and light quark fields, respectively.

To simplify the formulae we concentrate on mesonic transitions, i.e.
from a $b$-meson to a $c$-meson. The
generalisation to baryons is immediate. One may rewrite (\ref{eq:lsz})
in this case as
\bea
& & N_{b}N_{c}L\int d^{4}x d^{4}y
 e^{iP_{c}.x}e^{-iP_{b}.y} \nonumber \\
\; \; & & \langle 0 \vert
\bar{\chi}_{\sigma}\,^{\alpha}(\partial_{x})\bar{\psi}_{q}^{\sigma}(x)
\psi_{c\alpha}(x)J_{\mu}(0)\chi_{\beta}\,^{\rho}(\partial_{y})
\bar{\psi}_{b}^{\beta}(y)\psi_{q\rho}(y)\vert 0
\rangle \,.
\eea
Now any total differentiations with respect to $x$ and $y$, present in
$\chi(\partial_{y})$ and $\bar{\chi}(\partial_{x})$, can be shifted on to
the exponentials by
partial integration and, calling the subsequent projection operators
$\chi(v_{b})$ and $\bar{\chi}(v_{c})$, one can write the matrix element as
\bea
& & N_b N_c L(P_b,P_c) \int d^{4}x d^{4}y
 e^{iP_{c}.x}e^{-iP_{b}.y} \nonumber \\
 & & \; \;\langle 0 \vert
 \bar{\psi}_{b}(y).\chi(v_{b}).\psi_{q}(y)\bar{\psi}_{q}(x).\bar{\chi}
(v_{c}).\psi_{c}(x) J_{\mu }(0) \vert 0
\rangle \, , \label{eq:lsz1}
\eea
{}From now on (.) also indicates sum over Dirac indices, which should be
clear from the context. Since we will nearly always consider interpolating
fields inside reduction formulae with the relevant exponentials, we may
as well always replace $\chi(\partial)$ with $\chi(v)$ in the
interpolating field from the beginning. From now on we shall do so
unless otherwise stated.
Thus far we have followed the full theory in conventional form. Now we
apply the effective field theory.
 In order to do so, we need to assume, that in the complete theory, the
integral in (\ref{eq:lsz1}) will factor according to
\be
 \frac{1}{(P_{b}^{2}-M_{b}^{2})}
 \frac{1}{(P_{c}^{2}-M_{c}^{2})}
 \left( \eta^0 + \frac{1}{2m_c}\eta^1_c + \frac{1}{2m_b}\eta^1_b +
 O(\frac{1}{m_{Q}^2}) + \dots \right) \, , \label{eq:according}
\ee
 where $\eta^0$ is the term that arises in the lowest order effective
theory and $\eta_{c}^{1}$ and $\eta_{b}^{1}$ denote first order contributions.
 Let us naively apply the effective
 field theory. In
(\ref{eq:lsz1}) replace everywhere the heavy quark fields $\psi_{Q}$ with
the corresponding fields in the effective theory $Q(x)=
  W\left[ {\textstyle{x \atop v_{Q}}} \right] \tilde{Q}(x)$
to give
\bea
 && N_b N_c L(P_b,P_c) \int d^{4}x d^{4}y
e^{iP_{c}.x}e^{-iP_{b}.y}\nonumber\\
&&\;\;\;\;\;\;\; \langle 0 \vert
 \bar{\tilde{b}}(y).\chi(v_{b}).M(y,x).\bar{\chi}(v_{c}).%
\tilde{c}(x)\bar{\tilde{c}}(0).\Gamma_{\mu}.
 \tilde{b}(0)\vert 0\rangle , \label{eq:lsz2}
\eea
with
\be
  M(y,x)_{\beta } \ ^{\alpha }=
Tr\psi_{q\beta}(y) \bar{\psi}_{q}^{\alpha}(x)
  W\left[  {\textstyle{x \atop v_{c}}}
           {\textstyle{0 \atop v_{c}}}
           {\textstyle{0 \atop v_{b}}}
           {\textstyle{y \atop v_{b}}}\right]
\, , \nonumber
\ee
where the trace is over the colour indices. Here we have defined the
matrix product
\be
W\left[{\textstyle{x_{1}\atop v_{1}}}
       {\textstyle{x_{2}\atop v_{2}}}
       {\textstyle{x_{3}\atop v_{3}}}
       {\textstyle{x_{4}\atop v_{4}}}\right]
= W\left[{\textstyle{x_{1}\atop v_{1}}}\right]
  W\left[{\textstyle{x_{2}\atop v_{2}}}\right]^{-1}
  W\left[{\textstyle{x_{3}\atop v_{3}}}\right]
  W\left[{\textstyle{x_{4}\atop v_{4}}}\right]^{-1}\,.
\ee

We have used the fact that
the heavy quark propagators are diagonal in colour.
Note that $M(y,x)$ depends explicitly on $v_b$ and
$v_c $.
As the effective fields $\tilde{Q}$ decouple from the interactions,
(\ref{eq:lsz2}) factorises into
\bea
&-&N_b N_c L(P_b,P_c) \int d^{4}x d^{4}y
 e^{iP_{c}.x}e^{-iP_{b}.y} Tr \chi(v_{b})
\langle 0 \vert  M(y,x) \vert 0\rangle . \nonumber \\
 & & \; \; \bar{\chi}(v_{c})
\langle 0 \vert  \tilde{c}(x)\bar{\tilde{c}}(0)\vert 0\rangle  \Gamma _{\mu }
\langle 0 \vert  \tilde{b}(0)\bar{\tilde{b}}(y)\vert 0\rangle
\, . \label{eq:lsz3}
\eea
Set the momentum through $ \langle 0 \vert M(y,x) \vert 0\rangle
$ to be $q$ and $k$.
 Then the transition is easily expressed as
\bea
 & & N_b N_c L(P_b,P_c) \int d^{4}k d^{4}q
 Tr
 M(q,k)\nonumber \\
& & \; \; \bar{\chi}(v_{c})
 \frac{1}{\vslash_{c} v_{c}.(P_{c}-k )  - m_{c}}
 \Gamma _{\mu }
 \frac{1}{\vslash_{b} v_{b}.(P_{b}+q )  - m_{b}}\chi(v_{b})
 \, , \label{eq:lsz4}
\eea
where $M(q,k)$ is defined through the Fourier transform
\be
\langle 0\vert M(y,x)\vert 0\rangle=\int d^{4}q d^{4}k
e^{-iq.y}e^{-ik.x}M(q,k)\,,
\ee
and we have introduced the heavy quark propagator
\be
\langle 0\vert\tilde{Q}(x)\bar{\tilde{Q}}(y)\vert 0\rangle=
\int d^{4}p\frac{e^{-ip.(x-y)}}{\vslash v.p-m_{Q}}\,.
\ee
\subsection{Taking the on-shell limit}
To ensure that the transition (\ref{eq:lsz4})
 does not vanish there must be poles generated by the
integrals that match the zeros that arise from the factors outside the
integral. This is an assumption about bound states that needs to be made
in the full theory before the effective theory is invoked. However,
this need not be taken for granted within the context of the effective
field theory, as we have mentioned around
(\ref{eq:according});
 rather, due to the simplifications encountered
we may and will address this issue in some detail.

We are not free to take the limit in any fashion we choose. For example,
let us set $P_{Q}=M_{Q}v_{Q}+\epsilon_{Q}^{\perp }(v_{Q}) $, for some
infinitesimal $\epsilon_{Q} $, and
understand the limit to be
\bea
 && L(P_{Q})=\lim_{P_{Q}^{2}\rightarrow M_{Q}^{2}}
 (P_{Q}^{2}-M_{Q}^{2})=\lim_{\epsilon_{Q} \rightarrow 0} (\epsilon_{Q}^{\perp
})^{2}
 \, .
 \nonumber
\eea
(Recall that $\epsilon^{\perp\mu}(v)=\epsilon^{\mu}-v^{\mu}v.\epsilon $
so that $v.\epsilon^{\perp}=0$.)
The problem with taking the limit in this fashion is that $\epsilon_{Q}^{\perp
}$ cannot appear anywhere
in (\ref{eq:lsz4}) except in $L(P_{Q})$! The mass pole is not regulated in
this way. An immediate consequence is that we have no choice but to take
the limit along the $v_{Q}$ direction. We then take the limit in the
direction $P_{Q}=M_{Q}(1+\epsilon _{Q})v_{Q}$, so that
\be
 \lim_{P_{Q}^{2}\rightarrow M_{Q}^{2}}
 (P_{Q}^{2}-M_{Q}^{2}) = \lim_{\epsilon_{Q} \rightarrow 0} 2\epsilon_{Q}
M_{Q}^{2}
 \, . \label{eq:onshell1}
\ee

We can
leave the cancellation of the zeros implicit in (\ref{eq:lsz4}), but
nevertheless simplify its form.  The explicit form of $\chi(v)$, that
will be given in the next section,
satisfies $(\vslash - 1).\chi(v)= 0$.
 The propagators as contracted against
the wave functions then take the form
\be
(\vslash_{Q} v_{Q}.(P_{Q}-k )  - m_{Q})^{-1}\chi \rightarrow
 \frac{1}{v_{Q}.(P_{Q}-k)-m_{Q}  } \chi \, ,
\ee
so that the transition is quite generally
\bea
& &  N_b N_c L(P_b,P_c) \int d^{4}k d^{4}q
 \frac{1}{v_{c}.(P_{c}-k )  - m_{c}}\,
 \frac{1}{v_{b}.(P_{b}+q )  - m_{b}}
  \nonumber \\
& & \;\;\;\;\; \; Tr M(q,k)\bar{\chi}(v_{c}) \Gamma _{\mu}
 \chi(v_{b}) \, .
 \label{eq:lsz4b}
\eea

There is, however, a way to
take the limit so that the expected cancellation of the pole and zero may be
explicitly achieved. This requires that the heavy quark propagator
$(v_{Q}.(P_{Q}-k) - m_{Q})^{-1} $ exactly cancels the zero
$(P_{Q}^{2} - M_{Q}^{2})$. This requires that the Green's function $\langle 0
 \vert \tilde M(q,k)  \, \vert 0 \rangle $ is peaked, within the integral,
at  $v_{c}.k= \bar \Lambda _{c}$ and $-v_{b}.q=\bar \Lambda _{b}$, in the
heavy quark limit, where
$\bar \Lambda _{Q}=M_{Q}-m_{Q}$. Then, for example, the c-quark propagator
becomes
$(v_{c}.P_{c} - m_{c}-\bar \Lambda_{c})^{-1}$, exactly cancelling the
zero $(P_{c}^{2} - M_{c}^{2})$. These conditions are precisely the `on-shell'
assumptions
that we made in determining the heavy quark theory in the first place. We
assumed in
(\ref{eq:mompr}) that the projection of the quark momentum in the direction
of the hadron $4$-velocity is the quark mass and in this limit the hadron
mass. If this cancellation is accepted then (\ref{eq:lsz4}) becomes
\be
4N_b N_c M_{b}M_{c}\int d^{4}k
d^{4}q Tr M(q,k)
\bar{\chi}(v_{c})\Gamma _{\mu }\chi(v_{b})
\, . \label{eq:lsz4a}
\ee
It is not mandatory that the cancellation of the poles proceeds in this
way. For our purposes it is enough to notice that the denominators of the
heavy quark propagators depend only on $\bar{\Lambda}_{Q}$ and on
$\epsilon_{Q} M_Q$. However, we know (see Introduction) that to lowest order
$\bar{\Lambda}_{b} = \bar{\Lambda}_{c}= \bar{\Lambda}$. So that after the
limits are taken the only dependence will be on $\bar{\Lambda}$, which is
universal.

A problem that still remains is, how do the light degrees of freedom know
that there is such a momentum restriction (i.e. $v_{Q}.k \approx
\bar \Lambda$)? As we have noted the information that is lost in
going to the effective fields has to do with how the light degrees of
freedom respond to the heavy quark. The discusion thus far is then still not
complete. We need to ask ourselves about the applicability of the effective
theory in such a naive fashion. Certainly at the point of interaction with
the current the only important scale is that set by the transition, and the
heavy quark operators may be replaced with their effective counterparts. But
that still leaves the heavy quark operators that go into making up the
interpolating field.

We parameterise this ignorance
by the substitution
\be
\chi_{\alpha \alpha_{1} \dots \alpha_{n}} \rightarrow
\chi_{\alpha \beta_{1}  \dots \beta_{n}} B^{\beta_{1} \dots
\beta_{n}}_{\alpha_{1} \dots \alpha_{n} }(x) \, , \label{eq:genwf}
\ee
so that the spinor functions $B$ `correct' the behaviour of the light quarks
and soft gluons to remind them that they are in the presence of the heavy
quark, leaving the heavy quark label $\alpha$ `free'. This choice
corresponds to the physical picture presented in the Introduction that in
the heavy mass limit the
heavy quark spin is decoupled. These functions should be considered to
contain the nonperturbative
information (both in the coupling and in the mass), that is needed for a
bound state to emerge. One should expect that, while these are certainly
functions of the heavy mass, there may be no sense in which they can be
expanded in a $1/m_{Q}$ series. For example, such a $B$ might have a
dependence of the form
\be
\sqrt{m_{Q}/\mu^{3}}\exp{(-m_{Q}(v_{Q}.k)^{2}/ \mu^{3})} \, ,
\ee
which is a sum of derivatives of a delta function in a $1/m_{Q}$
expansion and which, consequently, may yield quite divergent integrals
term by term. This is reminiscent of the non-analytic behaviour one expects
in terms of the strong coupling in the confining region in QCD.

If one accepts, for the moment, the need to introduce the $B$-functions
into the definition of the wave function, then it is apparent that the
discussion on the normalization of the wave function needs to take this
into account. One would have to insert $B$ in the appropriate places in
(\ref{eq:nor41}), (\ref{eq:nor42}), (\ref{eq:nor45})
and (\ref{eq:nor46}).

 With the introduction of the $B$ factor, the form of the matrix element
(\ref{eq:lsz4a}) does not change but now $M(y,x)$ is defined as
\be
  M(y,x)_{\beta } \ ^{\alpha }=
TrB_{\beta}^{\delta}(y)\psi_{q\delta}(y)
 \bar{\psi}_{q}^{\gamma}(x)\bar{B}_{\gamma}^{\alpha}
  W\left[  {\textstyle{x \atop v_{c}}}
           {\textstyle{0 \atop v_{c}}}
           {\textstyle{0 \atop v_{b}}}
           {\textstyle{y \atop v_{b}}}\right]
\, . \nonumber
\ee

As $M(x,y)$ is in any case unknown, this `improvement' will not effect
any of our subsequent discussion on the number of form factors. One
diagonal contribution to the function $B$ is the correction coming from
the matching of HQET to QCD at one loop level for example (see Section 6).

\subsection{Bargmann-Wigner Wave Functions and Interpolating Fields}

In order to extract useful information from the overlap integral for
the transition (\ref{eq:lsz4}) we need to determine an appropriate set of
projectors $\chi$ that appear in the interpolating fields. It is well known
\cite{lu} that a wide variety of interpolating fields can be used to
represent a particular bound state within a reduction formula. We shall show in
this subsection that the Bargmann-Wigner wave functions are the natural
choice for these projectors in the heavy quark limit \cite{ht}.

As a first example consider a heavy vector meson.  An interpolating field
$\phi_{\mu}(x)$ for a
vector particle, with momentum  $P$, should satisfy the condition
\be
\langle 0\vert \phi_{\mu}(x)\vert P\rangle=\epsilon_{\mu}e^{-iP.x} \,,
\label{ipnc}
\ee
where the polarisation vector $\epsilon_{\mu}$ satisfies the
transversality condition $v.\epsilon=0$ and
$\epsilon^{*\mu}\epsilon_{\mu}=-1$.

Thus, we can take as
interpolating fields, for example, either
\be
\phi_{\mu}^{1}(x)=N_{1}T\bar{\psi}_{q}(x)\gamma_{\mu}^{\perp}\psi_{Q}(x)\,,
\label{ipfv1}
\ee
or
\be
\phi_{\mu}^{2}(x)=-N_{2}\frac{i\partial_{\nu}}{M_{Q}}T\bar{\psi}_{q}(x)
\gamma^{\nu}\gamma_{\mu}^{\perp}\psi_{Q}(x)\,.
\label{ipfv2}
\ee
The normalisation constants
\be
N_{1}=-\frac{1}{\langle 0\vert T\bar{\psi}_{q}(0)\epsslash^{*}\psi_{Q}(0)
\vert P\rangle}\,,
\ee
and
\be
N_{2}=\frac{1}{\langle 0\vert T\bar{\psi}_{q}(0)\vslash\epsslash^{*}\psi_{Q}(0)
\vert P\rangle}\,
\ee
follow from the normalisation of the polarisation vector.

Now one can use the equations of motion
\be
(i\Dslash-m)\psi=0
\ee
for the quark fields to obtain
\bea
& & -\frac{i\partial_{\nu}}{M_{Q}}T\bar{\psi}_{q}(x)
\gamma^{\nu}\gamma_{\mu}^{\perp}\psi_{Q}(x)\nonumber \\
& & \;\;\;\; =
\frac{(m_{Q}+m_{q})}{M_{Q}}[T\bar{\psi}_{q}(x)\gamma_{\mu}^{\perp}\psi_{Q}(x)
-2iT\bar{\psi}_{q}(x)\frac{\vec{D}_{\mu}^{\perp}}{m_{Q}+m_{q}}\psi_{Q}(x)]\,.
\eea

We see that the two interpolating fields differ by a term which can be
neglected in the heavy quark limit. Thus in the heavy quark limit, with
$m_{Q}+m_{q}=M_{Q}$, the two interpolating fields, $\phi^{1}_{\mu}$ and
$\phi^{2}_{\mu}$, are equal along with the important relation
\be
N_{1}=N_{2}=N\,.
\label{ir}
\ee
Note
that the term neglected, $D_{\mu}^{\perp}/M_{Q}$, is exactly  the kind of
term which was dropped in deriving the HQET. One can also look at this
result in another way.
$i\bar{\psi}_{q}(x)\frac{\vec{D}_{\mu}^{\perp}}{M_{Q}}\psi_{Q}(x)$ is in
fact another possible candidate for the interpolating field for the
vector meson but its overlap with the physical state becomes negligible
in the heavy mass limit. Note also that
$i\bar{\psi}_{q}(x)\frac{\vec{\partial}_{\mu}^{\perp}}{M_{Q}}\psi_{Q}(x)$
corresponds to a p-wave contribution to the vector state coming from the
anti-quark part of the heavy quark. In the zeroth order HQET recall that
the antiquark part of the heavy quark field is suppressed and thus the
overlap of this interpolating field with the physical heavy meson state
becomes very small in this limit.

The pseudoscalar case is simpler. One can show using equations of motion
that the two interpolating fields
\be
\phi^{1}(x)=P_{1}T\bar{\psi}_{q}(x)\gamma_{5}\psi_{Q}(x)\,,
\label{ipfps1}
\ee
and
\be
\phi^{2}(x)=-P_{2}\frac{i\partial_{\nu}}{M_{Q}}T\bar{\psi}_{q}(x)
\gamma^{\nu}\gamma_{5}\psi_{Q}(x)\,,
\label{ipfps2}
\ee
are equal along with the relation between the normalisation constants
\be
\frac{P_{1}}{P_{2}}=\frac{m_{Q}+m_{q}}{M_{Q}}\,.
\ee
Note again that, as in the vector case, in the heavy quark limit, where
$m_{Q}+m_{q}\rightarrow M_{Q}$, $P_{1}/P_{2}\rightarrow 1$.

Having established relationships between interpolating fields we now
address the question: what are the consequences of the freedom of choice
of interpolating fields for a matrix elements involving heavy mesons and
baryons?
As an example we shall consider an arbitrary matrix element involving a heavy
vector meson in the initial state
\be
{\cal M}= \langle P^{\prime}\vert J\vert P\rangle\,,
\label{ame}
\ee
where the state $\vert P\rangle$ represents an incoming heavy vector
meson, with momentum $P$ and mass $M_{Q}$ and the
final state is arbitrary as is the current $J$.
Now use the reduction theorem to write this matrix element as
\be
{\cal M}=\lim_{P^{2}\rightarrow M_{Q}^{2}}(P^{2}-M_{Q}^{2})\int d^{4}xe^{-iP.x}
\langle P^{\prime}\vert J\phi^{\dagger}(x)\vert 0\rangle\,,
\label{ame2}
\ee
The $\phi^{\dagger}(x)$ in (\ref{ame2})
$\epsilon^{\mu}\phi_{\mu}^{\dagger}$ where $\phi_{\mu}$ is given in eq.
(\ref{ipfv1}) or (\ref{ipfv2}). Thus one can write
\be
{\cal M}=\lim_{P^{2}\rightarrow M_{Q}^{2}}(P^{2}-M_{Q}^{2})N\int
d^{4}xe^{-iP.x}
\langle P^{\prime}\vert J\chi_{\alpha}\,^{\beta}(\partial_{x})
\bar{\psi}_{Q}^{\alpha}(x)\psi_{q\beta}(x)\vert 0\rangle\,,
\label{ame3}
\ee
where $\chi(\partial_{x})$ is either $\epsslash$ or
$\frac{i\dslash_{x}}{M_{Q}}\epsslash$ and correspondigly $N$ is either
$N_{1}$ or $N_{2}$.

The statement that one can use any one of the alternative interpolating
fields means that they should give the same result in the reduction
theorems. To see what this implies consider the second interpolating
field, i.e $\chi(\partial_{x})=\frac{i\dslash_{x}}{M_{Q}}\epsslash$ in the
matrix element (\ref{ame3}). Integrating by parts we can shift the $x$
derivative in $\chi$ onto $exp(-iP.x)$ to obtain
\be
{\cal M}=\lim_{P^{2}\rightarrow M_{Q}^{2}}(P^{2}-M_{Q}^{2})N_{2}\int
d^{4}xe^{-iP.x}
\langle P^{\prime}\vert J(\vslash\epsslash)_{\alpha}\,^{\beta}
\bar{\psi}_{Q}^{\alpha}(x)\psi_{q\beta}(x)\vert 0\rangle\,,
\label{ame4}
\ee
whereas with the other choice of the interpolating field we have instead
\be
{\cal M}=\lim_{P^{2}\rightarrow M^{2}}(P^{2}-M^{2})N_{1}\int d^{4}xe^{-iP.x}
\langle P^{\prime}\vert J(\epsslash)_{\alpha}\,^{\beta}
\bar{\psi}_{Q}^{\alpha}(x)\psi_{q\beta}(x)\vert 0\rangle\,.
\label{ame5}
\ee
Thus comparing we see that if the two expressions (\ref{ame4}) and
(\ref{ame5}) are to be equal, then $\vslash=N_{1}/N_{2}$ as an operator
inside the matrix element, at least in the mass shell limit for the
heavy meson. In general this is not much of a restriction as the $N_{i}$
are functions of the velocity $v$. However, we have shown above that, in
the heavy quark limit, the ratio of the normalisation constants is unity
and independant of the velocity, leading to $\vslash=1$. In fact we see
from (\ref{ame4}) and (\ref{ame5}) that this condition means that only
certain components of the quark fields entering in the interpolating
fields contribute to the matrix element, i.e those satisfying
\bea
\bar{\psi}_{Q}(x)\vslash&=&\bar{\psi}_{Q}(x) \nonumber \\
\vslash\psi_{q}(x)&=&-\psi_{q}(x)\,.\label{ame6}
\eea
in the meson mass shell limit.
In other words, in the rest frame of the {\em meson} on the mass shell,
only the quark part of the heavy quark field \underline{and}
surprisingly also only the antiquark part of the light quark field
contributes in the reduction formula. The same relation holds for the
pseudoscalar particle.
 It is obvious that this physical requirement will be enforced by taking
the interpolating field for the heavy vector meson, in the heavy quark
limit, to be
\bea
\phi_{\mu}(x)&=&\frac{1}{2}(\phi^{1}_{\mu}+\phi^{2}_{\mu})  \nonumber \\
&=&\frac{1}{2}N[T\bar{\psi}_{q}(x)\gamma_{\mu}^{\perp}\psi_{Q}(x)-
\frac{i\partial_{\nu}}{M_{Q}}T\bar{\psi}_{q}(x)
\gamma^{\nu}\gamma_{\mu}^{\perp}\psi_{Q}(x)]\,.
\eea
In fact even if we start with an arbitrary linear combination of the two
fields, only the sum survives in the heavy quark limit. With this choice
the matrix element picks up the correct projection operator
$\frac{1+\vslash}{2}$ to enforce the above condition (\ref{ame6}).

The above considerations turn out to be a general feature of the heavy
quark limit. Given a particular interpolating field
$NT\bar{\psi}_{q}{\bf \Gamma}\psi_{Q}$, then one can show, using the
equations of motion, that this is equal to the interpolating field
$N\frac{-i\partial_{\nu}}{M_{Q}}T\bar{\psi}_{q}\gamma^{\nu}{\bf
\Gamma}\psi_{Q}$, upto terms of order $O(\vec{D}^{\perp}/M_{Q})$. Here,
${\bf \Gamma}$ is some Dirac matrix, possibly with derivatives.
Thus
the natural choice for the projection operator $\chi(\partial)$
appearing in the interpolating fields for heavy mesons, eqs. (\ref{eq:int}) and
(\ref{eq:int2}), is\footnote{We have here introduced an explicit
$1/\sqrt{M_{Q}}$ factor in the projection operator to factor out the
heavy mass scale. This is related to the fact that our states are
normalised relativistically:
 $\langle P\vert P^{\prime}\rangle =
2E(2\pi)^{3}\delta^{3}(\vec{P}-\vec{P^{\prime}})$. }
\be
\chi_{\beta}\,^{\alpha}(\partial)=
\frac{1}{2\sqrt{M_{Q}}}[(1-\frac{i\dslash}{M_{Q}}){\bf
\Gamma}]_{\beta}\,^{\alpha}\,.
\label{bwpo}
\ee
Here, for example, ${\bf \Gamma}=\gamma_{5}$ or ${\bf \Gamma}=\epsslash$
for the pseudoscalar and vector meson respectively.
As we have noted earlier, the projection operator always appears in an
integral over the space-time coordinates along  with the relevant
exponential containing the momentum of the physical state, corresponding
to the interpolating field. Thus, by partial integration the differentiation
in the projection operator can be the shifted onto the exponential, so
that we can always replace the $\chi(\partial)$ of eq. (\ref{bwpo}) with
\be
\chi_{\beta}\,^{\alpha}(v)=
\frac{1}{2\sqrt{M_{Q}}}[(1+\vslash){\bf \Gamma}]_{\beta}\,^{\alpha}\,.
\label{bwpo2}
\ee
in all reduction formulae.

Although we shown that the above form of the projection operators is the
natural choice for the interpolating field for a heavy meson, we of
course have the freedom to choose this form also for a light meson. This
provides a uniform approach for both light and heavy mesons \cite{hlkkt}.
Similar considerations apply to the heavy and light baryon interpolating
fields, containing the projection operators
$\chi_{\alpha\beta\gamma}$.

One now recognises that these projection operators (\ref{bwpo2}) are
nothing but the well-known Bargmann-Wigner wave functions. Hence, we have
shown that the task of projecting out preferred
particle states using products of spin half fields can be achieved in an
elegant manner through the use of the so called Bargmann-Wigner
wave functions.

The Bargmann-Wigner wave functions are multi-spinor wave functions of some
given rank and symmetry type,
\be
\chi_{\alpha_{1} \dots  \alpha_{n}}(v) \, ,
\ee
which satisfy the free Dirac equation on all the labels
\bea
 (\vslash -1)_{\beta}\, ^{\alpha_{1}}\chi_{\alpha_{1} \dots
\alpha_{n}}(v)& =& 0 \, , \nonumber \\
  \dots \; \; \; \; \; \; \; \; \; \dots \; \; \; \; \; \; \; \; \; \dots
 & =& 0 \, , \nonumber \\
 (\vslash -1)_{\beta}\,^{\alpha_{n}}\chi_{\alpha_{1} \dots
\alpha_{n}}(v)& =& 0 \, . \label{eq:bw}
\eea
The reason these wave functions describe particles of a given spin is
amazingly simple \cite{hkt}. In the rest frame the Dirac equation can be seen
to be
a projection by the operator $(1+\gamma_{0})$. This means that the spin
labels take on not four independent values but two. Put another way, this
reduces $SO(3,1)$ down to one of its $SU(2)$ factors. After imposing the
Dirac equation on all the labels, the multi-spinor becomes a
product representation of $SU(2)$. On fixing to a given symmetry type one is
actually considering an irreducible representation of $SU(2)$ of
dimension $2s+1$.

For example, take $\chi(v)$ to be totally symmetric of rank $2s$
satisfying (\ref{eq:bw}). Then this is a wave function for a particle of
spin $s$ and even parity\footnote{Actually as defined only for
$S$-waves. To get
a handle on the $P$ and higher waves allow the $\chi(v,k)$ to depend on the
orbital
momentum $k$.\cite {arbsp}}. In the rest frame, $\gamma_{0}$ is
the parity operator for the spin $1/2$ Dirac spinor, tensor products of it
being the
parity operator for the multi-spinor.

A rank $2$ symmetric bispinor, for example, describes a spin one particle
as follows. The gamma matrices
may be split into a symmetric class, $\gamma_{\mu}C$ and $\sigma_{\mu
\nu}C$, and an antisymmetric class, $C$, $\gamma_{5}C$ and
$i\gamma_{\mu} \gamma_{5}C$. The symmetric bispinor is then a linear
combination
\be
\chi_{ \{ \alpha \beta \} } = \phi^{\mu}(\gamma_{\mu}C)_{  \alpha \beta
}  + \frac{1}{2}\phi^{\mu
\nu}(\sigma_{\mu \nu}C)_{  \alpha \beta  } \, .
\ee
The Bargman-Wigner equations (\ref{eq:bw}) impose the constraints that
\be
v^{\mu}\phi_{\mu} = 0 \, , \; \; \; i\phi_{\mu \nu} = v_{\mu} \phi_{\nu}
- v_{\nu} \phi_{\mu} \, ,
\ee
so that, on taking $\phi^{\mu}$ to be the polarisation tensor $\epsilon
^{\mu}$ the wave function takes the form
\be
\chi_{ \{ \alpha \beta  \} } =\frac{1}{2}[(1 + \vslash ) \epsslash C]_{ \alpha
\beta }
= [P_{+}  \epsslash C]_{ \alpha \beta }\, .
\ee

If one is interested, as we are, in using these multi-spinors to
describe baryons and pseudoscalar mesons, then the above example is not
enough. First, we may augment this with a rank $2$ anti-symmetric bispinor
$\chi_{ [\alpha \beta] }(v)$ describing a spin zero particle.
Expanding this out in the three possible
antisymmetric matrices and
imposing (\ref{eq:bw}) we find that
\be
\chi_{ [\alpha \beta] } = \frac{1}{2}[(1 + \vslash) \gamma_{5}C]_{ \alpha \beta
}
\, .
\ee
Secondly, as the
mesons are made out of quark anti-quark bound states the wave function
has one covariant and one contravariant index. The contravariant index
may be lowered with the help of the charge conjugation matrix $C$, so
that one obtains a covariant bispinor whose symmetry or anti-symmetry may be
declared unambiguously as above. We can then remove the charge
conjugation matrix by multiplying by $C^{-1}$ to obtain the projection
operators for the pseudoscalar and vector mesons
\be
\chi_{\alpha}\,^{\beta}=
\frac{1}{2\sqrt{M_{Q}}}[(1+\vslash)\gamma_{5}]_{\alpha}\,^{\beta}
\ee
and
\be
\chi_{\alpha}\,^{\beta}=
\frac{1}{2\sqrt{M_{Q}}}[(1+\vslash)\epsslash]_{\alpha}\,^{\beta}
\ee
respectively, as expected.

For the baryons our interest rests on the rank three trispinor $\chi_{\alpha
\beta \gamma}(v)$. After the Bargman-Wigner conditions are imposed each
label effectively ranges over two values, so that this wave function has
$2^{3}=8$ degrees of freedom. We wish to decompose this down to its
irreducible parts ($1/2 \oplus 1/2 \oplus 3/2$), under $SU(2)$. First set
\be
\chi_{\alpha \beta \gamma} = \chi_{ [\alpha \beta] \gamma} + \chi_{ \{ \alpha
 \beta \} \gamma} \, .
\ee
If we had not enforced the Bargman-Wigner conditions, then $\chi_{ [\alpha
\beta ] \gamma} $ would be reducible and the totally antisymmetric part
would have to be projected out, that is, we would have to also impose
\be
\chi_{ [\alpha \beta] \gamma} +\chi_{ [\beta \gamma] \alpha }
+\chi_{[\gamma  \alpha] \beta} =0  \, .
\ee
However, it is easy to see that after imposing the Bargman-Wigner
equations this tracelessness condition is identically satisfied.
$\chi_{ [\alpha \beta] \gamma} $ represents a spin half particle. On the
vother hand $\chi_{ \{ \alpha \beta \} \gamma}$ is reducible. To project
away the totally symmetric part we impose the condition
\be
\chi_{ \{ \alpha \beta \} \gamma} + \chi_{ \{ \beta \gamma \} \alpha} +
\chi_{ \{ \gamma \alpha \} \beta} = 0 \, . \label{eq:trace}
\ee
Let us keep denoting the rank three spinor of mixed symmetry that
satisfies this condition by $\chi_{ \{ \alpha \beta \} \gamma}$. It
represents a spin $1/2$ particle. We start with $6$ degrees of freedom
from which we exclude the $4$ coming from the totally symmetric
component, leaving $2$ components. The complete decomposition is then
\be
\chi_{\alpha \beta \gamma} = \chi_{ [\alpha \beta] \gamma} + \chi_{ \{ \alpha
 \beta \} \gamma} + \chi_{ \{ \alpha \beta  \gamma \} } \, . \label{eq:b}
\ee

The reason for explaining this at length is that it allows for a rather
nice description of the situation in the case of heavy baryons. Firstly, the
totally symmetric component describes a spin $3/2$ particle and so is suitable
for the $\Sigma^{*}$ and $\Omega^{*}$ baryons. Now for the heavy $\Lambda$ and
$\Xi$ baryons where the two light quarks are expected to form an
antisymmetric diquark $s=0$
state. If in (\ref{eq:b}) we let $\gamma$ be the index on the heavy quark leg
then
$\chi_{ [\alpha \beta] \gamma} $ is just what we are looking for, and it
has the form (following the same analysis as for the mesons), upto
normalisation,
\bea
\chi_{ [\alpha \beta] \gamma} &=& \frac{1}{2}[(1+\vslash) \gamma_{5}C]_{\alpha
\beta}
u_{\gamma} \, .
\label{eq:lambdawf}
\eea

For the heavy $\Sigma$ and $\Omega$ baryons the light diquarks are
expected to be in a symmetric $s=1$ state.  Still with $\gamma$ being the heavy
quark label, the other spin $1/2$ wave function $\chi_{ \{ \alpha \beta
\} \gamma}$ in (\ref{eq:b}) has the
two light quarks in an $s=1$ combination. This then is the correct
wave function for these baryons. There are two ways to get to an
explicit form for this wave function. In the first method we follow the
same route as we did for the mesons, namely we expand $\chi_{ \{ \alpha
\beta  \} \gamma} $ in terms of the symmetric gamma matrices in the
$\alpha$ and $\beta$ indices and impose the Bargmann-Wigner equations to
find, again upto overall normalisation,
\be
\chi_{ \{ \alpha \beta \} \gamma} = \frac{1}{2}\phi^{\mu}_{\gamma}
[(1 + \vslash) \gamma_{\mu}C]_{\alpha \beta} \, ,
\ee
with
\be
v_{\mu} \phi^{\mu}_{\gamma} = 0 \, \; \; ; (1-
\vslash)_{\alpha}\,^{\beta} \phi^{\mu}_{\beta}= 0 \, .
\label{eq:phimuga}
\ee

On imposing the tracelessness condition (\ref{eq:trace}), and after a
little algebra, eq. (2.81) becomes
\bea
\chi_{ \{ \alpha \beta \} \gamma} &=& \frac{1}{4}[(1+\vslash) \gamma^{\mu}
\gamma_{5}u]_{ \gamma}
[(1 + \vslash) \gamma_{\mu}C]_{\alpha \beta}
\nonumber \\
 &=& \frac{1}{2}[(\gamma^{\mu} +
v^{\mu})\gamma_{5}u]_{\gamma} [(1 + \vslash) \gamma_{\mu}C]_{\alpha
\beta} \, ,
\label{eq:siwf}
\eea
which is the wave function as put forward in \cite{g,g1}. Alternatively we
notice that a trace condition of the above form is easily solved in terms
of antisymmetric objects,
\be
\chi_{ \{ \alpha \beta \} \gamma} = \phi_{ [\alpha \gamma ] }
\phi_{\beta} + \phi_{ [\beta \gamma ] }
\phi_{\alpha} \, ,
\ee
which on also demanding (\ref{eq:bw}) is
\be
\chi_{ \{ \alpha \beta \} \gamma} = \frac{1}{2}[(1+\vslash)
\gamma_{5}C]_{\alpha
\gamma} u_{\beta} + \frac{1}{2}[(1+\vslash) \gamma_{5}C]_{ \beta \gamma}
u_{\alpha}
\, .
\ee
This is the form advanced in \cite{hkkt}. The equivalence of the various
forms of the wave functions was given in
\cite{hlkkt} in a slightly roundabout way. The derivation here shows
that from the Bargmann-Wigner (and group theoretic) point of view one
form is as natural as the other.

Likewise for the spin $3/2$ wave function there are two natural forms. By
expanding once more in the symmetric gamma matrices in the $\alpha$
and $\beta$ labels, imposing the Bargmann-Wigner equations and then
demanding total symmetry one finds, up to normalisation,
\bea
\chi_{ \{ \alpha \beta  \gamma \} } &=& u^{\mu}_{\gamma} [(1 + \vslash)
\gamma_{\mu}C]_{\alpha \beta} \,.
\label{eq:sistarwf}
\eea
and also that (\ref{eq:phimuga}) is satisfied with $u^{\mu}_{\gamma}$
 replacing $\phi^{\mu}_{\gamma}$.

The second (and obvious) approach is to begin with the expansion in
terms of the symmetric gamma matrices in any two of the labels and then add
symmetric permutations of these terms in all the labels. In this way we
obtain, upto normalisation,
\be
\chi_{ \{ \alpha \beta  \gamma \} } = [(1+\vslash) \gamma_{\mu}C]_{ \{
\alpha \beta \} } u^{\mu}_{\gamma} + [(1+\vslash) \gamma_{\mu}C]_{ \{
\beta \gamma \} } u^{\mu}_{\alpha} + [(1+\vslash) \gamma_{\mu}C]_{ \{
\gamma \alpha \} } u^{\mu}_{\beta} \, .
\ee
\subsection{Normalization to the lowest order}
Here we expand the interpolating field $\phi_{Q}$ and the physical state
$\Phi$ in terms of powers of the inverse heavy quark mass. $\phi_{Q}$ and
$\Phi$ appear in the defining relation for the normalization constant
$N_{Q}$, so that we are also led to an expansion of $N_{Q}$. Some equations are
spelled out in higher orders of $(1/m_{Q})$
 to exhibit the expansion, and as we have need
to refer to them at a later time, when we
deal with the normalization at $O(1/m_{Q})$. Using Eqs. (\ref{eq:psih}) we can
expand the interpolating fields $\phi_{Q}$,
Eqs. (\ref{eq:int}) - (\ref{eq:intb}), as
\be
 \phi_{Q} = \phi_0 +\frac{1}{2m_{Q}} \phi_1 + \mbox{ ... }  \, ,
 \label{eq:nor1}
\ee
where $\phi_0 $ is the interpolating field in the lowest approximation.
To the lowest part $\phi_0 $ we may associate a lowest order
term in the mass expansion of the physical state. That is, we set
\bea
 \vert \Phi \rangle &=&
 \vert \Phi^0 \rangle +\frac{1}{2m_{Q}}
 \vert \Phi^1 \rangle +\frac{1}{(2m_{Q})^2}
 \vert \Phi^2 \rangle
 + \mbox{ ...  .}
 \label{eq:nor2}
\eea
{}From (\ref{eq:lsz0}) we have for pseudoscalar mesons (the extension to
other mesons and baryons is immediate)
\be
 N_{Q} \langle 0 \vert \bar{\psi}_q (0)  \bar{\chi}(v) \psi_{Q} (0)
 \vert \Phi \rangle = 1 \, ,
 \label{eq:nor3}
\ee
so that we are led to
\be
 N_{Q} =  N_{Q}^{0} +\frac{1}{2m_{Q}} N_{Q}^{1}+ \mbox{ ...  ,}
 \label{eq:nor4}
\ee
where
\be
  N_{Q}^{0} \langle 0 \vert \bar{\psi}_q (0)  \bar{\chi} (v)  Q (0)
 \vert \Phi ^{0} \rangle = 1 \, ,
 \label{eq:nor41}
\ee
and
\bea
 \phi_0(x) &=&  N_{Q}^{0}
 \bar{\psi}_q (x)  \bar{\chi} (v) Q (x) \mbox{ , }
 \label{eq:nor42}
\eea
and similarly for the higher order terms. It is important to realise
that to zeroth order $N_{Q}^{0}$ is independant of the flavour of the
heavy quark.

For a pseudoscalar meson we
have shown above that we can take
$\chi(v)=\frac{1}{2\sqrt{M_{Q}}}(1+\vslash)\gamma_5$.
Take now $v=(1,\underline{0})$ to get
\be
 \frac{N_{Q}^{0}}{\sqrt{M_{Q}}}
 \langle 0 \vert \bar{\psi}_q (0) \gamma_0 \gamma_5  Q (0)
 \vert\Phi^{0}\rangle =1\mbox{ , }  \label{eq:nor45}
\ee
where we have used the fact that upto zeroth order the heavy quark field
is essentially $Q_{+}$, i.e. it satisfies $\gamma_{0}Q_{+}=Q_{+}$ in the
rest frame of the heavy meson.

We may relate the normalization back to the pseudoscalar decay constant
$f_P$,  for we have
\be
 \langle 0 \vert A_0(0)
 \vert P \rangle =  f_P M_{Q} \mbox{ , } \nonumber
\ee
so that
\be
 \langle 0 \vert \bar{\psi}_q (0) \gamma_0 \gamma_5 \tilde Q (0)
 \vert P^{0}\rangle =  f_{P}^{0} M_{Q} \mbox{ , } \label{eq:nor46}
\ee
where $f_{P}^{0}$ is the zeroth order decay constant.
Comparing with (\ref{eq:nor45}) we have the result
\be
f_{P}^{0} = \frac{1}{ N_{Q}^{0}\sqrt{M_{Q}}}\mbox{ . }
 \label{eq:nor5}
\ee
 Remembering that the zeroth order normalisation constant is independant
of the flavour, we have thus rederived the Voloshin- Shifman \cite{vs}
scaling law for the pseudoscalar decay constant.

\section{Transitions from the Bethe-Salpeter Point of view}
In order to make use of dynamical models of bound state wave functions it
is of practical advantage to cast the theory into a form which yields
the above transition as an overlap of wave functions for the two heavy
particle states. To do this, we first recall the definition
of the Bethe-Salpeter wave function, which is the covariant generalization
of the Schr\"odinger wave function.

For simplicity let us concentrate on two-quark bound  states for the
derivation of the connection between current transitions and wave functions.
The three-quark states are treated analogously. Consider a basis $\vert
P,a \rangle$  for heavy mesons consisting of eigenstates of the four-momentum
operator $P$,
with $ a $ denoting the eigenvalues of other commuting observables.
The definition of the heavy-light
Bethe-Salpeter wave function $\Phi^{Q}(P, x_1, x_2)$ that we take is
\bea
 \Phi^{Q}(P,x_1, x_2)_{\alpha}^{\hspace*{2mm}\beta} &=&
 \langle 0 \vert    \psi_{Q \alpha}(x_1)G(x_{1},x_{2})
\bar{\psi}^{\beta}_{q}(x_2) \vert P,a \rangle \nonumber \\
 \bar{\Phi}^{Q}(P, x_1, x_2)_{\beta} ^{\hspace*{2mm}\alpha}&=&
 \langle P,a \vert    \bar{\psi}_{Q}^{\alpha} (x_1)\bar{G}(x_{1},x_{2})
 \psi_{q \beta} (x_2)
 \vert 0 \rangle \label{eq:bswf}
\eea
where time-ordering is to be understood here and subsequently. $\alpha, \beta
$ are spinor
labels and $Q,q$ denote the heavy and light flavours. $G(x_{1},x_{2})$
and $\bar{G}(x_{1},x_{2})$ are colour matrices chosen to make the B-S
amplitudes gauge invariant. The canonical choice for them is the
path-ordered exponential
\be
P exp[ig\int_{x_{1}}^{x_{2}}A.dx]\,.
\ee
However we will not make an explicit choice yet.
In our case we are
interested in the $  \vert P, a \rangle
$  corresponding to a $b$-meson $  \vert \Phi_b \rangle $ or
 to a $c$-meson $
\vert \Phi_{c} \rangle $ \footnote{As all particle states that will
appear will have only one given velocity, there is no chance of confusion in
not
including the corresponding momentum explicitly in the kets.}.

The Bethe-Salpeter wave function is related to the two-body
propagator \cite{lu}\footnote{From now on we drop the Dirac indices. We shall
reinstate them whenever necessary. For example, ${\cal K}_{Q}(x_1,x_2:x_3,x_2)$
in (\ref{eq:k}) is a four spinor valued function ${\cal K}_{\rho
\delta}^{\alpha \beta}$.There is no sum over Dirac indices in this equation.}
\be
{\cal K}_{Q}(x_1,x_2;x_3,x_4) = \langle 0 \vert
  \psi_{Q}(x_1)G(x_{1},x_{2}) \bar{\psi}_{q}(x_2)
\bar{\psi}_{Q}(x_3)\bar{G}(x_{3},x_{4}) \psi_{q}(x_4) \vert 0 \rangle
 \label{eq:k}
\ee
by inserting a complete set of states between  $\bar{\psi}_{q}$ and $
\bar{\psi}_{Q}$. In this way we get
\be
{\cal K}_{Q}(x_1,x_2;x_3,x_4) = \sum_{\vert P,a \rangle }
\Phi^{Q}(P, x_1,x_2)
\bar{\Phi}^{Q}(P, x_3,x_4) ,(t_1,t_2\geq t_3,t_4),  \nonumber
\ee
where $ \Phi^{Q}(P)$ and $ \bar{\Phi}^{Q}(P)$ are the Bethe-Salpeter
amplitudes defined above.

To get at the $b \rightarrow c$ transition, begin with the Green's function
(c.f.
(\ref{eq:lsz1}))
\bea
& & {\cal G}(x_1,x_2;z;y_1,y_2) \nonumber \\
 & & = \langle 0 \vert \psi_{c}(x_{1})G(x_{1},x_{2})\bar{\psi}_{q}(x_{2})
 \bar{\psi}_{c}(z). \Gamma .\psi_{b}(z)
\bar{\psi}_{b}(y_{1})\bar{G}(y_{1},y_{2}) \psi_{q}(y_{2})
  \vert 0 \rangle
 \nonumber \\
 & & =  \sum_{\vert p,a \rangle ,\vert q,b \rangle }
 \langle 0 \vert \psi_{c}(x_{1})G(x_{1},x_{2})\bar{\psi}_{q}(x_{2})
  \vert p, a \rangle \langle p, a \vert
 \bar{\psi}_{c}(z). \Gamma. \psi_{b}(z)
  \vert q, b \rangle \, \cdot \nonumber \\
& &  \; \; \; \; \; \; \; \; \; \; \hspace*{2cm} \langle q, b \vert
 \bar{\psi}_{b}(y_{1})\bar{G}(y_{1},y_{2})\psi_{q}(y_{2})
  \vert 0
\rangle \, , \label{eq:bsc}
\eea
with $x_{10 }\geq x_{20 }\geq z_{0 }\geq y_{10 }\geq y_{20 }$ and
$\Gamma$ any matrix (for $(V-A)$ interactions it is
$\gamma_{\mu}(1-\gamma_{5}))$.
Both summations are over complete sets of states. We wish to project out
the two-particle bound states. To do this, one overlaps with the appropriate
Bethe-Salpeter wave functions and makes use of an orthonormality
condition to obtain \cite{lu}
\bea
&&\langle \Phi_{c} \vert \bar{\psi}_{c}(z) \Gamma \psi_{b}(z) \vert
\Phi_{b} \rangle =\nonumber \\
&&\int d^{4}x_{1} d^{4}x_{2} d^{4}y_{1} d^{4}y_{2}
\bar{\Phi}^{c}_{P_c}(x_1,x_2) {\cal
T}(x_1,x_2;z;y_1,y_2)
\Phi^{b}_{P_b}(y_1,y_2) \, , \label{eq:bsc1}
\eea
where ${\cal T}(x_1,x_2;z;y_1,y_2)$ is defined by
\bea
 && {\cal G}(x_1,x_2;z;y_1,y_2)=
\nonumber \\
 && \int  d^{4}x_{3} d^{4}x_{4} d^{4}y_{3} d^{4}y_{4} \,
{\cal K}_{c}(x_1,x_2;x_3,x_4) {\cal
T}(x_3,x_4;z;y_3,y_4) {\cal K}_{b}(y_3,y_4;y_1,y_2) \, ,\nonumber\\
&&
\label{eq:tran}
\eea
so that ${\cal T}(x_3,x_4;z;y_3,y_4)$ is two-particle irreducible on both the
$(x_3,x_4)$ and $(y_3,y_4)$ legs. ${\cal T}(x_3,x_4;z;y_3,y_4)$
is sometimes determined in lowest order perturbation theory to be
\be
{\cal T}(x_3,x_4;y_3,y_4;z) = \delta ^{(4)}(x_3-z)
 \delta ^{(4)}(y_3-z) S_q^{-1}(x_4,y_4) \otimes \Gamma \, , \label{eq:mn}
\ee
where $S_q(x_4,y_4)= \langle 0 \vert \psi_{q}(y_4)\bar{\psi}_{q}(x_4)
\vert 0 \rangle $ is the usual light quark propagator.
For our purposes we keep ${\cal T}(x_3,x_4;z;y_3,y_4)$ in the implicit form
(\ref{eq:tran}) for the present.

As mentioned previously, the
effective theory tells us nothing about confinement, so that in
particular it is inappropriate to use it in determining the Bethe-Salpeter
wave functions (\ref{eq:bswf}). It is in the region of interaction that the
effective theory may be used. That is, it is reasonable to expect that
the transition kernel ${\cal T}(x_1,x_2;z;y_1,y_2)$ may be accurately
determined
within a $1/m_{Q}$ expansion for large $m_{Q}$ and in a region where
there are no hard gluons.

${\cal T}(x_1,x_2;z;y_1,y_2)$ is to be calculated by using the defining
equations (\ref{eq:k}, \ref{eq:bsc}, \ref{eq:tran}) for the Greens
functions and evaluating them within the effective field theory.
Begin with the definition (\ref{eq:bsc}), replace the $\psi_{b}$ and
$\psi_{c}$ by the zeroth order fields $b$ and $c$ and use (\ref{eq:un}) to
go the uncoupled fields $\tilde{b}$ and $\tilde{c}$ to obtain
\bea
& & \hspace*{-6mm} {\cal G}(x_1,x_2;z;y_1,y_2) \nonumber \\
 & &  = \langle 0 \vert \psi_{c}(x_{1})G(x_{1},x_{2)}\bar{\psi}_{q}(x_{2})
   \bar{\psi}_{c}(z). \Gamma. \psi_{b}(z)
\bar{\psi}_{b}(y_{1})\bar{G}(y_{1},y_{2})\psi_{q}(y_{2})
  \vert 0
\rangle \nonumber \\
& &  = \langle 0 \vert
 W\left[  {\textstyle{x_{1} \atop
v_{c}}}\right]\tilde{c}(x_{1})
G(x_{1},x_{2})
\bar{\psi}_{q}(x_{2})
 \bar{\tilde{c}}(z). \Gamma.
W\left[  {\textstyle{z \atop v_{c}}}\right]^{-1}
W\left[  {\textstyle{z \atop v_{b}}}\right]  \nonumber \\
& & \; \; \; \tilde{b}(z)
\bar{\tilde{b}}(y_{1})W\left[  {\textstyle{y_{1} \atop
v_{b}}}\right]^{-1}
\bar{G}(y_{1},y_{2}) \psi_{q}(y_{2})
 \vert 0
\rangle \nonumber \\
& & = -\langle 0
\vert \tilde{c}(x_{1}) \bar{\tilde{c}}(z) \vert 0
\rangle
 \langle 0 \vert Tr\bar{\psi}_{q}(x_2)
  \Gamma  \psi_{q}(y_2)G(x_{1},x_{2})
  W\left[  {\textstyle{x_{1} \atop v_{c}}}
           {\textstyle{z \atop v_{c}}}
           {\textstyle{z \atop v_{b}}}
           {\textstyle{y_{1} \atop v_{b}}}\right]
\bar{G}(y_{1},y_{2})
\vert 0
\rangle. \nonumber\\
&&\;\;\;\;\;\;\;\;\;\;\;\;\;\;\;\langle 0 \vert
\tilde{b}(z)
\bar{\tilde{b}}(y_{1}) \vert 0 \rangle ,
\eea
where in the last step we have used the fact that the transformed heavy
quark fields, $b$ and $c$, decouple from the gluons. The trace is over
the colour indices.
Likewise we find for the two-body propagator, (\ref{eq:k}),
\bea
  {\cal K}_{b}(x_1,x_2;x_3,x_4) &=& \langle 0 \vert  \psi_{b}(x_{1})
G(x_{1},x_{2})\bar{\psi}_{q}(x_{2})
 \bar{\psi}_{b}(x_{3})\bar{G}(x_{3},x_{4}) \psi_{q}(x_{4})
  \vert 0
\rangle \nonumber \\
& =& -\langle 0 \vert  \tilde{b}(x_{1})
\bar{\tilde{b}}(x_{3})
\vert 0 \rangle
\langle 0 \vert Tr  \bar{\psi}_{q}(x_{2})
  \psi_{q}(x_{4})G(x_{1},x_{2})
\nonumber \\
& & \; \; \;  W\left[  {\textstyle{x_{1} \atop v_{b}}}\right]
  W\left[  {\textstyle{x_{3} \atop v_{b}}}\right]^{-1}\bar{G}(x_{3},x_{4})
 \vert 0 \rangle \, ,
\eea
where there is, again, a trace over the colour indices.
By inserting these into the expression (\ref{eq:tran}), one is able to
solve for ${\cal T}(x_1,x_2;z;y_1,y_2)$
\bea
{\cal T}(x_3,x_4;z;y_3,y_4)
&=&-\int d^{4}s_{1} d^{4}s_{2}
\delta^{4}(x_{3}-z) \delta^{4}(y_{3}-z) {\cal
S}^{-1}_{q}(v_c;x_{4},s_{1};x_{1},x_{3})
\nonumber \\
& & \;\;\; \tilde{{\cal G}}(x_{1},s_{1};z;y_{1},s_{2}) {\cal
S}^{-1}_{q}(v_b;s_{2},y_{4};y_{3},y_{1}) \, , \label{tau}
\eea
where
\bea
&&{\cal S}_{q}(v_{c};x_{2},x_{4};x_{1},x_{3})\nonumber\\
&& = \langle 0 \vert Tr\psi_{q}(x_{2})
\bar{\psi}_{q}(x_{4})G(x_{1},x_{2})
  W\left[  {\textstyle{x_{1} \atop v_{c}}}\right]
  W\left[  {\textstyle{x_{3} \atop v_{c}}}
   \right]^{-1}\bar{G}(x_{3},x_{4})
 \vert 0 \rangle \, ,
\eea
\bea
&&{\cal S}_{q}(v_{b};y_{4},y_{2};y_{3},y_{1})\nonumber\\
&& = \langle 0 \vert Tr\psi_{q}(y_{4})
\bar{\psi}_{q}(y_{2})G(y_{3},y_{4})
  W\left[  {\textstyle{y_{3} \atop v_{b}}}\right]
  W\left[  {\textstyle{y_{1} \atop v_{b}}}
   \right]^{-1}\bar{G}(y_{1},y_{2})
 \vert 0 \rangle \, ,
\eea

and
\bea
&&\tilde{{\cal G}}(x_{1},x_{2};z;y_{1},y_{2})\nonumber\\
&& = \langle 0 \vert Tr\bar{\psi}_{q}(x_{2})
    \psi_{q}(y_{2})G(x_{1},x_{2})
  W\left[  {\textstyle{x_{1} \atop v_{c}}}
           {\textstyle{z \atop v_{c}}}
           {\textstyle{z \atop v_{b}}}
           {\textstyle{y_{1} \atop v_{b}}}\right]\bar{G}(y_{1},y_{2})
  \vert  0 \rangle  \otimes \Gamma \, .\nonumber\\
&&
\eea
Again in the above equations the traces are over the colour indices.
The ${\cal S}^{-1}$ are defined through
\be
\int d^{4}x_{4}{\cal S}_{q}(v_{c};x_{2},x_{4};x_{1},x_{3})
{\cal S}_{q}^{-1}(v_{c};x_{4},s;x_{1},x_{3})=\delta^{4}(s-x_{2})
\ee
and
\be
\int d^{4}y_{4}{\cal S}_{q}^{-1}(v_{b};s,y_{4};y_{3},y_{1})
{\cal S}_{q}(v_{b};y_{4},y_{2};y_{3},y_{1})=\delta^{4}(s-y_{2})\,.
\ee
An important result follows from this.
We see that in the heavy quark effective theory one
can factorize the Green's function and so solve for  ${\cal T}$, a
situation that is not available in the full theory, except at the
perturbative level.

\subsection{The Form of the Bethe-Salpeter Wave Function}
A comparison of (\ref{tau})
with (\ref{eq:lsz4b}) or (\ref{eq:lsz4a})
allows us to infer the Bethe-Salpeter wave function of
the heavy meson to be of the form
\be
 \Phi^{Q\beta}_{\alpha}(x_{1},x_{2}) =
\chi_{\alpha}\,^{\delta}(v)A_{\delta}^{\beta}(x_{1},x_{2}) \, ,
\label{eq:hmbs}
\ee
or in momentum space
\be
 \Phi^{Q\beta}_{\alpha}(P,k) =
\chi_{\alpha}\,^{\delta}(v)A_{\delta}^{\beta}(P,k) \, ,
\label{eq:hmbsms}
\ee
for some multi-spinor function $A$. This may be obtained directly from
the LSZ representation of $\Phi^{Q}$ as follows. Writing
the state $\vert P,a \rangle$ in (\ref{eq:bswf}) in terms of the
interpolating field (\ref{eq:int2}), one sees that the B-S amplitude
for a heavy meson can be written as
\bea
&&\Phi_{\alpha}\,^{\beta}(x_{1},x_{2})\nonumber\\
&&=\chi_{\rho}\,^{\delta}(v)N\lim_{P^{2}\rightarrow M^{2}}
(P^{2}-M^{2}) \langle 0 \vert \psi_{Q\alpha}(x_{1})G(x_{1},x_{2})
\bar{\psi}_{q}^{\beta}(x_{2})
\nonumber \\
& & \;\;\; \int d^{4}y e^{-iP.y}\bar{\psi}_{Q}^{\rho}(y)\psi_{q\delta}(y)
\vert 0 \rangle\nonumber\\
&&= \chi_{\rho}\,^{\delta}(v)A_{\alpha
\delta}^{\rho \beta}(x_{1},x_{2})\,.
\label{hmbs1}
\eea
The form (\ref{eq:hmbs}) follows when we take the heavy quark fields,
$\psi_{Q}$, to be the corresponding fields  in the effective
theory\footnote{It follows from the discussion in the subsection on the
Bargmann-Wigner wave functions and interpolating fields that the form of
the B-S amplitude (\ref{hmbs1}) is valid also for light mesons, if we
take $Q$ to be a light quark. It only simplifies on going to the heavy
decoupled effective fields.}.
Thus when we transform to the
decoupled effective fields (c.f. \ref{eq:un}) we get
\bea
&&\Phi_{\alpha}^{Q\beta}(x_{1},x_{2})\nonumber\\
&&=\chi_{\rho}\,^{\delta}(v)N_{Q}\lim_{P^{2}\rightarrow M^{2}}
(P^{2}-M^{2}) \langle 0 \vert \tilde{Q}_{\alpha}(x_{1})
W\left[  {\textstyle{x_{1} \atop v}}\right]
G(x_{1},x_{2})\bar{\psi}_{q}^{\beta}(x_{2})\nonumber\\
&&.\int d^{4}y e^{-iP.y}\bar{\tilde{Q}}^{\rho}(y)
W\left[  {\textstyle{y \atop v}}\right]^{-1}\psi_{q\delta}(y)
\vert 0 \rangle\, .\nonumber\\
&&\label{eq:form}
\eea

Finally  the $\vslash$ in the decoupled heavy quark propagator is set to
unity (i.e. $\delta_{\alpha}^{\rho}$)
by the $\frac{1}{2}(1+\vslash)$ projector of the $\chi$, using
the same procedure as in deriving (\ref{eq:lsz4b}), leading to the form
of the B-S amplitude given in (\ref{eq:hmbs}) and (\ref{eq:hmbsms}).

The B-S amplitude (\ref{eq:hmbs}) (or \ref{eq:hmbsms}) has the property
that the heavy quark label $\alpha$ satisfies the Bargmann-Wigner
equation.(See footnote 4).
This single important fact leads to enormous simplification of matrix elements.

\subsection{Universal Form Factors and Symmetries of the Effective Theory}

Returning once more to the transition (\ref{eq:lsz4a}), we can write it in
the following compact form (it
makes no difference at this point if we use (\ref{eq:lsz4b}) rather than
(\ref{eq:lsz4a}) for the general features we are about to derive)
\be
\sqrt{M_{b}M_{c}}Tr \bar{{\bf \Gamma}}(\frac{1+\vslash_{c}}{2})
\Gamma_{\mu}(\frac{1+\vslash_{b}}{2}){\bf \Gamma}\Xi(v_{c},v_{b}) \, ,
\label{eq:bc}
\ee
with
\be
\Xi(v_{c},v_{b}) = 4N_{b}N_{c} \int d^{4}kd^{4}q M(q,k) \, \label{eq:bcx}.
\ee
Here ${\bf \Gamma} (\bar{{\bf \Gamma}})$ is either
$\gamma_{5}$ or $\epsslash$ ($\gamma_{5}$ or $\epsslash^{*}$)  depending on
whether we are dealing with a
pseudoscalar or vector particle in the initial (final) state.
Also, because of the heavy flavour
symmetry present in the zeroth order HQET, the normalisation constants
are equal upto $O(1/M)$ viz. $N_{b}=N_{c}=N$. We also note that $\Xi$ is
independent of the heavy meson masses as we have pulled out the mass
scales and therefore it is independent of flavour.

The most general form for $\Xi$ is
\be
\Xi(v_{b},v_{c}) = F_{0}(w) + \vslash_{c}F_{1}(w) +
 \vslash_{b}F_{2}(w) + \vslash_{b} \vslash_{c} F_{3}(w) \, ,
\ee
with $\omega=v_{b}.v_{c}$.
However, because of the projection operators $\frac{1+\vslash_{c}}{2}$
and $\frac{1+\vslash_{b}}{2}$, the matrices that go to
make up $\Xi$ may as well be replaced by the identity so that
(\ref{eq:bc}) takes the universal form
\be
\sqrt{M_{b}M_{c}} \xi(w) Tr \bar{{\bf
\Gamma}}(\frac{1+\vslash_{c}}{2})\Gamma_{\mu}
(\frac{1+\vslash_{b}}{2})\Gamma \,
\ee
with $\xi(w)=F_{0}-F_{1}-F_{2}+F_{3}$. This is a very strong result, for
it says that regardless of whether the transition is pseudoscalar to
pseudoscalar, pseudoscalar to vector or vector to vector,
there is only one form factor. The differences between these
various possibilities for transitions rests completely in the different
${\bf \Gamma}$'s  and $\bar{{\bf \Gamma}}$'s that appear in the trace.
This is what is referred to as the spin symmetry present in
the heavy quark effective theory. Thus doing the traces we have
\bea
\langle D(v_{c})\vert \bar{c}\gamma_{\lambda}b\vert B(v_{b})\rangle &=&
\sqrt{M_{B}M_{D}}\xi(w)(v_{b}+v_{c})_{\lambda}\nonumber\\
\langle D^{*}(v_{c})\vert \bar{c}\gamma_{\lambda}b\vert B(v_{b})\rangle &=&
\sqrt{M_{B}M_{D^{*}}}\xi(w)i\epsilon^{*\nu}v_{b}^{\rho}v_{c}^{\sigma}%
\epsilon_{\nu\rho\sigma\lambda}
\nonumber\\
\langle D^{*}(v_{c})\vert \bar{c}\gamma_{\lambda}\gamma_{5}b\vert
B(v_{b})\rangle &=&\sqrt{M_{B}M_{D^{*}}}\xi(w)[(1+w)\epsilon^{*}_{\lambda}
-v_{b}.\epsilon^{*}v_{c\lambda}]\nonumber\\
\langle D^{*}(v_{c})\vert \bar{c}\gamma_{\lambda}b\vert B^{*}(v_{b})\rangle &=&
\sqrt{M_{B^{*}}M_{D^{*}}}\xi(w)\{(v_{b}+v_{c})_{\lambda}\epsilon^{*}.\epsilon
-v_{c}.\epsilon\epsilon^{*}_{\lambda}-v_{b}.\epsilon^{*}\epsilon_{\lambda}\}
\nonumber\\
\langle D^{*}(v_{c})\vert \bar{c}\gamma_{\lambda}\gamma_{5}b
\vert B^{*}(v_{b})\rangle &=&\sqrt{M_{B^{*}}M_{D^{*}}}
\xi(w)i\epsilon^{*\nu}(v_{b}+v_{c})^{\rho}
\epsilon^{\sigma}\epsilon_{\nu\rho\sigma\lambda}
\label{mel}
\eea

Though the physics of this situation is clear, i.e. spin information is at
least of order $O(1/m_{Q})$, let us also predict this spin symmetry result
from a different more formal approach. Consider once more the transition as
expressed by (\ref{eq:trans}), and vary the $c$ quark field according to
the transformations (\ref{eq:symm}). As the action is invariant under
this change, one obtains a Ward identity
\be
\langle \Phi_{c}' \vert \bar{\psi}_{c} \Gamma \psi_{b} \vert \Phi_{b}
\rangle  = \langle \Phi_{c} \vert \bar{\psi}_{c} \gamma_{5} \epsslash  M_{0}
\Gamma \psi_{b} \vert \Phi_{b} \rangle \, , \label{eq:wi}
\ee
where the state $\langle \Phi_{c}' \vert$ is the one that corresponds
asymptotically to $\gamma_{5} \epsslash  M_{0} \chi$. On taking
$\epsilon _{\mu }$ to be the
polarisation vector for a spin one particle with four
velocity $v$, $M_{0}=1$ and $\chi$ to represent a pseudoscalar, we see
that the new state is a vector complete with the standard form of the
wave function! This allows us to relate the different $\chi$ factors, but
the reader may be wondering about the overall normalization $N_\Phi$.
Applying the spin symmetry argument to (\ref{eq:nor45}), shows that the
normalizations $N_P$, $N_V$ are the same at $O(1)$.

In this case then, (\ref{eq:wi}) relates the pseudoscalar
to vector transition with the pseudoscalar to pseudoscalar transition,
and the universal nature of heavy quark transitions is manifest.
Also the Bethe-Salpeter wave functions themselves can be shown to have a
universal multispinor factor $A$. For example, the vector meson
Bethe-Salpeter wave function $\Phi_{V}^{Q}$ is related to that of the
pseudoscalar meson $\Phi_{P}^{Q}$ via spin symmetry:
\bea
 \langle 0 \vert \gamma_5 \epsslash\tilde{Q}(x_1)
W\left[  {\textstyle{x_{1} \atop v}}
   \right]G(x_{1},x_{2}) \bar \psi_q (x_2)
  \vert P \rangle &=&
 \langle 0 \vert \tilde{Q}(x_1)W\left[  {\textstyle{x_{1} \atop v}}
   \right]G(x_{1},x_{2}) \bar \psi_q (x_2)
   \vert V \rangle \, , \nonumber \\
 \gamma_5 \epsslash \Phi_{P}^{Q} &=& \Phi_{V}^{Q}\, ,
\eea
where $\vert P \rangle$ and $\vert V \rangle$ denote pseudoscalar and
vector meson states, respectively.

\subsection{Normalization at $q^{2}_{max}$}
One immediate consequence of the heavy quark effective theory is the
normalization of the single form factors, $\xi(w)$, in the matrix
elements (\ref{mel}), at $q^{2}_{max}$ or $w=1$. This normalization
is achieved by comparing with the
conserved flavour charge number\footnote{In this equation and in the
rest of this subsection, the {\bf 1} on the right hand side stands
symbolically for $2E(2\pi)^{3}\delta^{3}(0)$ }
\be
\langle \Phi_{Q}(P) \vert \int d^{3}x \bar{\psi}_{Q}(x) \gamma_{0}
\psi_{Q}(x)  \vert \Phi_{Q}(P) \rangle  = {\bf 1} \, , \label{eq:psiist}
\ee
where $Q$ denotes a single heavy flavour. This
equation is exact in the
sense that it is  valid beyond the effective theory. Thus the zero
component of the vector part of the transition $Q \to Q $ at zero recoil
is absolutely normalized to any order in $(1/m_{Q})$.
This normalization is also valid if we take the limit that
$m_{Q} \to \infty $, where we can replace the heavy fields $\psi_{Q} (x)$
by their
effective fields $Q(x)$.
Now we prove, using the flavour symmetry of the action Eq.(\ref{eq:sumact}),
 that this normalization at $q^{2}_{max}$ extends also to heavy
flavour changing transitions  at $O(1)$. Let the unitary matrix $U$ of
Eq. (\ref{eq:utrafo}) be
\bea
 U &=& \left( \begin{array}{cc} \alpha & \beta  \\
 \gamma & \delta  \end{array} \right) \, , \nonumber
\eea
and the heavy quark in Eq. (\ref{eq:psiist}) be the charmed quark $c$.
Thus we consider the transformation
$ c \to \beta b + \delta c$,
$\bar {c}\to \bar \beta \bar{b} + \bar \delta \bar {c}$. From LSZ reduction
we learn that we can replace also the $c$ and $\bar{c}$ in the states by
the unitarily transformed ones to get
\bea
 {\bf 1} &=& \langle \Phi_{\beta b + \delta c}
\vert \int d^{3}x
(\bar \beta \bar{b}(x) + \bar \delta \bar{c}(x))
\gamma_{0}
(\beta b(x) + \delta c(x))
\vert \Phi_{\beta b + \delta c} \rangle
\nonumber \\
 &=& (\beta \bar \beta)^{2}\langle \Phi_{b }
\vert \int d^{3}x
\bar{b}(x) \gamma_{0} b(x) \vert \Phi_{b } \rangle
+ (\delta \bar \delta)^{2}\langle \Phi_{c }
\vert \int d^{3}x
\bar{c}(x)\gamma_{0}c(x)
\vert \Phi_{c } \rangle \nonumber \\
&+&(\beta \bar \beta)(\delta \bar \delta) \left( \langle \Phi_{b }
\vert \int d^{3}x
\bar{b}(x) \gamma_{0} c(x) \vert \Phi_{c } \rangle
+\langle \Phi_{c }
\vert \int d^{3}x
\bar{c}(x)\gamma_{0}b(x)
\vert \Phi_{b } \rangle \right)\,.\nonumber\\
&&
\eea
As $U$ is unitary, $(\beta \bar \beta)+(\delta \bar \delta)=1$. Using this
and the fact that $q^{2}_{max}= (M_b-M_c)^{2}$ is below threshold, so that the
amplitude $\langle \Phi_{c }
\vert \int d^{3}x
\bar{c}(x)\gamma_{0} b(x)
\vert \Phi_{b }
 \rangle $ is strictly real, we conclude that also
the zero component of the vector part of the transition $b\to c$ is
absolutely normalized at $q^{2}_{max}$ to $O(1)$, that is
\be
 \langle \Phi_{c }(P)
\vert \int d^{3}x
\bar {c}(x)\gamma_{0} b(x)
\vert \Phi_{b }(P) \rangle ={\bf 1}\,.
\ee
Comparing with the first of the equations (\ref{mel}) leads to the wellknown
result, $\xi(1)=1$, to $O(1)$

As the action (\ref{eq:sumact}) is invariant under the transformations
(\ref{eq:symm}), we have also a manifest spin symmetry
of heavy quark transitions,
as exhibited in (\ref{eq:wi}) at $O(1)$.
Using this we obtain absolute normalizations of spin related transitions.
For example, the pseudoscalar to vector transition is normalized by taking
 $\Gamma =\epsslash \gamma_5 \gamma_0 $ with $\epsilon =(0,0,0,1)$ and
 $v=(1,\underline{0})$ in (\ref{eq:wi}). Thus, we have
\bea
& & \langle V_{c }(P)
\vert \int d^{3}x
\bar {c}(x)\gamma_{3}\gamma_{5}b(x)
\vert P_{b }(P) \rangle \nonumber\\
 && \;\;\;\;=\langle P_{c }(P)
\vert \int d^{3}x
\bar {c}(x)\gamma_{0} b(x)
\vert P_{b }(P)\rangle   ={\bf 1}
\eea
using the fact that $\gamma_{0}$ acts on $(1+\gamma_{0})/2$  to give
$(1+\gamma_{0})/2$.

In Section 5 we will derive the remarkable fact that this normalization
holds also to $O(1/m_{Q})$.

\section{Effective theory to higher orders}
In this Section we will have a look at some of the interesting features
and consequences of the effective theory at $O(1/m_{Q})$. As the
derivation of the higher order form of the effective theory is no more
difficult than the derivation of the first order form we will see
firstly how to derive all the higher order terms, in the effective theory.
The $O(1/m_{Q})$ term is determined explicitly. Our aim then, in the
rest of this Section, is to apply the formalism we have just derived to
first order in the heavy quark mass.

\subsection{Derivation of Higher Order Action}

Let us write the action in terms of the fields $\tilde{Q}$ (or $Q$)
as
\be
S_{Q} = S^{0}_{Q}(v) + \sum_{n=1}^{\infty}S^{n}_{Q}(v)/(2m_{Q})^{n}
 \, \label{hqet}
\ee
where $S^{0}_{Q}(v)$ was given in (\ref{eq:bn}) and derived from the
standard action in (\ref{eq:bn0}). In order to determine the
$S^{n}_{Q}(v)$ we need to substitute (\ref{eq:psih})
into (\ref{eq:ac}) and keep terms up to $O(1/(2m_{Q})^{n})$. To simplify
the formulae we define
\bea
\Dslash^{\perp}& =& \Dslash - \vslash v.D \nonumber \\
&=& \gamma^{\mu} D_{\mu}^{\perp} \, ,
\eea
which has the useful property that
\be
\{\Dslash^{\perp},\vslash \} = 0 \, .
\ee
$D_{\mu}^{\perp}$ is the covariant derivative transverse to the
velocity.

The action (\ref{eq:ac}) is then
\bea
S_{Q} &=& \int  \bar{Q}
e^{\textstyle i (\Dslash^{\perp}) /(2m_{Q})} (i \Dslash -m_{Q}) e^{\textstyle
 i (\Dslash^{\perp}) /(2m_{Q})} Q  \nonumber \\
&=& \int  \bar{Q}
e^{\textstyle i (\Dslash^{\perp} ) /(2m_{Q})} (i \Dslash^{\perp}
+i\vslash v.D -m_{Q})
 e^{\textstyle
 i (\Dslash^{\perp}) /(2m_{Q})} Q \, . \label{eq:effact}
\eea
On setting $S^{n}_{Q}(v_{Q}) = \bar{Q}{\cal
O}_{n}Q$,
from (\ref{eq:effact}) we can find the ${\cal O}_{n}$ to be
\be
{\cal O}_{n} = \frac{2^{n}n}{(n+1)!} (i\Dslash^{\perp})^{n+1}
 + \sum_{k=0}^{n} \frac{(i\Dslash^{\perp})^{k}}{k!} i\vslash
v.D \frac{(i\Dslash^{\perp})^{n-k}}{(n-k)!}\, .
\ee

\vspace{1cm}
\noindent \underline{Exercise:} Check this equation.
\vspace{1cm}

This is rather too implicit. As an example we now show how to put $ {\cal
O}_{1}$ in a conventional form,
\bea
{\cal O}_{1} &=& -(\Dslash^{\perp})^2  -
\{\Dslash^{\perp} , \vslash v.D \}
\nonumber \\
&=&- (\Dslash^{2} - (v.D)^{2} )
\nonumber \\
&=&  -(D^2-(v.D)^2)+\frac{g}{2}
 \sigma _{\mu \nu} F^{\mu \nu} \, ,
\eea
where
\be
\sigma _{\mu \nu} = \frac{i}{2}[\gamma_{\mu},\gamma_{\nu}]
\mbox{ and } [D^{\mu},D^{\nu}]=-ig F^{\mu \nu} \, .
\ee
The action up to first order is then
\bea
\hspace*{-6mm}
S_{Q} &=& \int  \bar{Q} (i \vslash  v.D-m_{Q}) Q -
\bar{Q} \left( (D^2-(v.D)^2) - \frac{g}{2}
 \sigma _{\mu \nu} F^{\mu \nu} \right) \frac{Q}{2m_{Q}}+ \dots \, .
\label{eq:sfirord}
\eea

Returning to the question of why does the Foldy-Wouthuysen
transformation have the form that it does, we note a tell-tale feature
already present at $O(1/m_{Q})$. This is that there are no derivatives
in ${\cal O}_{1}$ in the $v$ direction. This is important for if there
were such a derivative, there then would arise terms in diagrams, where
this operator was inserted, that would go like $v.P = m_{Q}$ in the
numerator thus raising this $O(1/m_{Q})$ correction to $O(1)$. Quite
generally the appearance of $\Dslash^{\perp}$ in the transformations
ensures that no derivatives in the direction of $v$ appear at any order
in the $1/m_{Q}$ expansion. One can see this quite easily from the
equation (\ref{eq:effact}). There the only term containing such a
derivative is
\be
e^{\textstyle i (\Dslash^{\perp} ) /(2m_{Q})} i\vslash v.D
 e^{\textstyle
 i (\Dslash^{\perp}) /(2m_{Q})}=e^{\textstyle i (\Dslash^{\perp} ) /(2m_{Q})}
iv.D
 e^{\textstyle
-i (\Dslash^{\perp}) /(2m_{Q})}\vslash\,.
\ee
We now use the identity
\be
e^{B}Ae^{-B}= A +
\frac{1}{n!}\sum_{n=1}^{\infty}\underbrace{[B,[B,...[B,A]]...]]}_{n\;
times}
\ee
with $B=i(\Dslash^{\perp})/2m_{Q}$ and $A=iv.D$. In this case
$[B,A]=\frac{ig}{2m_{Q}}\gamma_{\mu}v_{\nu}F^{\mu\nu}$. Thus to all orders
one does not get derivatives in the $v.D$ direction.

The operators ${\cal O}_{n}$ are appropriate for the $Q$
fields. Equivalent operators $\tilde{{\cal O}}_{n}$ may be found for the
$\tilde{Q}$ fields from
\be
S^{n}_{Q}(v_{Q}) = \bar{Q}{\cal
O}_{n}Q = \bar{\tilde{Q}}\tilde{{\cal O}}_{n}\tilde{Q} \, .
\ee
The required relationship is
\be
\tilde{{\cal O}}_{n}(x) =
  W\left[  {\textstyle{x \atop v}}
            \right] ^{-1}
 {\cal O}_{n}(x)
  W\left[  {\textstyle{x \atop v}}
           \right]
 \, ,
\ee
and as the $\tilde{{\cal O}}_{n}$ only involve covariant derivatives,
this relationship may also be expressed as
\be
\tilde{{\cal O}}_{n}(A) = {\cal O}_{n}(A') \, ,
\ee
where the gauge field $A'$ is given in (\ref{eq:Ap}) and satisfies
$v.A'=0$.

In order to get an action which completely disentangles the high and low
frequency states (i.e. $Q_{+}$ from $Q_{-}$) requires another set of
transformations. We will not need that action and leave this as an
exercise to the interested reader (see \cite{kt1}).

\subsection{Normalization of the interpolating field at $O(1/m_{Q})$}
 From the normalization of the interpolating field (\ref{eq:nor3}), the
expansion of $N_{Q}$ (\ref{eq:nor4})
and of the physical state (\ref{eq:nor2}), and taking the action
correction $S_1$ into account, we get for mesons
\bea
  1 &=& \left( N_{Q}^{0}+\frac{1}{2m_{Q}}N_{Q}^{1}\right)
 \langle 0 \vert \bar \psi_q \chi \left(
 1+i \frac{\Dslash^{\perp}}{2m_{Q}} \right)\tilde{Q} \vert \Phi^{0}
 +\frac{1}{2m_{Q}}\Phi^{1} \rangle \nonumber \\
 && + N_{Q}^{0} \langle 0 \vert \bar \psi_q \chi \tilde{Q}
  \frac{S_1}{2m_{Q}} \vert \Phi^{0} \rangle  \, .
 \label{eq:norcorre}
\eea
Here $N_{Q}^{1}$ can be read off to be
\bea
  N_{Q}^{1} &=& -(N_{Q}^{0})^{2} \left( \langle 0 \vert \bar \psi_q \chi
 i \Dslash^{\perp} \tilde{Q} + \psi_q \chi \tilde{Q}  S_1
 \vert \Phi^{0} \rangle
 + \langle 0 \vert \bar \psi_q \chi \tilde{Q} \vert \Phi^{1} \rangle
 \right) \, .
\eea
If there are $B$-corrections (see Sec. 3.2), they must also be incorporated in
(\ref{eq:norcorre}).
\subsection{$\frac{1}{m_{Q}}$ corrections to form factors}
In this subsection we consider the $O(1/m_{Q})$ corrections to the zeroth
order meson transition form factor $\xi(w)$.
Here we will establish that there are no $0(1/m_{Q})$ corrections to the
normalization, at zero recoil, of the form factor that we found in Section 4.2.
The strategy is the following. One relates, once more, the vector part
of the transition at equal velocity to the $c$-number charge. As this
has already been normalized to $1$ at $0(1)$, any corrections at
$0(1/m_{c})$ necessarily vanish. This, together with the spin symmetry,
yields the desired result. We give two proofs of this result. The first
is purely diagrammatic, while the second is rather more explicit. In the
second method we explicitly calculate the $O(1/m_{c})$ corrections to the
mesonic matrix elements of the $b\rightarrow c$ vector and axial transition
currents away from the zero recoil point and show that they vanish at
zero recoil. The $O(1/m_{b})$ correction can be evaluated in an analogous way.

There are three possible sources of corrections that need to be taken
into account. One stems from the extra term that appears at this order in
the action, namely $S^{1}_{Q}$. The second correction is due to the fact
that the fields $\psi_{c}$ that appear in the reduction formulae are not
just the lowest order fields $c$ but rather $(1 + i\Dslash
^{\perp}_c/(2m_{c})) c$. The last contribution can come from
perturbative corrections to the $B$-functions.

{}From the methods of proof it will be apparent that {\em both} $1/m_{c}$
and $1/m_{b}$ corrections vanish at threshold. We therefore only address the
$0(1/m_{c})$ corrections directly.

\subsubsection{Diagrammatic Proof}
This proof was given in \cite{kt1} and elaborated upon in \cite{kt2}. Let
all of the corrections be thought of as insertions into the lowest order
diagram and represent these insertions by a cross. The cross stands for
either a momentum kick or a source of glue or both. For the $b
\rightarrow c$ transition this is exhibited in fig. 2. Note that
the correction at the vertex is only on the $c$-quark side. Also there
should be (dressed) gluon lines running between all the
quark lines\footnote{In the gauge $v_{b}.A =0$ there are no gluons at
all connecting the heavy quarks. In a general gauge these do contribute, but
always in the form of Wilson lines.}, which for ease of visualization are
not drawn. The corresponding charm charge
normalization diagrams that should sum to zero are exhibited in fig. 3.
 There are two crosses on the vertex corresponding to the corrections on
either side of the vertex.

The charm charge normalization diagrams may be expressed as twice those
of fig. 2  where once more at the vertex the cross is a correction on
the right hand side only. This doubling up occurs because we may read the
diagrams in any way we choose as the momentum flow is the same on either
side and $Tr (\gamma_{\mu_{1}} \dots \gamma_{\mu_{n}}) = Tr
(\gamma_{\mu_{n}}  \dots \gamma_{\mu_{1}})$.  But these last diagrams
are just those of the zero component of the vector part of the $b
\rightarrow c$ transition. Hence that component of the transition must
vanish. To show that the axial component also vanishes, one invokes the
fact that the spin symmetry also implies a normalization condition
(see Section 2.7).

For this proof to work, we need to make the assumption that there is a
sensible $1/m_{Q}$ expansion of the $B$-function (see Sec. 3.2). We have no
control over
this term, and it would require some `modelling' to ascertain its form.
This lies outside the scope of these lectures and we take as a working
assumption, that such an expansion exists.

\subsubsection{$\frac{1}{m_{c}}$ correction to vector and axial transition
matrix elements}
In this subsection we follow the proof given by Luke \cite{l}. Let us first
look at the correction coming from the fact that,
upto $O(1/m_{c}$, in the reduction formulae we do not have just the field
$c$ but $(1+i\Dslash_{c}^{\perp}/(2m_{c}))c$. This
firstly leads to a modification in the currents
\be
\bar{c}\Gamma b\rightarrow\bar{c}\Gamma b
-\frac{1}{2m_{c}}\bar{c}i\stackrel{\leftarrow}{\Dslash_{\perp}}
\Gamma b\,,\label{vercorr}
\ee
where the $\perp$ subscript is with respect to $v_{c}$.
Secondly the field corrections will also enter in the interpolating fields
(\ref{eq:nor1}). These correspond to corrections to the
physical state (\ref{eq:nor2}).
The total field
corrections to the $\chi \tilde{c} \bar {\tilde c} \Gamma$ (when one
transforms to the free fields) portion of the LSZ
reduction formula are thus
\bea
 && \frac{i}{2m_c}\left( \bar{\chi}(v_{c})
 \Dslash ^{\perp}(x) \tilde c (x) \bar {\tilde c}(z)\Gamma
 -\bar{\chi}(v_{c})
 \tilde c (x) \bar {\tilde c}(z)
 \stackrel{\leftarrow}{\Dslash ^{\perp }}(z)\Gamma \right) \mbox{ , }
 \eea
the first term being an interpolating field correction, the second is the
correction to the current (vertex correction) already written in eq.
(\ref{vercorr}).
The charm propagator is made up of $\vslash_{c} $ and the identity matrix,
while $\bar{\chi}(v_{c}) $ projects onto the $(1+\vslash_{c} )$
component. One may then move
the $\Dslash ^{\perp}$ to the right hand side of the first term without
engendering any new gamma matrix structure. So as far as the
 gamma matrix structure is concerned both terms are of the same type. One
can do better, and establish that the wave function correction vanishes
completely when one goes to the mass pole. This is not needed for our
subsequent analysis; but the reader might like to keep in mind that for all
intents and purposes, $\vert \Phi ^1 \rangle =0 $ in
(\ref{eq:nor2}). Thus we will only consider the vertex correction, i.e.
the correction due to the modified currents. Let us now calculate this
for the $B\rightarrow D$ and $B\rightarrow D^{*}$ transitions. Thus we
begin by considering the following matrix element
\be
-\frac{1}{2m_{c}}\langle D(v_{c})\vert
(\bar{c}i\stackrel{\leftarrow}{\Dslash}_{\perp}\Gamma_{\mu} b)
(0)\vert B(v_{b})\rangle\,,
\ee
where $\Gamma_{\mu}=\gamma_{\mu}(1-\gamma_{5})$.
To evaluate this consider first the matrix element
\be
\langle D(v_{c})\vert
(\bar{c}i\stackrel{\leftarrow}{D}_{\lambda}\Gamma_{\mu}b)
(0)\vert B(v_{b})\rangle\,.
\ee
{}From our previous analysis, in Sections {\bf 3} and {\bf 4}, we see
that, because $\stackrel{\leftarrow}{D}_{\lambda}$ does not contain any
gamma matrices, this matrix element can be written immediately as
\bea
&&\sqrt{M_{B}M_{D}}Tr\gamma_{5}(\frac{1+\vslash_{c}}{2})\Gamma_{\mu}
(\frac{1+\vslash_{b}}{2})\gamma_{5}
\left[Av_{b\lambda}+Bv_{c\lambda}+C\gamma_{\lambda}\right]\nonumber\\
&&=\sqrt{M_{B}M_{D}}(v_{b}+v_{c})_{\mu}(Av_{b\lambda}+Bv_{c\lambda})
\nonumber \\
& & \;\;\;-C[g_{\mu\lambda}(1-w)+v_{c\mu}v_{b\lambda}+v_{c\lambda}v_{b\mu}]
-C4i\epsilon_{\mu\lambda\rho\sigma}v_{c}^{\rho}v_{b}^{\sigma}\, ,
\label{corr1}
\eea
where $A,B,C$ are unknown fuctions of $w=v_{b}.v_{c}$.
Now multiply by $v_{c}^{\lambda}$ and use the equation of motion
\be
\bar{c}iv_{c}.\stackrel{\leftarrow}{D}=-m_{c}\bar{c}\vslash_{c}
\ee
to write
\be
-m_{c}\langle D(v_{c})\vert
\bar{c}\vslash_{c}\Gamma_{\mu}b\vert B(v_{b})\rangle
=\sqrt{M_{B}M_{D}}(Aw+B-C)(v_{b}+v_{c})_{\mu}\,.
\ee
On the left hand side we now notice that $\vslash_{c}$ will be hit by
the projector $\frac{1+\vslash}{2}$, on going to the free ($\tilde{c}$)
field,  and therefore can be set to unity.
Recall the discussion at the end of Section 3.1 where the projector has
already been used to remove the $\vslash_{c}$ from the c-quark
propagator.
Hence we get
\be
-m_{c}\langle D(v_{c})\vert
\bar{c}\Gamma_{\mu}b\vert B(v_{b})\rangle
=\sqrt{M_{B}M_{D}}(Aw+B-C)(v_{b}+v_{c})_{\mu}\,.\label{corr2}
\ee
Comparing with the first of equations (\ref{mel}), we get
\be
Aw+B-C=-m_{c}\xi\,.\label{corr3}
\ee
We also have, upto $O(1/m_{Q})$,
\bea
\langle D(v_{c})\vert i\partial_{\lambda}
(\bar{c}\Gamma_{\mu}b)\vert B(v_{b})\rangle&&\nonumber\\
&=&(P_{b}-P_{c})_{\lambda}\vert
\bar{c}\Gamma_{\mu}b\vert B(v_{b})\rangle\nonumber\\
&=&(M_{B}v_{b\lambda}-M_{D}v_{c\lambda})(v_{b}+v_{c})_{\mu}\xi(w)\,.
\label{corr4}
\eea
The left hand side of this equation can now be written as
\be
\langle D(v_{c})\vert
(\bar{c}i\stackrel{\leftarrow}{D}_{\lambda}
\Gamma_{\mu}b)\vert B(v_{b})\rangle
+
\langle D(v_{c})\vert
(\bar{c}
\Gamma_{\mu}i\stackrel{\rightarrow}{D}b)\vert B(v_{b})\rangle\,.
\label{corr5}
\ee
The first term in the above expression we have already evaluated in eq.
(\ref{corr1}). Hence multiplying eq. (\ref{corr4}) by $v_{b}^{\lambda}$
and using the equations of motion
$iv_{b}.Db=m_{b}\vslash_{b}b$ and the same argument we
used to derive eq. (\ref{corr2}) we arrive at
\be
A+Bw-C=(M_{B}-m_{b})\xi(w)-M_{D}w\xi(w)\,.\label{corr5a}
\ee
Comparing with eq. (\ref{corr3}), we get
\bea
A(1+w)-C&=&\bar{\Lambda}\xi(w)\,,\nonumber\\
A-B&=&M_{D}\xi(w)\,,\label{corr6}
\eea
where $\bar{\Lambda}=(M_{B}-m_{b})=(M_{D}-m_{c})$.

Now let us finally look at the vertex correction
\bea
&&-\frac{1}{2m_{c}}\langle D(v_{c})\vert
\bar{c}i\stackrel{\leftarrow}{\Dslash}_{\perp}\Gamma_{\mu}b
\vert B(v_{b})\rangle \nonumber\\
&&\;=-\frac{1}{2m_{c}}\langle D(v_{c})\vert
\bar{c}i\stackrel{\leftarrow}{D}_{\lambda}\gamma^{\lambda}_{\perp}
\Gamma_{\mu}b
\vert B(v_{b})\rangle\nonumber\\
&&\;=-\frac{1}{2m_{c}}\sqrt{M_{b}M_{c}}Tr\gamma_{5}\frac{(1+\vslash_{c})}{2}
\gamma_{\perp}^{\lambda}\Gamma_{\mu}\frac{(1+\vslash_{b})}{2}\gamma_{5}
(Av_{b\lambda}+Bv_{c\lambda}+C\gamma_{\lambda})\,.\nonumber\\
&&
\eea
Here the functions $A,B$ and $C$ are the same as in eq. (\ref{corr1}).
We see immediately that only the vector transition current contributes
and the trace can be worked out to give
\be
-\frac{1}{2m_{c}}\sqrt{M_{b}M_{c}}(v_{b}-v_{c})_{\mu}\left(A(1+w)-3C\right)\,,
\ee
which, on using eq. (\ref{corr6}), becomes
\be
\frac{1}{2m_{c}}\sqrt{M_{b}M_{c}}(v_{b}-v_{c})_{\mu}\left
(2A(1+w)-3\bar{\Lambda}\xi(w)\right)\,.\label{corr7}
\ee

The second kind of correction comes from the fact that we have to
include the effects of the new terms which appear in the action at order
$O(1)/m_{c}$
\be
S_{1}=\frac{1}{2m_{c}}\int Q\left((iD)^{2}-(iv.D)^{2}+\frac{g}{2}
\sigma_{\mu\nu}F^{\mu\nu}\right)\,.
\ee
Inserting this in the path integral leads to a correction equal to
\bea
&\frac{i}{2m_{c}}\langle D(v_{c})\vert \int
d^{4}xT\left(\bar{c}(iD_{\perp})^{2}c\right)(x)(\bar{c}\Gamma_{\mu}b)(0)
\vert B(v_{b})\rangle &\nonumber\\
&+\frac{i}{2m_{c}}\langle D(v_{c})\vert \int
d^{4}xT\left(\bar{c}g\sigma_{\nu\lambda}F^{\nu\lambda}c\right)(x)%
(\bar{c}\Gamma_{\mu}b)(0)
\vert B(v_{b})\rangle\label{corr8}
\eea
Using the reduction formulae and going to the free (tilde) fields, as in
Sec. 3 we
find, since $(iD_{\perp})^{2}$ does not include $\gamma$ matrices, that the
first term will lead to the trace
\bea
&\sqrt{M_{b}M_{c}}\frac{1}{2m_{c}}Tr\gamma_{5}\frac{(1+\vslash_{c})}{2}%
\Gamma_{\mu}\frac{(1+\vslash_{b}}{2}\gamma_{5}\eta(w)&\nonumber\\
&=\sqrt{M_{b}M_{c}}\frac{1}{2m_{c}}(v_{b}+v_{c})_{\mu}\eta(w)&
\,,\label{corr10}
\eea
where $\eta(w)$ is an unknown function. This simple form appears because
the $\vslash_{c}$ appearing in the two free c-quark propagators can be
put to unity because of the projector $\bar{\chi}(v_{c})$ appearing in
the reduction formula.

In the second term we follow the same procedure but we see immediately
that because of the $\sigma_{\nu\lambda}$, the projector
$\bar{\chi}(v_{c})$ arising from the reduction cannot act directly on
the second of the free c-quark propagators. However we recall
(\ref{hqprop1}) that near the heavy quark mass pole the main
contribution in the propagator
comes from
\be
i\frac{\frac{1}{2}(1+\vslash_{c})}{v_{c}.p_{c}-m_{c}}\,.
\ee
Thus the second term leads to the following trace
\be
\sqrt{M_{b}M_{c}}\frac{i}{4m_{c}}Tr\gamma_{5}\frac{(1+\vslash_{c})}{2}
\sigma_{\nu\lambda}\frac{(1+\vslash_{c})}{2}\Gamma_{\mu}
\frac{(1+\vslash_{b}}{2}\gamma_{5}G^{\nu\lambda}\,.
\ee
It is easy to see that the only independant form of $G^{\nu\lambda}$ is
\be
G^{\nu\lambda}=Dv_{b}^{\nu}\gamma^{\lambda}+E\gamma^{\nu}\gamma^{\lambda}\,,
\ee
leading to the correction
\be
\sqrt{M_{b}M_{c}}\frac{1}{2m_{c}}[D(w-1)+3E](v_{b}+v_{c})_{\mu}\,,
\ee
where $D$ and $E$ are unknown functions of $w$.

Thus the total $O(1/m_{c})$ correction to the $B\rightarrow D$ decay is
\be
\sqrt{M_{b}M_{c}}\frac{1}{2m_{c}}\left[\left(\eta+D(w-1)+3E\right)
(v_{b}+v_{c})_{\mu}+\left(2A(1+w)-3\bar{\Lambda}\xi\right)(v_{b}-v_{c})_{\mu}
\right]\,.\label{bdcor}
\ee

We can follow the same procedure for the other s-wave matrix
elements to obtain finally the following set of matrix elements for
flavor changing currents upto $O(1/m_{c})$.
\bea
& & \langle D(v_{c})\vert \bar{c}\gamma_{\mu}b\vert
B(v_{b})\rangle\nonumber\\
&&
=\sqrt{M_{b}M_{c}}\left[\left(\xi(w)+\frac{1}{2m_{c}}\{\eta(w)+(w-1)D(w)%
+3E(w)\}\right)(v_{b}+v_{c})_{\mu}\right.\nonumber\\
&&\left.+\frac{1}{2m_{c}}\{2(1+w)A(w)-3\bar{\Lambda}\xi(w)\}
(v_{b}-v_{c})_{\mu}\right]\,,\\
&&\langle D^{*}(v_{c})\vert \bar{c}\gamma_{\mu}b\vert
B(v_{b})\rangle \nonumber\\
&&=\sqrt{M_{b}M_{c}}\left[(1+\frac{\bar{\Lambda}}{2m_{c}})\xi(w)+
\frac{1}{2m_{c}}(\eta(w)-E(w))\right]i\epsilon^{*\nu}v_{b}^{\rho}v_{c}^{\sigma}
\epsilon_{\nu\rho\sigma\mu}\,,\\
&&\langle D^{*}(v_{c})\vert \bar{c}\gamma_{\mu}\gamma_{5}b\vert
B(v_{b})\rangle \nonumber\\
&&=\sqrt{M_{b}M_{c}}\left[\epsilon^{*}_{\mu}\left\{\xi(w)\left((1+w)
-\frac{(1-w)\bar{\Lambda}}{2m_{c}}\right)
+\frac{1}{2m_{c}}(1+w)(\eta(w)-E(w))\right\}\right.\nonumber\\
&&-v_{b}.\epsilon^{*}v_{c\mu}\left\{\xi(w)(1+\frac{\bar{\Lambda}}{2m_{c}})
+\frac{1}{2m_{c}}(\eta(w)-2A(w)+D(w)-E(w))\right\}\nonumber\\
&&\left.+v_{b}.\epsilon^{*}v_{b\mu}\frac{1}{2m_{c}}(2A(w)+D(w))\right]\,.
\\
\label{corrbdm}
\eea
Here $\epsilon^{*\nu}$ is the polarisation vector of the $D^{*}$ meson.

To determine the normalisation at zero recoil it is necessary to
consider the neutral current matrix elements, which are governed by the same
set of form factors because of the heavy flavour symmetry. Following the
same procedure as for the charged currents, we get
\bea
&&\langle D(v_{2})\vert \bar{c}\gamma_{\mu}c\vert
D(v_{1})\rangle\nonumber\\
&& \;\;\;=M_{c}\left[\xi(w)+\frac{2}{2m_{c}}\left(\eta(w)
+(w-1)D(w)+3E(w)\right)\right](v_{1}+v_{2})_{\mu}\,.\\
&&\langle D^{*}(v_{2})\vert \bar{c}\gamma_{\mu}c\vert
D(v_{1})\rangle\nonumber\\
&&\;\;\;=M_{c}\left[(1+\frac{4\bar{\Lambda}}{2m_{c}})
\xi(w)\right.\nonumber\\
&&\;\;\;+\left.\frac{1}{2m_{c}}\left(2\eta(w)+2(1+w)A(w)-(1-w)D(w)-2E(w)%
\right)\right]
i\epsilon^{*\nu}v_{1}^{\rho}v_{2}^{\sigma}\epsilon_{\nu\rho\sigma\mu}\,.\\
&&\langle D^{*}(v_{2})\vert \bar{c}\gamma_{\mu}\gamma_{5}c\vert
D(v_{1})\rangle\nonumber\\
&&\;\;\;=M_{c}\left[\epsilon^{*}_{\mu}\left\{\xi(w)\left((1+w)-4(1-w)
\frac{\bar{\Lambda}}{2m_{c}}\right)\right.\right.\nonumber\\
&&\;\;\;+\left.\frac{1}{2m_{c}}\left(2(1-w^{2})A(w)
-(1-w^{2})D(w)+2(1+w)E(w)+2(1+w)\eta(w)\right)\right\}\nonumber\\
&&\;\;\;-v_{1}.\epsilon^{*}v_{2\mu}\left\{\xi(w)\left(1+
\frac{4\bar{\Lambda}}{2m_{c}}\right)
+ \frac{1}{2m_{c}}\left(2\eta(w)-2(2+w)A(w)
+wD(w)+2E(w)\right)\right\}\nonumber\\
&&\;\;\;+\left.v_{1}.\epsilon^{*}v_{2\mu}\frac{1}{2m_{c}}%
(2A(w)+D(w))\right]\,.\\
&&\langle D^{*}(v_{2})\vert \bar{c}\gamma_{\mu}c\vert
D^{*}(v_{1})\rangle\nonumber\\
&&\;\;\;=M_{c}\left[\left(\xi(w)+\frac{2}{2m_{c}}(\eta(w)-E(w))\right)
(v_{1}+v_{2})_{\mu}\epsilon_{1}.\epsilon^{*}_{2}\right.\nonumber\\
&&\;\;\;+\frac{1}{2m_{c}}\left(2A(w)-D(w)\right)(v_{1}+v_{2})_{\mu}
v_{1}.\epsilon^{*}_{2}v_{2}.\epsilon_{1}\nonumber\\
&&\;\;\;+\left\{-\xi(w)
-\frac{1}{2m_{c}}\left(2\eta(w)+2(1+w)A(w)+(1-w)D(w)-2E(w)\right)\right\}
\epsilon^{*}_{2\mu}v_{2}.\epsilon_{1}\nonumber\\
&&\;\;\;+\left.\left\{-\xi(w)-\frac{1}{2m_{c}}\left(2\eta(w)+2(1+w)A(w)%
+(1-w)D(w)
-2E(w)\right)\right\}\epsilon_{1\mu}v_{1}.\epsilon_{2}^{*}\right]\,.\nonumber\\
&&
\eea
Here $v_{1}$ and $v_{2}$ are the velocities of the incoming and outgoing
$D$-mesons, respectively and the $\epsilon_{1}$ and $\epsilon^{*}_{2}$
are the corresponding polarisation vectors.

We observe immediately that all the $O(1/m_{c})$ corrections vanish in
both the flavour changing and neutral matrix elements at the zero recoil
point $w=1$, except those terms proportional to $\eta(w)$ and $E(w)$. Next we
recall that the neutral vector current is normalised at $w=1$ since at this
point, $v_{1}=v_{2}=v$, it is a symmetry current corresponding to the
conserved $c$-quark number. Thus
\be
\langle D(v)\vert \bar{c}\gamma_{0}c\vert
D(v)\rangle=2M_{c}v_{0}=2M_{c}v_{0}\left(1+\frac{1}{2m_{c}}(\eta(1)+3E(1))
\right)
\ee
and
\be
\langle D^{*}(v)\vert \bar{c}\gamma_{0}c\vert
D^{*}(v)\rangle=2M_{c}v_{0}=2M_{c}v_{0}\left(1+\frac{1}{2m_{c}}(2\eta(1)
-2E(1))\right)\,.
\ee
Here we have used the fact that $\xi(1)=1$. Thus one gets the
normalisation condition
\be
\eta(1)=E(1)=0\,.
\ee
Thus we arrive at the conclusion that all corrections to the zeroth
order result vanish at the symmetry point $w=1$.

One can calculate the baryon transition matrix elements upto
$O(1/m_{c})$, in a similar
manner, using the baryon Bethe-Salpeter amplitudes developed earlier.
Without going into the details we give the results, upto $O(1/m_{c})$, for
the transition
$\Lambda_{b}\rightarrow\Lambda_{c}$ \cite{hkkt}, \cite{mannel3}, \cite{bb1},
\cite{bb2}, \cite{iw348}, \cite{g1}.
\bea
&&\langle \Lambda_{c}(v_{c})\vert \bar{c}\gamma_{\mu}b\vert
\Lambda_{b}(v_{b})\rangle\nonumber\\
&&\;\;\;=\bar{u}(v_{c})\gamma_{\mu}u(v_{b})\left\{\xi_{\Lambda}(w)+%
\frac{1}{2m_{c}}
\left(\eta_{\Lambda}(w)+\bar{\Lambda}\xi_{\Lambda}(w)\right)\right\}\nonumber\\
&&\;\;\;\;-\frac{1}{2m_{c}}v_{b\mu}\frac{\bar{\Lambda}\xi_{\Lambda}}{1+w}%
\bar{u}(v_{c})u(v_{b})
\eea
and
\bea
&&\langle \Lambda_{c}(v_{c})\vert \bar{c}\gamma_{\mu}\gamma_{5}b\vert
\Lambda_{b}(v_{b})\rangle\nonumber\\
&&\;\;\;=\bar{u}(v_{c})\gamma_{\mu}\gamma_{5}u(v_{b})\left\{\xi_{\Lambda}(w)
+\frac{1}{2m_{c}}\left(\eta_{\Lambda}(w)
-\frac{(1-w)\bar{\Lambda}\xi_{\Lambda}(w)}{1+w}\right)\right\}\nonumber\\
&&\;\;\;\;-\frac{1}{m_{c}}\frac{\bar{\Lambda}\xi_{\Lambda}(w)}{1+w}v_{b\mu}
\bar{u}(v_{c})\gamma_{5}u(v_{b})\,.
\eea
Here, as in the mesonic case, $\xi_{\Lambda}(w)$ and $\eta_{\Lambda}(w)$
are unknown functions normalised as follows, at the zero recoil point:
\bea
\xi_{\Lambda}(1)&=&1,\nonumber\\
\eta_{\Lambda}(1)&=&0\,.
\eea

\section{Renormalisation and the Relationship with QCD}

So far we have considered how QCD is related to the HQET theory at
the classical level. That is we have fixed the tree level coefficients
in (\ref{hqet}) by relating the theories by a Foldy-Wouthuysen
transformation. When it comes to renormalising the theories a
discrepency arises. One obvious difference is that the higher order
operators that appear on the right hand side of (\ref{hqet}) are
non-renormalisable. Due to infinities that the insertion of such operators
entails the tree level coefficients will all need to be renormalised. Indeed
as the operators in the
expansion have increasing mass dimension, the higher the order in $1/m_{Q}$ one
goes to the worse the divergence becomes. In order to renormalise
therefore, one will need to perform an infinite number of `experiments'.
That problem aside, for the moment, even the HQET$_{0}$ differs from QCD
once loops are taken into account.

It might come as a surprise that this is so, as we have derived the
HQET from QCD, and one might have expected that knowing the
renormalisation constants of QCD would be enough. The differences arise
because the transformations used do not respect the regularisation that
is needed to define QCD in the first place. To see how this situation
comes about consider the integral
\be
\int d^{4}p \frac{1}{(p^{2}-m_{Q}^{2})^{2}}
\ee
which, when regularised with dimensional regularisation, apart from a
pole has a logarithmic behaviour $\ln{m_{Q}/\mu}$. Now let us perform
the integral by expanding about the mass shell $p= m_{Q}v +k$,
\be
\int d^{4}k \frac{1}{4m_{Q}^{2}(v.k)^{2}}\left(1-\frac{k^{2}}{m_{Q}v.k} +
\dots \right) = 0 \label{sexp} \, .
\ee
(We have used the fact that $\int d^{n}k k^{m}=0$ in dimensional
regularisation). The
discrepancy lies in the fact that, while at tree level we can keep the
$k$ (and gluonic momenta) bounded by $m_{Q}$, we are unable to do so in
loops. Similar manipulations show that even for convergent
integrals one gets a mismatch. Notice that the type of terms that appear
in the series expansion are of the form of the higher order operators
that appear in the HQET.

Is this the death knell for the HQET? No not at all. This is a common
situation in effective field theories. The naive effective field theory
will correctly reproduce the physics at some scale (in our case the
long-distance physics, $k \ll m_{Q}$) but will require corrections to
reproduce the physics at other scales (here the short distance physics
or for typical momenta $k \geq m_{Q}$).
Indeed one can give a systematic treatment of the short-distance
corrections. As the effective theory is QCD in disguise, the
`experiments' that are needed to fix the renormalisation constants in
front of every term on the right hand side of (\ref{hqet}) are simply a
comparison of the effective theory and QCD up to the given order at some
scale where we will demand equality between the two theories.
We will come to terms with what this means presently, but
first we wish to establish, for external momenta $k \ll m_{Q}$, that
at leading order QCD and HQET$_{0}$ are indeed simply related.

\subsection{Renormalisability of the HQET$_{0}$}

The lowest order effective theory, HQET$_{0}$, has a power counting
renormalisable action
(essentially as the counting is the same as for QCD). In principle we can stop
here and conclude the theory is renormalisable. There is a quicker way to
arrive at this, namely that, in the gauge $v.A=0$, the heavy quark decouples
altogether from the rest of the theory. The heavy quark is free,
while the rest of the theory is renormalisable in this gauge.

If the reader
does not particularly like this gauge, then he or she can work in a covariant
gauge, again with free heavy quarks but with the inclusion of the Wilson lines
(see section \ref{fci}). Wilson lines are basic as observables in QCD
and better be renormalisable. Indeed this was postulated for such operators
of smooth loops by Polyakov \cite{pol} and proved shortly after by
\cite{dv} and \cite{bns}. A review of these matters may be found in
\cite{dorn}.

\subsubsection{Wavefunction Renormalisation of the heavy quark}
Let us calculate the wavefunction renormalisation in the Feynman gauge,
as we will have need of it later on. From now on quantities with asterix
superscripts are defined in HQET$_{0}$. The same objects without the
asterix are the equivalents in QCD. We wish to evaluate (with $p=mv+k$)
\bea
& & -\frac{(1+\vslash)}{2}g^{*2}\mu^{\epsilon}T^{a}T^{a}\int
\frac{d^{n}q}{(2\pi)^{n}}v_{\mu}\frac{1}{\vslash
v.(p+q)-m_{Q}}v^{\mu}\frac{1}{q^{2}} \nonumber \\
& & = - \frac{(1+\vslash)}{2} \frac{4}{3}g^{*2}\mu^{\epsilon} \int
\frac{d^{n}q}{(2\pi)^{n}}
\frac{1}{v.(k+q) q^{2}} \, .
\eea

One now makes use of the identity,
\be
\frac{1}{a^{n}b}= \int_{0}^{\infty} d\alpha \frac{\alpha^{n-1 }}{(a\alpha
+b)^{n+1}} \label{id}
\ee
to rewrite the self energy as
\bea
& & - \frac{(1+\vslash)}{2} \frac{4}{3}g^{*2}\mu^{\epsilon}\int_{0}^{\infty}
d\alpha \int
 \frac{d^{n}q}{(2\pi)^{n}} \frac{1}{(\alpha v.(q+k) + q^{2})^{2}}
\nonumber \\
& &= - \frac{(1+\vslash)}{2} \frac{4}{3}g^{*2}\mu^{\epsilon}
\frac{i}{(4\pi)^{2-\epsilon/2}} \Gamma(\epsilon/2) \int_{0}^{\infty} d
\alpha (\alpha^{2}/4- \alpha v.k)^{-\epsilon/2} \, .
\eea
In order to perform the last integral one scales
\be
\alpha \rightarrow -4v.k \alpha
\ee
and then changes variables to
\be
z = \frac{1}{1 + \alpha}
\ee
to arrive at
\bea
& & - \frac{(1+\vslash)}{2} \frac{4}{3}g^{*2}\mu^{\epsilon}
\frac{i}{(4\pi)^{2-\epsilon/2}} \Gamma(\epsilon/2) \int_{0}^{1}dz
(1-z)^{-\epsilon/2}z^{\epsilon-2}\nonumber \\
&=& - \frac{(1+\vslash)}{2} \frac{4}{3}g^{*2}\mu^{\epsilon}
\frac{i}{(4\pi)^{2-\epsilon/2}} (-4v.k)^{1-\epsilon} \Gamma(1-\epsilon/2)
\Gamma(\epsilon-1) \, .
\eea

The singular part is easily extracted from this expression. It is
\be
-i \frac{(1+\vslash)}{2} \frac{16}{3}\frac{g^{*2} \mu^{\epsilon}}{16
\pi^{2}}  v.k \frac{1}{\epsilon}
\ee
from which we see that the wave function renormalisation is
\be
Z_{Q} = 1 + \frac{16}{3}\frac{g^{*2} \mu^{\epsilon}}{16 \pi^{2}}
\frac{1}{\epsilon} \, . \label{ren}
\ee

The renormalised inverse propagator is, therefore,
\be
\Gamma_{(2,0)}^{*}=i \frac{(1+\vslash)}{2} v.k \left( 1 -
\frac{16}{3}\frac{g^{*2}}{16 \pi^{2}} \ln{\mu} + \dots \right) \, .
\ee
Now the anomalous dimension for the heavy quark $\gamma_{Q}^{*}$ can be
determined from the renormalisation group equation
\be
\left( \mu \frac{\partial}{\partial \mu} + \beta(g^{*})
\frac{\partial}{ \partial
g^{*}} + 2 \gamma_{Q}^{*} \right)\Gamma_{(2,0)}^{*} = 0 \, . \label{rge2pt}
\ee
The beta function is of order $g^{*3}$ and so we can ignore it
for present purposes to obtain
\be
\gamma_{Q}^{*} =  \frac{8}{3}\frac{g^{*2}}{16 \pi^{2}} \, . \label{anomQ}
\ee

Before closing this section let us return to the observation that what
we are really calculating are expectation values of Wilson loops. The
heavy quark wavefunction renormalisation should be the composite
operator renormalisation of the Wilson loops and consequently the
anomalous dimension (\ref{anomQ}) ought to be the anomalous dimension of
the Wilson loops with end points. Indeed it is as has been shown by
\cite{kr}. The reader should bear this older literature in mind. Many of
the calculations performed in the context of HQET$_{0}$ could be
extracted from past work on Wilson loop renormalisation.

\vspace{1cm}
\noindent \underline{Exercise:} Prove (\ref{id}) and fill in the details
of the calculation leading to (\ref{ren}).

\noindent \underline{Exercise:} What of the renormalisation of the
$\frac{(1-\vslash)}{2}$ component of the propagator?

\subsection{Comparison of the HQET$_{0}$ and QCD }

In the previous section we calculated the one loop anomalous dimension
of the heavy quark in the lowest order HQET$_{0}$. On the other hand, the
one loop anomalous dimension for a quark in QCD is,
\be
\gamma_{Q} = -\frac{4}{3}\frac{g^{2}}{16 \pi^{2}} \, . \label{anomq}
\ee
So there is a
mismatch between the QCD and the HQET$_{0}$ predictions. While
renormalisability
for the HQET$_{0}$ is not an issue, it is its relationship with QCD
which requires some elaboration.

In the sequel, for simplicity, we will assume that the theory of interest
is that of the heavy quark coupled to the glue, with no other matter
present. The general situation can be dealt with along similar lines but
is rather more involved.

\subsubsection{Comparison of Gluonic $1$PI Diagrams}
One of the important features of the HQET at lowest order is that there are
no heavy quark loops at all. This is easy to see. Heavy quark loops correspond
to the fermionic determinant, which is gauge invariant, and therefore we can
work in the
gauge $v.A=0$ to conclude that the determinant is gauge field independent.
Consequently there are no heavy fermion loops. An alternative way of saying
this is that, in the rest frame, the heavy quark can only propagate forward
in time, leading again to no loops. So how does this square with QCD?

In fact what is relevant here is a very important result due to Appelquist and
Carazzone \cite{ac}(see also \cite{s}). Consider an $n$ point function
with only external gluon legs whose
momenta are of the order of $k$ ($k<<m_{Q}$). These authors tell us
that, in renormalised QCD, for such diagrams, those which contain a heavy
quark loop are supressed by at least a factor of $k/m_{Q}$ relative to those
which do not contain such a loop. The detailed proof of this theorem for
a simpler model plus a useful guide to the literature may be found in
the book \cite{c}.

\vspace{1.5cm}
\noindent \underline{Convergent Diagrams}
\vspace{1cm}

In order to get a feel for this part of the Appelquist
Carazonne theorem, consider a diagram with $n \geq 5$ where all the external
gluon lines are connected to the heavy fermion loop and where all
sub-diagrams are
convergent. One has the generic expression (we drop $\gamma$-matrices,
traces and so on),
\bea
& & \int (\prod_{i=1}^{F} d^{4}q_{i}) \delta (\sum_{i=1}^{F} q_{i})
\Gamma^{F}(q_i)
(\prod_{i=1}^{F} \frac{1}{q_{i}^{2}}) \times \nonumber \\
& & \; \; \; \; \; \; \; \;
\int d^{4}p (\prod_{i=1}^{F}
\frac{1}{p-\sum_{j=1}^{i}q_{j}-m_{Q}}) (\prod_{i=1}^{n}
\frac{1}{p+\sum_{j=1}^{i}k_{j}-m_{Q}}) \, ,
\eea
where $\Gamma^{F}(q_{i})$ is the gluonic $1$PI $F$-point function, (Fig.4).

Since the degree of divergence of $\Gamma^{F}$ is $4-F$, and as the
subdiagrams are all to be convergent, we require $F \geq 5$ as well.
Scale the integration variables in this formula as $q_{i} \rightarrow
m_{Q}q_{i}$
and $p \rightarrow m_{Q}p$ and note that, as everything in sight is
convergent, the
scaling law of $\Gamma$ is $\Gamma^{F}(m_{Q}q_{i})=m_{Q}^{4-F}
\Gamma^{F}(q_{i})$. The
scaling argument shows us that the diagram behaves as
\be
c_{0}m_{Q}^{4-n} + c_{1}k m_{Q}^{3-n} + \dots
\ee
for some finite constants $c_{j}$. Such an $n$-point function without a heavy
quark loop would, by dimensional analysis, go like
\be
ck^{4-n}
\ee
validating the claim. The essence of the scaling argument that we have used is
simply that, because of the presence of the mass in the fermionic loop, we can
set the external momenta to zero without generating a divergence.

At the other extreme, consider the situation where none of the external lines
land on the heavy quark loop, (Fig.5). This time we wish to get a handle on
\be
\int (\prod_{i=1}^{F} d^{4}q_{i}) \delta (\sum_{i=1}^{F} q_{i})
\Gamma^{n+F}(k,q_i)
(\prod_{i=1}^{F} \frac{1}{q_{i}^{2}}) \int d^{4}p (\prod_{i=1}^{F}
\frac{1}{p-\sum_{j=1}^{i}q_{j}-m_{Q}})  \, .
\ee
Suppose that $5 \leq F <n$. In this case the heavy quark loop behaves like
$m_{Q}^{4-F}$ as long as the loop momenta $q$ are small. In any case contract
the loop to a point and consider the reduced diagram. The reduced diagram's
overall degree of divergence is $(F-n)<0$, so that it superficially
converges. All
the subdiagrams converged and so we may conclude (by Weinberg's theorem) that
the reduced graph converges. The shrinking of the loop to a point is
the $m_{Q} \rightarrow \infty$ limit of the original diagram, so that we find
that such graphs behave as $m_{Q}^{4-F} k^{F-n}$, which are once more
supressed with respect to diagrams without fermion loops. When $(F-n)\geq
0$, one estimates this integral by looking at the regions
where all the integration momenta are small (relative to $m_{Q}$) which
is the dominant contribution (again by application of Weinbergs
theorem). One can show that such diagrams behave as $m_{Q}^{4-n}$. These
are clearly supressed.

\vspace{1.5cm}
\noindent \underline{Divergent Diagrams}
\vspace{1cm}

We now have to address the question of power counting divergent diagrams.
According to Appelquist and Carazzone, once one renormalises the basic
divergent graphs then their inclusion into other diagrams is subdominant. The
argument for this is that, heuristically, the subtracted diagrams (at the scale
$\mu \ll m_{Q}$) behave as (here $\Lambda$ is a cutoff)
\be
[\ln{\frac{\Lambda}{m_{Q}}} + O(\frac{k}{m_{Q}})] -
[\ln{\frac{\Lambda}{m_{Q}}} + O(\frac{\mu}{m_{Q}})] = O(\frac{k}{m_{Q}},
\frac{\mu}{m_{Q}})\,. \label{est}
\ee
An example of this behaviour (in dimensional regularisation) is afforded
by the one loop correction to the gluon self energy due to a heavy quark
loop. This is
\be
\Pi_{\mu \nu}(k) = (k_{\mu}k_{\nu}-k^{2} \eta_{\mu \nu}) \Pi(k)
\ee
where
\be
\Pi(k) =\frac{2}{3} \frac{g^{2}}{16 \pi^{2}}(\frac{2}{\epsilon} - \gamma -
6\int_{0}^{1}dx \, x(1-x) \ln{ \left[ \frac{m_{Q}^{2}+ k^{2}x(1-x)}{4 \pi
 \mu^{2} }\right] } ) \, .
\ee
Once we subtract $\Pi(k^{2})-\Pi(\mu^{2})$ and substitute into diagrams,
the contributions are bound to be convergent (this is the essence of the
renormalisation programme), as the high $k$ behaviour is tempered. On the
other hand, for $k \ll m_{Q}$, the (subtracted) logarithm will behave like
\be
\frac{k^{2}}{m_{Q}^{2}}\,,
\ee
as in (\ref{est}).

Once these, and similarly subtracted gluonic three and four point
functions, are considered in the above expressions (with $F\leq 4$), one
obtains the estimates
\be
k^{4-F} \times O(\frac{k}{m_{Q}})\; {\textstyle or}\;
O(\frac{\mu}{m_{Q}}) \, ,
\ee
again establishing the suppression of such heavy quark contributions.

\vspace{1.5cm}
\noindent \underline{The Relationship between QCD and HQET$_{0}$}
\vspace{1cm}

We have thus shown that, in a particular (mass dependent) subtraction scheme,
the $n$-point light particle $1$PI diagrams $\Gamma_{n}(g,m_{Q},m,\mu)$ in
QCD and the $n$-point light particle $1$PI diagrams
$\Gamma_{n}^{*}(g^{*},m^{*},\mu)$ in the HQET$_{0}$ are related by the
following simple relationship which holds to order $1/m_{Q}$ and for $\mu \ll
m_{Q}$,
\be
\Gamma_{n}(g,m_{Q},m,\mu,k) = \Gamma_{n}^{*}(g^{*} ,m^{*},\mu,k) \, .
\label{rge1}
\ee
(Here we have reintroduced
all of the other light, relative to the heavy quark in question, particles).

The couplings which appear on the right hand side of this equation are
those that would be obtained in QCD with the heavy quark omitted. One
immediate and very important consequence of all of this is that the
beta function of QCD at scales below the heavy quark mass goes over to
the beta function of QCD without the heavy quark. This is the idea behind
grand unified theories.

There is a technical aside that should be made
and that is that Appelquist and Carazzone work in the Landau gauge. The
reason for this is that in any other covariant gauge the gauge parameter
itself needs to be renormalised. The theorem has been carried over to
other covariant gauges. We have already seen that the gauge $v.A=0$ can,
in this respect, be problematic.

\vspace{1cm}
\noindent\underline{Cautionary Remark:} It is worth emphasising that one
should not blindly apply the decoupling theorem. Here is a
counterexample. Consider the divergence of the axial current
$\bar{Q}\gamma_{5} \gamma_{\mu}Q$ at one loop. The anomaly is mass
independent and does not vanish no matter how heavy the heavy quark is.

When choosing the renormalistion prescription we do not wish to introduce
any extra $m_{Q}$ dependence into the renormalised couplings. It is thus
good policy to employ some $MS$ scheme for which the renormalisation
constants are mass independent. We can relate the $m_{Q}$ dependent
scheme of Appelquist and Carazzone to an $MS$ scheme in the following way.
Let the unrenormalised,
but dimensionaly regularised, $n$-point light
particle $1$PI diagrams be
$\Gamma_{U,n}(g_{0},m_{Q0},m_{0},\mu,\epsilon)$.
There relationship to the renormalised
$\Gamma_{i,n}(g_{i},m_{Qi},m_{i},\mu)$,
determined in two schemes $i=1,2$, (one might like to think of $i=1$ as the
mass dependent subtraction scheme used by Appelquist and Carazzone, while
$i=2$ is some $MS$ scheme) is
\be
\Gamma_{U,n}(g_{0},m_{Q0},m_{0},\mu,\epsilon) = {\cal
Z}_{i}^{n/2}\Gamma_{i,n} (g_{i},m_{Qi},m_{i},\mu) \, .
\ee
By construction, the $\Gamma_{i,n}(g_{i},m_{Qi},m_{i},\mu)$ are finite and
are related by the finite ratio
\be
\Gamma_{2,n} (g_{2},m_{Q2},m_{2},\mu)  = \left( \frac{{\cal Z}_{1}}{ {\cal
Z}_{2}}\right)^{n/2} \Gamma_{1,n} (g_{1},m_{Q1},m_{1},\mu) \, .
\ee
Similar formulae hold for the HQET$_{0}$ $n$-point functions as well
\be
\Gamma_{2,n}^{*} (g_{2}^{*},m_{2}^{*},\mu) = \left(\frac{ {\cal Z}_{1}^{*}}{
{\cal Z}_{2}^{*}}\right)^{n/2}
\Gamma_{2,n}^{*}
(g_{1}^{*},m_{1}^{*},\mu)\,.
\ee
Now starting from (\ref{rge1}) one can pass to an $MS$ scheme on both
sides to obtain (we have switched notation so that the $MS$ scheme
couplings are to be understood and we hope our laxness causes no
difficulties),
\be
\Gamma_{n}(g,m_{Q},m,\mu,k) = Z(g,m_{Q},\mu)^{n/2}
\Gamma_{n}^{*}(g^{*} ,m^{*},\mu,k)\,,
\label{rge}
\ee
for a certain function $Z(g,m_{Q},\mu)$ which is given in terms of the
finite renormalisation constants ${\cal Z}_{i}$ and ${\cal Z}_{i}^{*}$.

\vspace{1cm}
\noindent \underline{Exercise:} What is the relationship between
$Z(g,m_{Q},\mu)$ and the renormalisation constants ${\cal Z}_{i}$ and ${\cal
Z}_{i}^{*}$?

\subsubsection{Comparison of Green Functions with $2$ Heavy quark legs}

One would hope that a simple relationship like (\ref{rge}) holds also for
external heavy quark legs and indeed it is true. Here, however, we should
specify that the momentum running through the heavy quark line is
``almost'' on shell. That is we should take that momentum to be
$m_{Q}v_{\mu} +k_{\mu}$, where $k \ll m_{Q}$. This will ensure
that we can pass to the HQET$_{0}$ (we need this condition even at tree
level). Given the discusion above, we will take it for granted that
there is no need to take into account heavy quark loops and that the
renormalisation of the light part of the theory has been accomplished.
Again we should distinguish the case of convergent versus
divergent diagrams. Also, given that the HQET$_{0}$ is
renormalisable, it is apparent that we only need to renormalise the heavy
quark propagator and the heavy quark-heavy quark-gluon vertex, as we
must do also in QCD. The following argument, which gives us our sought
for relationship is a variant of that due to Feinberg \cite{f} and to
Grinstein \cite{grin}.

\vspace{1.5cm}
\noindent \underline{Convergent Diagrams}
\vspace{1cm}

Consider within QCD, Green functions $G_{2,n}$, with two external
heavy quark lines. Denote by $G_{2,n}^{*}$ the same Green functions
evaluated in HQET$_{0}$. To simplify life later on we take it for granted that
there is a projection of $(1 + \vslash)/2$ on these both in QCD and in
HQET$_{0}$. Set $n \geq 2$. The overall degree of divergence of
such Green functions is $(1-n) < 0$, so that, apart from subdiagrams
which are divergent, these Green functions would be convergent. Consider
any convergent diagram contributing to such a $G_{(2,n)}$.

According to \cite{f}, the integrals of such
diagrams are dominated by the region where the loop momenta satisfy $q_{i}
\le K$, where $K \ll m_{Q}$ is the typical momentum flowing through any of
the external gluon lines or $k$. If this is the case, then one can
substitute all the heavy quark QCD propagators with the corresponding
HQET$_{0}$ ones,
\be
\frac{1}{\pslash + \qslash -m_{Q}} \rightarrow  \frac{1}{\vslash v.(p+q)
-m_{Q}} \, . \label{psub}
\ee
The two diagrams will then agree up to subdominant contributions from
wich we may conclude that for such convergent diagrams
$G_{(2,n)}=G_{(2,n)}^{*}$.

As an example of the argument presented in \cite{f} consider the diagram
fig. 6. We wish to determine the behaviour of
\be
\frac{(1 + \vslash)}{2} \int d^{4}q\gamma_{\mu} \frac{1}{\pslash +
\qslash -m_{Q} } \gamma_{\nu}
\frac{1}{q^{2}} \frac{1}{(q+k_{1})^{2}} \frac{(1 + \vslash)}{2}  \, .
\label{intt}
\ee
Split the region of integration into a region where $q^{2}$ is less than
$K^{2 }$ and one greater than $K^{2}$, so that symbolically we have
\be
\int_{-\infty}^{\infty}d^{4}q = \int_{0}^{K}d^{4}q +
\int_{K}^{\infty}d^{4}q \, .
\ee
In the integral $\int_{0}^{K}$ the momentum $q$ is small relative to
$m_{Q}$ and may be ignored in the QCD quark propagator. Furthermore, up
to subdominant terms, the QCD propagator becomes the HQET$_{0}$
propagator. So
this part of the integration coincides in the two theories.

It is
easy to see that the integral $\int_{sm_{Q}}^{\infty}d^{4}q$, on the
otherhand,
is itself subdominant for $s \le 1$ and $sm_{Q} >K$. In this region one can
set $k=k_{1}=0$ in
(\ref{intt}) and change variables to $q \rightarrow m_{Q}q$, to find
\be
\int_{sm_{Q}}^{\infty} d^{4}q\gamma_{\mu} \frac{1}{\vslash m_{Q}+
\qslash -m_{Q}} \gamma_{\nu}
\frac{1}{q^{4}} \rightarrow \frac{1}{m_{Q}}\int_{s}^{\infty}
d^{4}q\gamma_{\mu}
\frac{1}{\vslash + \qslash -1} \gamma_{\nu} \frac{1}{q^{4}} \, .
\ee
This is clearly subdominant and shows us that the integration momenta
which lie somewhat above $K$ are irrelevant.

\vspace{1.5cm}
\noindent \underline{Divergent Diagrams}
\vspace{1cm}

What of $G_{2,n}$ for $n=0,1$? Focus on the one loop correction to the
three point functions $\Gamma_{(2,1)}$ and $\Gamma_{(2,1)}^{*}$. Within
dimensional regularisation the integrals are finite and so up to
$O(1/m_{Q})$ they agree. However, both integrals diverge as we approach
four dimensions and need counterterms. Unfortunately there is no
guarantee that the counterterms preserve the relationship. If we
differentiate with respect to the residual momentum or the external
gluon momentum, the differentiated diagrams are finite. Then the previous
arguments imply that
\bea
\frac{\partial}{\partial k_{\mu}} \Gamma_{(2,1)}& =&
\frac{\partial}{\partial  k_{\mu}} \Gamma_{(2,1)^{*}} + O(k/m_{Q},q/m_{Q})
\nonumber \\
\frac{\partial}{\partial q_{\mu}} \Gamma_{(2,1)} &=&
\frac{\partial}{\partial q_{\mu}} \Gamma_{(2,1)^{*}} +
O(k/m_{Q},q/m_{Q}) \, .
\eea
So, apart from counterterms, the two $\Gamma_{(2,1)}$ are in agreement. Now
the counterterms are of the form $a\Gamma_{(2,1)}^{0}$ and
$a^{*}\Gamma_{(2,1)}^{* \,0}$, where the superscript $0$ indicates the
tree diagram, and $a$ and $a^{*}$ are infinite constants. Now one
chooses $a$ and $a^{*}$ to ensure equality (up to subdominant terms).
However, such a choice is bound to be $m_{Q}$ dependent. So at this
point we have
\be
\Gamma_{(2,1)}^{R} = \Gamma_{(2,1)}^{*\,R}
\ee
where $R$ denotes the renormalised Green functions.

\vspace{1.5cm}
\noindent \underline{The Relationship Including Two Heavy Quark Lines}
\vspace{1cm}

We can pass from this $m_{Q}$ dependent renormalization scheme to some
mass independent scheme (as we saw above) by multiplying the Green
functions by some finite renormalisation constant, so that
\be
\Gamma_{(2,1)}^{R}(p,q;\mu) = C(m_{Q}/\mu,g_{s})Z(m_{Q},\mu,g_{s})
\Gamma_{(2,1)}^{*\,R}(k,q; \mu) \, .
\ee
After this long discourse we finally
arrive at the generalisation of (\ref{rge}) that we sought:
\be
\Gamma_{(2,n)}(g,m_{Q},m,\mu,k) =C(m_{Q}/\mu,g) Z(g,m_{Q},
\mu)^{n/2}
\Gamma_{(2,n)}^{*}(g^{*},m^{*},\mu,k)\,. \label{rge2}
\ee

The reader may be surprised that we have by-passed a discusion of the
heavy quark two point function altogether. The reason for this is that,
in deriving (\ref{rge2}), we only need to know about the wave function
renormalisation to get the multiplicative factors. These are fixed
between the two and three point functions by gauge invariance.
Notice also that we have an equation for $C(m_{Q}/\mu,g)$. That is, by
applying the renormalisation group both to the left and the right of
(\ref{rge2}), with $n=0$, (and ignoring beta function contributions) we obtain
\bea
\mu \frac{\partial C}{\partial \mu}& =& -2(\gamma_{Q}-\gamma^{*}_{Q})C
\nonumber \\
&=& \frac{g^{*2}}{4\pi^{2}}C  \, . \label{C}
\eea

\subsection{Matching}

The HQET is constructed to reproduce correctly the low energy behaviour
of QCD. That is, in any diagram where the external momenta are much
smaller than the mass of the heavy quark one can pass from QCD to HQET,
remembering to multiply by the relevant finite renormalisation
constants. This means, in particular, that we can hope that the mismatch
should not depend at all on very low energy physics, such as infrared
singularities, physical cuts etc. However, as we pass to lower and lower
energies perturbation theory becomes more and more unreliable and the
equality of QCD and HQET more and more difficult to ascertain. For
example there may be non-perturbative effects that `see' the quark
masses. These questions aside, it is apparent that there is a range of
energies for which the HQET and QCD will agree to the desired accuracy
in inverse powers of $m_{Q}$.

Mismatches between QCD and HQET certainly begin to set in as one
approaches scales of $m_{Q}$, for in this regime the decoupling
theorems are no longer applicable. In order to get agreement between QCD
and the HQET one `matches' them at $\mu = m_{Q}$. These matching
conditions manifested themselves in (\ref{rge}), where the extra
multiplicative renormalisation constants arose to take into account hard
gluon exchange, missed by the decoupling theorems. One extends this to
other operators. This means that if we have some operator $A$ and
calculate in QCD or in HQET$_{0}$ then the relationship is
\be
\langle A(m_{Q}) \rangle =
C_{0}(m_{Q},\mu) \langle A_{0}(\mu) \rangle_{0} +
\frac{C_{1}(m_{Q},\mu)}{2m_{Q}}\langle A_{1}(\mu)
\rangle_{0} + \dots\,. \label{ope}
\ee
The notation is that on the righthand side one is working within QCD,
while on the left hand side all expectation values are taken with
respect to HQET$_{0}$ and the subscript $j$ on the operators indicates
that they are the operators together with the field and
action corrections at order $(m_{Q})^{-j}$. The Wilson co-efficients
$C_{j}(m_{Q},\mu)$ are defined by this relation.

The idea that one should match operators in this way goes back to
Voloshin and Shifman \cite{vs}.See also \cite{pw}. The following example
is the one originally worked out in \cite{vs}.

\subsubsection{Matching for $\bar{q}\Gamma Q$}
Let us see that the Wilson coefficients are calculable in practice. We
write
\be
\langle  \bar{q} \Gamma \psi_{Q}  \rangle =
C_{0}^{\Gamma}(m_{Q},\mu)  \langle \bar{q} \Gamma Q
\rangle_{0} + \dots\,. \label{curr}
\ee
The matrix $\Gamma$ can be either $\gamma_{\mu}$ or $\gamma_{\mu}
\gamma_{5}$. We wish to get a handle on $C_{0}^{\Gamma}(m_{Q}/\mu,
g^{*}_{s})$. Act on both sides of (\ref{curr}) with $d/d\mu$, to obtain
\be
\mu\frac{dC_{0}^{\Gamma}}{d\mu}= (\gamma_{\Gamma}-\gamma_{\Gamma}^{*})
 C_{0}^{\Gamma}\,,
\ee
where $\gamma_{\Gamma}$ and $\gamma_{\Gamma}^{*}$ are the anomalous
dimensions of the currents $\bar{q} \Gamma \psi_{Q}$ and $\bar{q} \Gamma
Q$ respectively.

In QCD the operators $\bar{q} \Gamma \psi_{Q}$, with
massless quarks and $\Gamma=\gamma_{\mu}$, are conserved currents, and
are partially conserved currents when $\Gamma=\gamma_{\mu}\gamma_{5}$.
They have vanishing anomalous dimensions. The currents correspond to flavour
symmetry in QCD. Adding mass terms for the quarks breaks the symmetry
softly. This means that the ultraviolet properties of the theory are not
changed. In loop integrals the presence of masses certainly effects the
soft momentum region of integration, but they play no role in the
ultraviolet region. Consequently, the currents remain finite and at most
require a finite renormalisation. Thus, even if the quarks have mass,
the anomalous dimension is zero. We are left with
\be
\mu\frac{dC_{0}^{\Gamma}}{d\mu}= -\gamma_{\Gamma}^{*} C_{0}^{\Gamma}
 \, . \label{dif}
\ee

The reason that there may, nevertheless, be a non-zero anomalous dimension
for the effective theory operator $\bar{q}\Gamma Q$ is that in the HQET it
does not correspond to a (partially) conserved current. One can see this
by noting that the heavy quark kinetic term is totally different to the
usual light quark kinetic term and so there is no hope of a symmetry at this
point. Consequently it is not protected from renormalisation. In practice
such a renormalisation is required, as in QCD the currents exhibit a
logarithmic dependance on
$m_{Q}$ which diverges as $m_{Q} \rightarrow \infty$, that is it
diverges as we pass to the HQET.

The solution to (\ref{dif}) has been discussed in other lectures at this
school and is
\be
C_{0}^{\Gamma}(\mu,g^{*}(\mu))= \exp{\left( - \int_{\bar{g}^{*}
(\mu_{0})}^{g^{*}(\mu) }dg'\frac{\gamma_{\Gamma}^{*}(g')}{\beta(g')}
\right) } C_{0}^{\Gamma}(\mu_{0},\bar{g}^{*}(\mu_{0}))\,, \label{coef}
\ee
where $\bar{g}$ is the running coupling defined by
\be
\mu'\frac{d \bar{g}(\mu')}{d\mu'} = \beta(\bar{g}(\mu') )
\ee
and with initial condition
\be
\bar{g}(\mu)=g^{*} \, .
\ee
Take $\mu_{0}=m_{Q}$ in (\ref{coef}) to give
\be
C_{0}^{\Gamma}(m_{Q}/\mu,g^{*}(\mu))= \exp{\left( - \int_{\bar{g}^{*}
(m_{Q})}^{\bar{g}(\mu) }dg'\frac{\gamma_{\Gamma}^{*}(g')}{\beta(g')}
\right) } C_{0}^{\Gamma}(\mu_{0},\bar{g}^{*}(m_{Q})) \, . \label{coeff}
\ee
In order to determine the Wilson coefficient we need to know two things,
\bea
&(i)& \;\;\;\;  \gamma_{\Gamma}^{*}(g') \; \; \; \; {\textstyle and}
\nonumber \\
&(ii)& \;\;\;\; C_{0}^{\Gamma}\left( 1,\bar{g}(m_{Q}) \right)\,.
\eea
Both may be calculated perturbatively provided that $\mu$ and $m_{Q}$
are large enough to ensure that $\bar{g}(\mu)$ and $\bar{g}(m_{Q})$ are
small (i.e. $\Lambda \ll \mu \ll m_{Q}$).

\vspace{1.5cm}
\noindent \underline{$(i)$ The Anomalous Dimension $\gamma_{\Gamma}^{*}$}
\vspace{1cm}

We can determine the anomalous dimension of the current by
demanding that the renormalisation group equation
\be
\left( \frac{\partial}{\partial \mu} + \beta(g) \frac{\partial}{\partial
g} +  \gamma^{*}_{Q} + \gamma_{q} -
\gamma_{\Gamma}^{*}\right)\Gamma_{\Gamma }^{*} = 0 \label{rge3pt}
\ee
holds. We also know from (\ref{anomQ}) and (\ref{anomq}) that
\be
\gamma^{*}_{Q} = \frac{8}{3}\frac{g^{2}}{16 \pi^{2}}, \; \;
\gamma_{q}= -\frac{4}{3}\frac{g^{2}}{16 \pi^{2}}
\ee
and therefore we only need to determine the pole part of $\Gamma_{\Gamma
}^{*}$ to work out $\gamma_{\Gamma}^{*}$. The diagram of interest,
fig. 7, corresponds to
\be
-i \frac{4}{3}g^{*2} \mu^{\epsilon} \int \frac{d^{n}l}{(2\pi)^{n}}
\frac{\vslash (\lslash + \qslash) \Gamma}{l^{2}(l+q)^{2}v.(l+k)} \, .
\ee
One can set $k=0$ without engendering an infrared divergence, and as we
are only interested in the pole behaviour, we do so. The $q_{\mu}$ in the
numerator gives a convergent integral and therefore we ignore it.
Furthermore, the pole part of the $l_{\mu}$ integral must be proportional to
$v_{\mu}$, and thus it is
given by
\be
 -i \frac{4}{3}g^{*2} \mu^{\epsilon} \int \frac{d^{n}l}{(2\pi)^{n}}
\frac{v.l \Gamma}{l^{2}(l+q)^{2}v.l} = \frac{8}{3}\frac{g^{*2}}{16\pi^{2}}
\Gamma\, .  \frac{1}{\epsilon}\,.
\ee
We have then
\be
\Gamma_{\Gamma}^{*}= \left( 1 +
\frac{8}{3}\frac{g^{*2}}{16\pi^{2}}\ln{\mu} \right)\Gamma \, .
\ee
Plugging this into (\ref{rge3pt}) (and once more ignoring the $\beta$
term) we obtain
\be
\gamma_{\Gamma}^{*} = 4 \frac{g^{*2}}{16 \pi^{2}} \, .
\ee

\vspace{1.5cm}
\noindent \underline{$(ii)$ The Wilson Coefficient $C_{0}^{\Gamma}$}
\vspace{1cm}

At tree level, QCD and the HQET agree. So we have
\be
C_{0}^{\Gamma}(1,\bar{g}(m_{Q})) = 1 + O(\bar{g}(m_{Q})^{2})\,,
\ee
which will suffice for our needs. We only want to work at the level of
leading logarithms.

Recall that
\be
\beta(g) = g \left( -b_{0}\frac{g^{2}}{16 \pi^{2}} + \right. b_{1}
\left.\left( \frac{g^{2}}{16 \pi^{2}} \right)^{2} \right)\,,
\ee
with $b_{0}= 11- \frac{2}{3}n_{f}$, where $n_{f}$ is the number of
`active' quarks. Below the $b$-quark, but above the $c$-quark, threshold
we have $n_{f}= 4$ etc..
Putting all the threads together we can determine the exponent in
(\ref{coef})
\bea
& & -\int_{\bar{g}(m_{Q})}^{\bar{g}(\mu)}dg'\frac{g'^{2}}{16 \pi^{2}}[-b_{0}
 \frac{g'^{3}}{16 \pi^{2}}]^{-1} \nonumber \\
& =& \frac{4}{b_{0}} \int_{\bar{g}(m_{Q})}^{\bar{g}(\mu)} \frac{dg'}{g'}
\nonumber \\
&=& \frac{4}{b_{0}} \ln{ \frac{\bar{g}(\mu)}{\bar{g}(m_{Q})}   } \, .
\eea
We thus arrive at
\bea
C_{0}^{\Gamma} &=& \exp{ \left(\frac{4}{b_{0}} \ln{
\frac{\bar{g}(\mu)}{\bar{g} (m_{Q})}
} \right)  } \nonumber \\
& = & \left( \frac{\bar{\alpha}_{s}(m_{Q}) }{\bar{\alpha}_{s}(\mu)
}\right) ^{a}
\eea
where
\be
\bar{\alpha}_{s}= \frac{\bar{g}^{2}}{4\pi}
\ee
and
\be
a= -\frac{2}{b_{0}} \, .
\ee

If one wants more precision one can go to the next to leading
logarithms approximation. For this one would need to calculate $\gamma_{
 \Gamma}^{*}$ to the next order.

\vspace{1.5cm}
\noindent \underline{Exercise:} Determine C from (\ref{C}).

\appendix
\section{Symmetries of the Heavy Quark Action}
In this appendix we give the most general symmetries of the heavy quark
action.There are 16 symmetries, 8 ``vector" and 8 ``axial vector" types.
In order to exhibit this we need to introduce some notation. Suppose the
dimension of space-time is $d$. Pick a $\gamma$ matrix algebra for this
space and let the $\gamma$ matrices be $n\times n$ matrices. As a basis
for the $\gamma$ choose $\vslash$ and $\gamma^{\perp}$. Any $n\times n$
may be expanded in terms of a basis $\Gamma$ generated by the $\gamma$.
The basis we choose is such that $\Gamma=\Gamma_{+}\otimes \Gamma_{-}$
where
\be
\{\vslash,\Gamma_{+}\}=0
\ee
and
\be
[\vslash,\Gamma_{-}]=0\,,
\ee
depending solely on whether there are an odd or an even number of
$\gamma^{\perp}$ in the product used to define the given element of
$\Gamma$. Clearly $\Gamma_{+}$ and $\Gamma_{-}$ each have dimension
$n^{2}/2$. The symmetries of $\bar{\tilde{Q}}\vslash v.\partial\tilde{Q}$
are
\bea
\delta\tilde{Q}&=&(\alpha^{+}\Gamma_{+}+\alpha^{-}\Gamma_{-})%
\tilde{Q}\nonumber\\
\delta\bar{\tilde{Q}}&=&\bar{\tilde{Q}}(\alpha^{+}\Gamma_{+}-\alpha^{-}%
\Gamma_{-})
\eea
Notice that \underline{all} the $\Gamma_{+}$ symmetries are ``chiral",
though, obviously, none of them is anomalous.
\newpage
\section{Figure Captions}
\begin{description}
\item[Figure 1] Born diagram for the electron-proton potential
\item[Figure 2] Insertion of $O(1/m_{c})$ corrections in a $b \rightarrow c$
transition
\item[Figure 3] Insertion of $O(1/m_{c})$ corrections in a $c \rightarrow
c$ transition
\item[Figure 4] Gluon n-point function where all the external gluon lines
are connected to a fermion loop
\item[Figure 5] Gluon n-point function where none of the external gluon
lines land on the heavy quark loop
\item[Figure 6] Convergent four vertex integral
\item[Figure 7] 1-loop heavy light vertex correction
\end{description}


\begin{thebibliography}{99}
\bibitem{hmw}
 K. Hagiwara, A.D. Martin, and M.F. Wade, Phys. Lett. {\bf B228} (1989) 144;
 Nucl. Phys. {\bf B327} (1989) 569; M. Suzuki, Nucl. Phys. {\bf B258} (1985)
553.
\bibitem{hkt} F. Hussain, J.G. K\"orner and G. Thompson, Ann. Phys. {\bf 206}
(1991) 334.
\bibitem{iw232} N. Isgur and M.B. Wise, Phys. Lett. {\bf B232} (1989) 113.
\bibitem{vs} M.B. Voloshin and M.A. Shifman, Sov. J. Nucl.Phys. {\bf 45 (2)}
(1987) 292; ibid. {\bf 47 (3)} (1988) 511.
\bibitem{eh} E. Eichten and B. Hill, Phys. Lett. {\bf B234} (1990) 511.
\bibitem{f} F.L. Feinberg, Phys. Rev. {\bf D17} (1978) 2659.
\bibitem{cl} W.E. Caswell and G.P. Lepage, Phys. Lett. {\bf B167} (1986) 437.
\bibitem{g}  H. Georgi, Phys. Lett. {\bf B240} (1990) 447.
\bibitem{iw66} N. Isgur and M.B. Wise, Phys. Rev. Lett. {\bf 66} (1991) 1130.
\bibitem{iw43} N. Isgur and M.B. Wise, Phys. Rev. {\bf D43} (1991) 819.
\bibitem{iw237} N. Isgur and M.B. Wise, Phys. Lett. {\bf B237} (1990) 527.
\bibitem{falk} A.F. Falk, H. Georgi, B. Grinstein and M.B. Wise,
Nucl. Phys. {\bf B343} (1990) 1.
\bibitem{mannel0} T. Mannel, W. Roberts and Z. Ryzak, Harvard preprint
HUTP-91/A017 (1991).
\bibitem{mannel1} T. Mannel, W. Roberts and Z. Ryzak,  Phys. Lett. {\bf B254}
(1991)
274.
\bibitem{mannel2} T. Mannel, W. Roberts and Z. Ryzak,  Phys. Lett. {\bf B259}
(1991)
485.
\bibitem{mannel3} T. Mannel, W. Roberts and Z. Ryzak, Nucl. Phys. {\bf B355}
(1991) 38.
\bibitem{bb1} C.G. Boyd and D.E. Brahm, Phys. Lett. {\bf B254} (1991) 468.
\bibitem{g1} H. Georgi, Nucl. Phys. {\bf B384} (1991) 293.
\bibitem{grin} B. Grinstein, Nucl. Phys. {\bf B339} (1990) 253.
\bibitem{iw348} N. Isgur and M.B. Wise, Nucl Phys. {\bf 348} (1991) 276.
\bibitem{l} M.E. Luke, Phys. Lett. {\bf B252} (1990) 447.
\bibitem{bb2} C.G. Boyd and D.E. Brahm, Phys. Lett. {\bf B257} (1991) 391.
\bibitem{kt1} J.G. K\"orner and G. Thompson, Phys. Lett. {\bf B264} (1991)
185.
\bibitem{kt2} J.G. K\"orner and G. Thompson, Mainz preprint,
MZ-TH/91-35.
\bibitem{hkstw} F. Hussain, J.G. K\"orner, K. Schilcher, G. Thompson and
Y.L. Wu, Phys. Lett. {\bf B249} (1990) 295.
\bibitem{hlkkt} F. Hussain, D. Liu, M. Kr\"amer, J.G. K\"orner and S.
Tawfiq, Nucl. Phys. {\bf B370} (1992) 259.
\bibitem{arbsp} F. Hussain, J.G. K\"orner and G. Thompson, ICTP preprint
IC/93/314 and MZ-TH/93-23.
\bibitem{hkkt} F. Hussain, J.G. K\"orner, M. Kr\"amer and G. Thompson,
Z. Phys. {\bf C51} (1991) 321.
\bibitem{Neub} M. Neubert, preprint CERN-TH.7225/94, lectures presented
at TASI-93, Boulder, Colorado, 1993.
\bibitem{grinrev} B. Grinstein, {\em Lectures on Heavy Quark Effective
Theory}, preprint HUTP-91/A040, SSCL-Preprint-17, to be published in
``High Energy Phenomenology", Proceedings of the Workshop, Mexico City,
1-12 July 1991, R. Huerta and M.A. P\'{e}rez, eds. World Scientific
Publishing Co., Singapore.
\bibitem{bn} F. Bloch and A. Nordsieck, Phys. Rev. {\bf 52} (1937) 54.
\bibitem{bs} N. Bogoliubov and D.V. Shirkov, {\bf Introduction to the Theory
of Quantised Fields}, Interscience Publishers, John Wiley and Sons, Inc.
(N.Y., 1959).
\bibitem{ef} E. Eichten and F.L. Feinberg, Phys. Rev. {\bf D23} (1981) 2724.
\bibitem{lu} D. Lurie, {\bf Particles and Fields}, Interscience Publishers,
John Wiley and Sons, Inc. (N. Y., 1968).
\bibitem{ht} F. Hussain and G. Thompson, Phys. Lett. {\bf B335} (1994) 205.
\bibitem{wsb} M. Wirbel, B. Stech and M. Bauer, Z. Phys. {\bf C29} (1985) 637.
\bibitem{FGL} A. Falk, B. Grinstein and M. Luke, Nucl. Phys. {\bf B357}
(1991) 185.
\bibitem{ew}E. Witten, Nucl. Phys. {\bf B104} (1976) 445.
\bibitem{pol} A. Polyakov, Nucl. Phys. {\bf B164} (1980) 171.
\bibitem{dv} V. Dotsenko and S. Vergeles, Nucl. Phys. {\bf B169} (1980)
527.
\bibitem{bns} R. Brandt, F. Neri and M. Sato, Phys. Rev. {\bf D24}
(1981) 879.
\bibitem{dorn} H. Dorn, Fortschr. Phys. {\bf 34} (1986) 11.
\bibitem{kr} G. Korchemsky and A. Radyushkin, Phys. Lett. {\bf B279}
(1992) 359.
\bibitem{ac} T. Appelquist and J. Carazzone, Phys. Rev. {\bf D11} (1975)
2856.
\bibitem{s} K. Symanzik, Commun. Math. Phys. {\bf 34} (1973) 7.
\bibitem{c} J. Collins, {\bf Renormalization}, C.U.P. Cambridge (1984).
\bibitem{pw} H.D. Politzer and M.B. Wise, Phys. Lett. {\bf B206} (1988)
681; Phys. Lett. {\bf B208} (1988) 504.



\end{thebibliography}
\end{document}